\documentclass[12pt]{JHEP3}
\usepackage{graphicx}
\usepackage{dcolumn}
\usepackage{bbm}
\usepackage{graphics}
\usepackage{amssymb}
\usepackage{amsmath}
\usepackage{url}

\newcommand{\ba}{\begin{eqnarray}}
\newcommand{\ea}{\end{eqnarray}}
\newcommand{\be}{\begin{equation}}
\newcommand{\ee}{\end{equation}}
\newcommand{\bea}{\begin{eqnarray}}
\newcommand{\eea}{\end{eqnarray}}
\newcommand{\beq}{\begin{equation}}
\newcommand{\eeq}{\end{equation}}
\newcommand{\beqar}{\begin{eqnarray}}
\newcommand{\eeqar}{\end{eqnarray}}
\newcommand{\beqars}{\begin{eqnarray*}}
\newcommand{\eeqars}{\end{eqnarray*}}
\newcommand{\bc}{\begin{center}}
\newcommand{\ec}{\end{center}}
\newcommand{\ben}{\begin{enumerate}}
\newcommand{\een}{\end{enumerate}}
\newcommand{\bit}{\begin{itemize}}
\newcommand{\eit}{\end{itemize}}
\newcommand{\bw}{\begin{widetext}}
\newcommand{\ew}{\end{widetext}}

\newcommand{\dd}{\mbox{d}}
\newcommand{\p}{\mrm{p}}
\newcommand{\lsim}{\lesssim}
\newcommand{\gsim}{\gtrsim}

\newcommand{\ie}{\emph{i.e.~}}

\newcommand{\eg}{\emph{e.g.~}}

\newcommand{\ab}{a_{\mathrm{b}}}

\newcommand{\mP}{m_{_\mathrm{Pl}}}

\newcommand{\lP}{\ell_{_\mathrm{Pl}}}
\newcommand{\mrm}[1]{\mathrm{#1}}
\newcommand{\mcl}[1]{\mathcal{#1}}
\newcommand{\mfk}[1]{\mathfrak{#1}}

\newcommand{\lb}{\left(}
\newcommand{\rb}{\right)}
\newcommand{\lsb}{\left[}
\newcommand{\rsb}{\right]}
\newcommand{\lcb}{\left\{}
\newcommand{\rcb}{\right\}}
\newcommand{\real}{\mathbbm{R}}

\newcommand{\entier}{\mathbbm{N}}

\newcommand{\varphib}{\varphi_{\mrm{b}}}

\newcommand{\mpl}{m_{_{\mrm{Pl}}}}

\newcommand{\order}[1]{\mathcal{O}\left(#1\right)}

\newcommand{\Posc}{\mcl{P}_{\zeta}^{\mrm{osc}}}
\newcommand{\nuT}{\nu_{_{\mrm T}}}

\newcommand{\PT}{\mcl P_{_{\mrm{T}}}}

\title{Observational signatures of a non-singular bouncing cosmology}

\author{Marc Lilley\\APC, Universit\'e Paris 7, 10 rue Alice Domon et L\'eonie Duquet, 75205 Paris Cedex 13, France\\ E-mail: \email{lilley@apc.univ-paris7.fr}}

\author{Larissa Lorenz \\ Institute of Mathematics and Physics, Centre for Cosmology,
Particle Physics and Phenomenology, Louvain University,
2 Chemin du Cyclotron, 1348 Louvain-la-Neuve, Belgium \\ E-mail: \email{larissa.lorenz@uclouvain.be}}

\author{S\'ebastien Clesse \\ Service de Physique Th\'eorique, Universit\'e Libre de Bruxelles, CP225, Boulevard du Triomphe,
1050 Brussels, Belgium \\ and \\ Institute of Mathematics and Physics, Centre for Cosmology,
Particle Physics and Phenomenology, Louvain University,
2 Chemin du Cyclotron, 1348 Louvain-la-Neuve, Belgium\\ E-mail: \email{seclesse@ulb.ac.be}}

\abstract{We study a cosmological scenario in which inflation is preceded by a bounce. In this scenario, the primordial singularity, one of the major shortcomings of inflation, is replaced by a non-singular bounce, prior to which the universe undergoes a phase of contraction. Our starting point is the bouncing cosmology investigated in Falciano \emph{et al.} (2008), which we complete by a detailed study of the transfer of cosmological perturbations through the bounce and a discussion of possible observational effects of bouncing cosmologies.  We focus on a symmetric bounce and compute the evolution of cosmological perturbations during the contracting, bouncing and inflationary phases.   We derive an expression for the Mukhanov-Sasaki perturbation variable at the onset of the inflationary phase that follows the bounce.  Rather than being in the Bunch-Davies vacuum, it is found to be in an excited state that depends on the time scale of the bounce.  We then show that this induces oscillations superimposed on the nearly scale-invariant primordial spectra for scalar and tensor perturbations.   We discuss the effects of these oscillations in the cosmic microwave background and in the matter power spectrum.  We propose a new way to indirectly measure the spatial curvature energy density parameter $\Omega_{\mcl K}$ in the context of this model.} 
\keywords{cosmology, bounce, inflation, CMB}

\begin{document}
\section{Introduction}
The flatness, homogeneity, isotropy and monopole problems of  standard Big Bang cosmology are easily resolved by invoking a primordial epoch of inflation~\cite{Guth:1980zm,Mukhanov:1981xt,Linde:1981mu,Linde:1983gd}. This framework also explains the origin of today's cosmic microwave background (CMB) anisotropies and large scale structure with compelling naturalness and ease.  It traces these anisotropies back to the quantum fluctuations of the scalar field responsible for inflation. As the universe grows exponentially, these perturbations are stretched outside the Hubble radius at early times, become classical, source the temperature anisotropies of the CMB and provide the seeds for structure formation. In addition, all pre-existing super-Hubble fluctuations are conveniently stretched outside the Hubble radius today and can be safely ignored.  The statistical properties of the CMB and of the large scale structure of the universe in the observable range are then directly related to the initial quantum state of sub-Hubble primordial fluctuations at the onset of inflation.  The initial quantum state is unambiguously defined since the wavelength of all perturbations observable today was much shorter than the Hubble scale at the onset of inflation so that perturbations start their evolution in the adiabatic vacuum.  With this initial state prescription, inflation yields a primordial spectrum of density fluctuations, which, when evolved through the radiation and matter dominated epochs of standard cosmology, yields a CMB temperature anisotropy angular power spectrum and a matter power spectrum compatible with observations~\cite{Komatsu:2010fb}, lending strong support to the inflationary hypothesis, and any model of the primordial universe lacking this unambiguous initial state prescription generally suffers from a lack of predictability.

Inflation however raises two new issues and leaves one unsolved. Firstly, the identity and origin of the scalar field that drives inflation remain, for lack of candidates within the Standard Model (SM) of particle physics, entirely unknown despite the fact that, in recent years, significant progress has been made by implementing inflation into a high energy framework beyond the SM.  For example, in string theory, although scalar degrees of freedom abound, the derivation of their potentials remains challenging and few candidate fields are suitable for inflation without (a more or less significant amount of) finetuning.

Secondly, the inflationary mechanism used to generate classical perturbations from the stretching and amplification of quantum fluctuations leads to a conceptual issue of self-consistency. As spacetime expands during inflation, the physical wavelength of any given perturbation increases proportionally to the scale factor of the universe. Conversely, if we follow a perturbation backwards in time, its wavelength becomes shorter, possibly decreasing below the Planck length $\lP$ if the duration of inflation exceeds some 60 to 70 $e$-folds.  However, close to the Planck scale, quantum gravity corrections become relevant.  This is the well-known {\it trans-Planckian problem} of inflation. It has been studied both theoretically and numerically~\cite{Martin:2003kp,Martin:2003sg,Martin:2004iv,Martin:2004yi,Martin:1999fa,Sriramkumar:2004pj} by introducing modified dispersion relations or imposing non-vacuum initial conditions on the perturbations or by introducing a new cut-off scale of order of the Planck mass in the inflationary mechanism.  Although they remain phenomenological, these models generically predict oscillations superimposed onto the standard (approximately scale invariant) spectrum of primordial  fluctuations. In Refs.~\cite{Martin:2003sg,Martin:2004iv} it was shown that these oscillations then carry through to the CMB temperature anisotropy and the matter density spectra.

Thirdly, the usual models of inflation, even in the context of eternal inflation~\cite{Borde:1994ai,Borde:2001nh}, are not past complete, and inevitably contain a time-like singularity in the past at which the universe is of vanishing size.  On the contrary, higher order curvature terms in the Einstein-Hilbert action~\cite{Utiyama:1962sn,Utiyama:1962fs,Nariai:1971st,Fulling:1974zr,Macrae:1981ic,Barrow:1983rx,Wands:1993uu,Carloni:2005ii}, non-minimal coupling of matter fields to gravity~\cite{Gunzig:2000kk,Saa:2000ik,Abramo:2002rn,Carloni:2005ii} and the low energy effective actions of some string theories~\cite{Antoniadis:1993jc,Easther:1996yd,Brustein:1997cv,Cartier:1999vk,Foffa:1999dv,Fabris:2002pm,Tsujikawa:2003pn,KOST01,Khoury:2001zk,Kachru:2002kx,Easson:2007fz} do allow \emph{non-singular} cosmological background solutions.  In these, instead of shrinking to zero size, the universe experiences a bounce: starting out large, the universe undergoes a contracting phase until it reaches a finite minimal size, after which an expanding phase occurs.  The existence of bouncing solutions in these high energy effective theories warrants the study of simpler {\it classical} non-singular cosmological models, in which background and perturbations are completely understood and tractable. This simpler class includes models with minimally coupled scalar fields~\cite{Falciano:2008gt,Starobinsky:1980te,Martin:2003sf,Cartier:2003jz}, scalar fields with non-standard kinetic terms~\cite{Abramo:2007mp} or in geometries that depart from the Friedmann-Lema\^itre-Robertson-Walker (FLRW) geometry, \ie Bianchi or Kantowski-Sachs spacetimes~\cite{Solomons:2001ef}. Another possibility is to consider unconventional forms of matter~\cite{Peter:2001fy,Peter:2003rg,Cai:2007qw,Cai:2007zv}, or a combination of radiation and scalar field matter~\cite{Peter:2002cn}. For an extensive review, see \eg Refs.~\cite{Novello:2008ra,Lilley09}.

Although bouncing cosmologies have most often been discussed as alternatives to inflation, see \eg Refs.~\cite{Gasperini:2002bn,Shtanov:2002mb,KOST01,Kallosh:2001ai}, here we shall see both the contracting and the non-singular bouncing phase as cosmological epochs that connect to inflation. In this paper, the cosmological singularity is replaced by a classical bounce which can entirely be described within General Relativity (GR), and after which inflation takes place as usual.
Moreover, we will mostly focus, for computational simplicity, on a \emph{symmetric} bounce, meaning that the rate of contraction (up to a minus sign) and its duration are the same as in the inflationary phase that follows. The transition to standard Big Bang cosmology after inflation through a phase of reheating remains unchanged from the standard case without bounce.

To determine the viability of a non-singular cosmology, it is important to investigate the stability and attractor properties of the background cosmology and the necessary level of fine-tuning required in order to obtain a bounce~\cite{Falciano:2008gt,Starobinsky:1980te,Barrow:1980en}.  At the perturbative level, it is important to understand the transfer of cosmological perturbations through the bouncing phase, whether there can exist a (robust) prescription to choose the initial conditions for cosmological perturbations at some stage in their evolution, and finally, whether such bouncing models have distinctive observational signatures.

In this paper, we focus on the latter three issues, using the background cosmology of Ref.~\cite{Falciano:2008gt}. This model relies on a dynamical scalar field with a symmetric potential of the small field type and a positive spatial curvature term.  This is the minimal possible setup in order to obtain a bounce but violate the least number of energy conditions~\cite{Falciano:2008gt,MolinaParis:1998tx}. 

This paper is organized as follows. We first give a detailed overview of the background cosmology in Sec.~\ref{sec:background}, and an introductory discussion of the evolution of cosmological perturbations through the bouncing phase in Sec.~\ref{sub:preliminaries}.   We then derive, in Sec.~\ref{sub:uevol}, the expression of the Bardeen variable at the onset of inflation at first order in slow roll by solving its mode equation prior to and after the bounce and by then applying the standard matching procedure at the bounce.  We compute the corresponding form of the initial state of the mode functions of the Mukhanov-Sasaki variable describing scalar perturbations at the onset of inflation (Secs~\ref{sub:utov} and \ref{sub:initialstate}).  This initial state is an excited state rather than the vacuum state, and it is shown that it can result in the presence of oscillations at either {\it leading} or {\it subleading} order in the primordial spectrum of scalar perturbations, see Sec.~\ref{sub:primordialspectrum}.   Similar results are obtained for tensor perturbations in Sec.~\ref{sec:tensors}.  While the amplitude of the oscillations depends only on the initial state of the cosmological perturbations prior to the bounce, the frequency of oscillation is given by a new time scale, which is discussed in Sec.~\ref{sec:bouncingtimescale}.  In Sec.~\ref{sec:bouncingtimescale}, we also discuss a new possibility to indirectly measure the spatial curvature energy density parameter $\Omega_{\mcl K}$ in the context of the model proposed.  Finally, in Sec.~\ref{sec:obs}, we write down the form of the angular spectrum of CMB temperature anisotropies at low multipoles, commenting on the appearance of slow oscillatory features on large angular scales, and also provide numerical examples, using the \texttt{CAMB} code~\cite{Lewis:1999bs}, of the angular power spectra of CMB temperature anisotropies and the spectrum of matter density fluctuations. In Appendix \ref{sec:zetaevol}, the evolution of the curvature perturbation in the contracting phase is discussed in some more detail, and some numerical examples of symmetric and asymmetric bouncing cosmologies are provided.
\section{Background cosmology}
\label{sec:background}
\subsection{Preliminaries}
We consider a homogeneous and isotropic FLRW universe with line element
\be
\dd s^2=a^2\lb\eta\rb\lsb -\dd\eta^2+\frac{\dd r^2}{1-\mcl Kr^2}+r^2\lb \dd\theta^2+\sin^2\theta\dd\phi^2\rb\rsb\,,
\label{eq:flrw}
\ee
where $a(\eta)$ is the scale factor as a function of conformal time $\eta$ (where $\dd\eta=\dd t/a$, with $t$ the cosmic time). The constant parameter $\mcl K$ describes the curvature of spatial sections and can always be rescaled such that $\mcl K=0$, $\pm 1$.  In standard inflation, one usually sets $\mcl K=0$. For the non-singular model studied in this paper, however, we shall require $\mcl K=+1$ (see below).  For a perfect fluid, with equation of state $P=w\rho$, where $\rho$ and $P$ and $w$ are the energy density, pressure and equation of state parameter, respectively, the Friedmann equations derived from the Einstein equations and the conservation of the stress energy tensor read
\bea
\kappa\lb \rho+P\rb&=&\frac{2}{a^2}\lb \mcl H^2-\mcl H'+\mcl K\rb\,,
\label{eq:f1}\\
\kappa\lb\rho+3P\rb&=&-\frac{6}{a^2}\mcl H'\,,
\label{eq:f2}\\
\rho'+3\mcl H\lb \rho+P\rb&=&0\,,
\label{eq:ce}
\eea
where $\kappa\equiv8\pi/\mpl^{2}$ with $\mpl$ the Planck mass, and where $\mcl H\equiv a'/a$ is the (conformal) Hubble parameter, with a prime denoting a derivative with respect to $\eta$.

Consider a bounce occuring at conformal time $\eta=0$ (or equivalently $t=0$).  At the bounce, the scale factor $a$ reaches its (finite) minimum, $\ab$, while the Hubble parameter evaluated at the bounce, $\mcl H_{\mrm b}$, is equal to zero and its derivative with respect to conformal time, $\mcl H'_{\mrm b}$, is positive.  For a perfect fluid, the null and strong energy conditions derived from the Penrose-Hawking singularity theorems \cite{Hawking:1969sw} read
\be\label{eq:energyconditions}
\rho+P\geq 0\,,\qquad\rho+3P\geq 0\,.
\ee
It can be seen from Eq.~(\ref{eq:f1}) that the null energy condition is preserved at the bounce only if $\mcl K>\mcl H'_{\mrm{b}}$.  Recast in terms of cosmic time, this inequality reads $\mcl K/a^2>\dot H_{\mrm{b}}$ where $\dot H_{\mrm{b}}$ is the derivative of the (cosmic) Hubble parameter with respect to $t$ evaluated at the bounce.  Hereon, we set $\mcl K=+1$ and this condition turns into a condition on the combination $\ab^2\dot H_{\mrm{b}}$ [see Eq.~(\ref{eq:ab_and_M})].  From Eq.~(\ref{eq:f2}), one finds that the strong energy condition must necessarily be violated in order to obtain a bounce.   Finally, note that the spatial curvature term $\mcl K/a^{2}$ dominates the dynamics at the time of the bounce.  An inflationary phase is therefore required afterwards in order to drive the spatial curvature contribution to the energy content of the universe down to a level compatible with observations~\cite{Dunkley:2008ie}.
\subsection{The de Sitter cosmological bounce}\label{sub:desitterbounce}
Let us first consider the de Sitter solution.  If $w=-1$, the energy content of the universe is in the form of a cosmological constant.  For $\mcl K=0$, one has
\be
a\lb t\rb\propto e^{Ht},
\label{eq:aflat}
\ee
where the (cosmic) Hubble parameter $H\equiv\dot{a}/a$ is a positive constant so that the scale factor of the universe increases monotonically.  Re-expressing Eq.~(\ref{eq:aflat}) using the conformal time coordinate, one has
\be
a(\eta)=-\frac{1}{H\eta}\qquad\text{and}\qquad\mcl H=aH=-\frac{1}{\eta}\,.
\ee
For $t\rightarrow \infty$, $a(t)\rightarrow \infty$ so that $\eta\rightarrow 0$ while if, formally, one lets the scale factor become arbitrarily small at early times, then $\eta\rightarrow -\infty$ in the past.  In this case, one thus has $\eta\,\in\,\, ]-\infty,0\,[$.  On the other hand, in a $\mcl K=+1$ de Sitter universe, one has
\be
a(t)=\ab\cosh\lb\frac{t}{\ab}\rb\qquad\text{and}\qquad H(t)=\frac{1}{\ab}\tanh\lb\frac{t}{\ab}\rb\,,
\label{eq:adSt}
\ee
For $t<0$, one then has $H<0$ and the universe contracts exponentially. At $t=0$, the {\it dimensionful} scale factor $a(t)$ reaches its minimum $\ab$, and the Hubble parameter vanishes.  For $t>0$, on the other hand, $H>0$ and the universe inflates. Re-written in conformal time, one finds
\be
a(\eta)=\ab\sec\lb\eta\rb\qquad\text{and}\qquad \mcl H(\eta)=\tan \lb\eta\rb\,,
\label{eq:adSeta}
\ee
so that the range of $\eta$ for a de Sitter bounce is $\eta\in\,\,]-\pi/2,\pi/2\,[$.
\subsection{Scalar field-driven cosmological bounce}\label{subsec:quasidS}
If instead of a simple cosmological constant with $w=-1$, the energy content of the universe is dominated by a scalar field $\varphi$ with energy density and pressure
\be\label{eq:rhoandp}
\rho=\frac{\varphi'^2}{2a^2}+V(\varphi)\qquad\textnormal{and}\qquad P=\frac{\varphi'^2}{2a^2}-V\lb\varphi\rb\,,
\ee
it follows from Eqs~(\ref{eq:f1}) and (\ref{eq:f2}) that
\be
\kappa\lb\rho+P\rb=\kappa\frac{\varphi'^2}{a^2}=\frac{2}{a^2}\lb \mcl H^2-\mcl H'+\mcl K\rb\,,
\label{eq:F1}\\
\ee
and
\be
\kappa\lb\rho+3P\rb=-\frac{6}{a^2}\mcl H'=-2\kappa V\lb\varphi\rb\lsb 1-\frac{\varphi'^2}{a^2V\lb\varphi\rb}\rsb\,,
\label{eq:F2}
\ee
while from Eq.~(\ref{eq:ce}) one has
\be
\varphi''+2\mcl H\varphi'+a^2\frac{\dd V}{\dd\varphi}=0\,.
\label{eq:KG}
\ee
As stated earlier, at the bounce, $\mcl H_{\mrm b}=0$ and $\mcl H'_{\mrm b}\geq 0$ and if $\mcl K>\mcl H_{\mrm{b}}'$, the null energy condition is verified.  This also implies $\varphi'\in\real$ so that ghost fields are avoided.  Again, the strong energy condition is not preserved at the bounce and implies $\ab^2V\lb\varphi_{\mrm{b}}\rb>\varphi_{\mrm{b}}'^2$ in Eq.~(\ref{eq:F2}).  If, as required in order to satisfy the observed flatness of spatial sections today, an inflationary phase lasting $N_{\mrm{inf}}=60$ to 80 $e$-folds is to take place after the bouncing phase, one requires the usual slow roll conditions,
\be
a^2V\lb \varphi\rb\gg\varphi'^2,\qquad \mcl H\varphi'\gg\varphi''\,,
\label{eq:conformalinfconds}
\ee
to be verified for $N_{\mrm{inf}}$ $e$-folds.  It is therefore seen that the strong energy condition is consistently violated throughout both the bouncing and the inflationary phase.  If in addition, the cosmology is taken to be symmetric or close to symmetric, then a deflationary phase prior to the bounce is required.  The strong energy condition is then similarly violated in part of the contracting phase as well.

The cosmological evolution of the early universe is conveniently described in terms of a set of {\it horizon flow functions} defined by~\cite{Schwarz:2001vv}
\be
\epsilon_1=1-\frac{\mcl H'}{\mcl H^2}\,,\qquad \epsilon_{i+1}=\frac{\dd \ln |\epsilon_i|}{\dd N}\quad (i\geq 1)\,,
\label{eq:hff}
\ee
which during inflation are slowly varying functions of time with $\epsilon_i\ll 1$ for all $i$.  The first three horizon flow functions are related to the usual slow roll parameters $\epsilon$, $\delta$ and $\xi$ by
\be
\epsilon=\epsilon_1,\qquad\delta=\epsilon_1-\frac{1}{2}\epsilon_2,\qquad \xi=\frac{1}{2}\epsilon_2\epsilon_3\,,
\label{eq:defsrparams}
\ee
so that the functions $\epsilon_1$ and $\epsilon_2$ correspond to the inequalities in Eqs.~(\ref{eq:conformalinfconds}) while the third is related to the time derivative of $\epsilon$ and $\delta$ through its definition in terms of $\xi$.  The field $\varphi$ is in the slow-roll regime as long as $\epsilon_{i}\ll 1$ for all $i$ and inflation ends when $\epsilon_{1}\simeq 1$.

In the remainder of the paper, we shall work at first order in slow-roll, meaning that the $\epsilon_{i}$ parameters \emph{themselves} are considered constant and terms of $\order{\epsilon_{i}^{2}}$ or higher are dropped.  In the context of a symmetric bounce, the slow-roll contracting (deflationary) phase is obtained by imposing the same conditions on the $\epsilon_{i}$'s but with $\mcl H<0$.

In the standard inflationary picture, the horizon flow functions express the deviation from a pure de Sitter expansion in regions where the spatial curvature term is negligible.  Here in fact, {\it both} the deflationary and inflationary background evolutions away from the bouncing phase itself can be fully described in terms of the $\epsilon_i$'s.  In the bouncing phase itself, where spatial curvature is important, the horizon flow functions are ill-defined\footnote{In fact, regardless of spatial curvature, in a bouncing cosmology with a bounce occuring at say $t=0$, the horizon flow functions are well-defined only in the separate contracting ($t<0$) and expanding ($t>0$) domains but not globally since $\mcl H(N)$ is not bijective throughout and since $N$ diverges at the bounce.}, but it is also possible to write down a generalized form of the scale factor describing deviations from a purely de Sitter bounce~\cite{Martin:2003sf}.  To describe the scale factor of a quasi de Sitter bouncing universe, we simply rewrite Eq.~(\ref{eq:adSeta}) in the generalized form
\be
a=\ab\cosh\lb \omega t\rb,
\ee
where $\omega$ is some dimensionful scale characterizing the deviation from a de Sitter bounce for which $\omega=1/\ab$.  In conformal time, one has 
\be
a\lb\eta\rb=\ab\sec\lb\frac{\eta}{\eta_{\mrm{c}}}\rb,
\label{eq:aqdS}
\ee
where $\eta_{\mrm{c}}=1/(\omega \ab)$ is a {\it dimensionless} conformal time scale.  For a {\it symmetric} bounce (where contracting and expanding phases last for the same number of $e$-foldings $N$) a Taylor expansion of Eqs.~(\ref{eq:F1}--\ref{eq:KG}) and (\ref{eq:aqdS}) around $\eta=0$ yields~\cite{Martin:2003sf,Falciano:2008gt}
\beq
\omega^2=\kappa V(\varphi_{\mrm{b}})-\frac{2\mcl K}{\ab^2}\,,
\label{eq:omega2}
\eeq
while the first Friedmann equation gives
\beq
\varphi_{\mrm{b}}'^2=\frac{6\mcl K}{\kappa}-2\ab^2V(\varphi_{\mrm{b}})\,,
\label{eq:varphib}
\ee
where everywhere in these expressions the subscript ``b'' denotes quantities evaluated at the bounce. Eqs.~(\ref{eq:omega2}) and (\ref{eq:varphib}) imply that
\be
\omega^2=\frac{\mcl K}{\ab^2}-\frac{\kappa}{2}\frac{\varphi_{\mrm{b}}'^2}{\ab^2}=\frac{\mcl H_{\mrm{b}}'}{\ab^2}\,.
\ee
If $\varphi_{\mrm{b}}'=0$, $\omega=1/\ab$ and the de Sitter solution is recovered.  The null energy condition (\ref{eq:energyconditions}) is preserved provided $\varphi_{\mrm{b}}'>0$, so that $\omega < \ab^{-1}$.

Let us discard the possibility that a bounce occurs when $\varphi_{\mrm{b}}'=0$, in which case, for any bounce, the potential for the scalar perturbations diverges at the bounce, see Eq.~(\ref{eq:uevol}) and Eq.~(\ref{eq:upot}).  Then, once again under the assumption of a symmetric bounce, a bouncing solution is obtained only if~\cite{Martin:2003sf,Falciano:2008gt}
\be
\varphi_{\mrm{b}}=0\,,\qquad\varphib''=0\,,\qquad V(\varphi_{\mrm{b}})=V_0\,,\qquad\left.\frac{\dd V}{\dd\varphi}\right|_{\mrm{b}}=0\,,\qquad\left.\frac{\dd^2 V}{\dd\varphi^2}\right|_{\mrm{b}}\le 0\,.
\label{eq:bouncereqs}
\ee
Note that these conditions can be expected to hold approximately for a (slightly)  \emph{asymmetric} bounce as well.  The simplest renormalizable potential that satisfies the conditions~(\ref{eq:bouncereqs}) and is bounded from below reads
\be
V\lb\varphi\rb=M^4\lsb 1-\lb\frac{\varphi}{\mu}\rb^2\rsb^2\,,
\label{eq:potential}
\ee
with $M$ and $\mu$ \emph{a priori} free parameters.  Given this form of potential, the phase of inflation that follows the bounce falls into the class of  small field models, for which the accelerated expansion sets in near the origin $\varphi=0$.  Requiring that both $\varphi_{\mrm{b}}'$ and $\omega$ be real further yields the constraint
\be\label{eq:ab_and_M}
\ab^2=(\Upsilon/\kappa) M^{-4}\,,\qquad 2<\Upsilon<3\,.
\ee
\subsection{Scales and variables}\label{sec:numdefs}
Having established the background cosmology, let us now briefly recall the relevant observable range of scales and orders of magnitude of the parameters of the potential that will be used in the numerical work and observational predictions in Secs \ref{sec:bouncingtimescale} and \ref{sec:obs}, and in Appendix \ref{sec:zetaevol}.  The Hubble scale and curvature radius today are given by
\be
\ell_H=\frac{1}{H_0}\qquad\text{and}\qquad\ell_{\mrm{c}}=\frac{a_0}{\sqrt{\mcl K}}\,.
\ee
Given that the Hubble scale today is $H_0=100h\,\text{km}\,\text{s}^{-1}\,\text{Mpc}^{-1}$ with the  reduced Hubble parameter $h\simeq0.7$, one has $\ell_H=3h^{-1}\,\text{Gpc}$.  Furthermore, for $\mcl K=+1$, given that $|\Omega_{\mcl K}|\leq 10^{-2}$~\cite{Komatsu:2010fb} and using $|\Omega_{\mcl K}|=\mcl K/(a_0H_0)^2$, one finds
\be
\frac{\ell_H^2}{a_0^2}\leq10^{-2}\,,
\label{eq:Omega0}
\ee
so that $a_0\geq 30h^{-1}\,\text{Gpc}$.

Although part of the motivation for studying non-singular cosmologies is the presence of bounce-inducing corrections to GR (or other bounce-inducing mechanisms) arising in some high energy theories, in this work, we shall remain in the domain of applicability of GR and quantum field theory.  We therefore set $\ab\sim 10^6\ell_{\mrm{Pl}}$.  From the first Friedmann equation, one can see that this corresponds to an order of magnitude $M\simeq\mcl O(10^{-3}\mpl)$ for the mass parameter of $V(\varphi)$, consistent with small field inflationary model values for $M$~\cite{Martin:2006rs}, see Eq.~(\ref{eq:Omagpara}) but also Eq.~(\ref{eq:ab_and_M}) where the range of allowed $M$ parameter values is given.  Using these values for $a_{0}$ and $a_{\mrm b}$, the number of $e$-foldings of expansion since the bounce then is $N=\ln(a_0/a_{\mrm{b}})\gsim 125$.  The entire range of observable scales today, given in terms of the corresponding physical wavenumber $k_{\mrm{phys}}$, is
\be
10^{-3}\,h\mrm{Mpc}\lsim k_{\mrm{phys}}\lsim 10^3\,h \mrm{Mpc}.
\ee
For $a_0=30\,h^{-1}\mrm{Gpc}$, the range of comoving wavenumbers is then $10^2\lsim k \lsim 10^8$. Finally, in order to obtain $60$ to $80$ $e$-foldings of inflation, the parameters of the potential~(\ref{eq:potential}) must be of the order of the following small field inflation fiducial values~\cite{Martin:2006rs}
\be
M\simeq\order{10^{-3}\mP}\,,\qquad\mu\simeq\order{\mP}\,.
\label{eq:Omagpara}
\ee
\section{Scalar perturbations}
\label{sec:perturbations}
\subsection{Scalar perturbation variables}
\label{sub:preliminaries}
We shall now study the evolution of cosmological perturbations around the homogeneous and isotropic background cosmology discussed in the previous section.  We mainly consider the scalar part of the metric perturbations (given by the gauge invariant gravitational Bardeen potential $\Phi$) in the presence of scalar field perturbations $\delta\varphi$, but also give, although in less detail, the result obtained for tensor perturbations.  In longitudinal gauge, the scalar part of the perturbed metric reads~\cite{Mukhanov:1990me}
\be
\dd s^2=a^2\lb\eta\rb\lsb-\lb 1+2\Phi\rb\dd\eta^2+\lb 1-2\Phi\rb\gamma_{ij}^{(3)}\dd x^i\dd x^j\rsb,
\label{eq:metric}
\ee
where $\gamma_{ij}^{(3)}$ is the background metric of the spatial sections.  Expressing the scalar parts of the perturbed Einstein equations in terms of $\Phi$ and the density perturbation variable $\delta\varphi$ and combining them in the appropriate way yields a second order differential equation for the modes $\Phi_k$~\cite{Mukhanov:1990me}
\be
\Phi_k''+2\lb\mcl H-\frac{\varphi''}{\varphi'}\rb\Phi_k'+\lsb k^2-4\mcl K+2\lb \mcl H'-\mcl H\frac{\varphi''}{\varphi'}\rb\rsb\Phi_k=0.
\label{eq:Phi-eofm}
\ee
The wavenumber $k$ is the eigenvalue of the Laplace-Beltrami operator on positively curved spatial sections.  It is therefore a function of an integer $n$, and is given by $k=\sqrt{n\lb n+2\rb}$ with $n=0$ and $n=1$ corresponding to gauge modes~\cite{ Lifshitz:1963ps}.  As is well-known, it is possible to define the variable $u$ related to $\Phi$ through the relation~\cite{Mukhanov:1990me}
\be
\Phi=\frac{\kappa}{2}(\rho+P)^{1/2}u=\frac{\kappa}{2}\frac{\varphi'}{a}u,
\label{eq:phitou}
\ee
in terms of which the mode equation (\ref{eq:Phi-eofm}) reduces to
\be
u_k''+\lsb k^2-V_u(\eta)\rsb u_k=0,
\label{eq:uevol}
\ee
where
\be
V_u(\eta)=\frac{\theta''}{\theta}+3\mcl{K}\lb 1-c_s^2\rb,\qquad\theta=\lb\frac{3}{\kappa}\rb^{1/2}\frac{\mcl H}{a\varphi'},
\label{eq:Vueta}
\ee
and
\be
c_{\mrm{s}}^2=\frac{\delta P}{\delta\rho}=-\frac{1}{3}\lb 1+2\frac{\varphi''}{\mcl H\varphi'}\rb.
\ee
The quantity $c_{\mrm{s}}$ can in some regimes be interpreted as the velocity of sound.
The explicit form of $V_u(\eta)$ reads
\be
V_u(\eta)=\mcl{H}^2+2\lb \frac{\varphi''}{\varphi'}\rb^2-\frac{\varphi'''}{\varphi'}-\mcl{H}'+4\mcl{K},
\label{eq:upot}
\ee
which is manifestly regular when $\mcl H=0$.  Another possible choice of variable is given by the generalized form of the Mukhanov-Sasaki variable, valid for arbitrary $\mcl K$ and which we shall denote $\tilde v$.  The modes of the variable $\tilde v$ are related to $\Phi_k$, in Newtonian gauge, by the relation~\cite{Martin:2003sf,Garriga:1999vw,Hwang:2002fp}
\be
\tilde v_k=-\frac{a}{\chi_k}\lb\delta\varphi_k+\frac{\varphi'}{\mcl H}\Phi_k-\frac{2\mcl K}{\kappa\mcl H\varphi'}\Phi_k\rb,
\label{eq:MSV}
\ee
where
\be
\chi^2_k=1-3\mcl K\frac{1-c_{\mrm{s}}^2}{k^2}\,.
\label{eq:chics2}
\ee
The gauge-invariant combination of the modes $\Phi_k$ and $\delta\varphi_k$ defining the generalized Mukhanov-Sasaki variable in Eq.~(\ref{eq:MSV}) yields an equation of motion for $\tilde v_k$ which is reminiscent of that for $u_k$,
\be
\tilde v_k''+\lsb k^2-V_{\tilde v}(\eta)\rsb \tilde v_k=0,
\label{eq:MSevol}
\ee
where
\be
V_{\tilde v}(\eta)=\frac{\tilde z_k''}{\tilde z_k}+3\mcl K(1-c_{\mrm{s}}^2)\qquad\text{and}\qquad \tilde z_k=a\frac{\varphi'}{\mcl H\chi_k}.
\label{eq:defvpot}
\ee
The variable $\tilde v$ has the additional property that it appears naturally as the canonical variable in the total action for cosmological perturbations from which Eq.~(\ref{eq:MSevol}) is derived and, as such, should be the starting point for quantization and for the choice of the quantum initial conditions seeding the growth of cosmological structures.  It can however be seen from Eqs.~(\ref{eq:MSV}-\ref{eq:MSevol}) that using the combination defining $\tilde v$ in Eq.~(\ref{eq:MSV}) to construct the relevant variable for the study of the evolution of cosmological perturbations is inappropriate in the present context.  The use of $\tilde v$ across the bounce is forbidden because the variable itself, $c_{\mrm{s}}^2$ and the term $\tilde z_k''/\tilde z_k$ are either ill-defined or divergent when $\mcl H=0$~\cite{Martin:2003sf,Brandenberger:2007by,Alexander:2007zm,Cai:2007zv}.  Furthermore, the potential $V_{\tilde v}(\eta)$ is $k$-dependent so that defining an asymptotically flat spacetime for high frequency modes is ambiguous.  From the expression for $\chi_k$, and setting $\mcl K=+1$, there would also appear to be divergences whenever $c_{\mrm{s}}^2=1-k^2/3$~\cite{Martin:2003sf}.  This equality can be expressed as
\be
3-\frac{k^2}{2}=1-\frac{\varphi''}{\mcl H\varphi'}=\delta,
\label{eq:divcs2}
\ee
where $\delta$ is the second slow roll parameter defined in terms of the horizon flow functions in Eq.~(\ref{eq:defsrparams}).  In the context of small field models and away from $\mcl H=0$, the slow roll parameter $\delta$ verifies $\delta<0$ and $|\delta|\ll 1$, so that the condition (\ref{eq:divcs2}) is clearly not a problem.  It is therefore possible to retain the generalized forms $\tilde v_k$ and $\tilde z_k$ far from $\mcl H=0$.  In such regions, the modes $\tilde v_k$ are everywhere well-defined.  Furthermore, in the regions where both $\mcl K/a^2\ll 1$ and where slow-roll applies, the {\it cosmic} Hubble parameter $H$ is slowly varying and different from zero.  One may therefore neglect the terms proportional to $\mcl K$, and use the approximation $\chi_k\simeq 1$.  The expressions for the modes $\tilde v_k$ and for the $k$-dependent function $\tilde z_k$ then reduce to their standard ($\mcl K=0$) forms $v_k$ and $z$, with $z$ now $k$-independent.  A quantum interpretation of $v$ is then possible and the standard quantization procedure on $v_k$ can in principle be achieved in the usual way~\cite{Mukhanov:1990me}.  Furthermore, in these regions, the modes of the variables $u$ and $v$ may be related to one another in the standard way.
\be
v_k=\lb\frac{3}{\kappa}\rb^{1/2}\theta\lb\frac{u_k}{\theta}\rb'\qquad\text{and}\qquad k^2u_k=-z\lb\frac{v_k}{z}\rb'.
\label{eq:urelatev}
\ee
Using these relations,  it is therefore possible to prepare an initial quantum state by choosing the mode functions $v_k$ of the variable $v$ in the regions $\eta<0$ where modes are sub-Hubble and spatial curvature is negligible but study the evolution of cosmological perturbations through the entire bouncing phase using $u$.  The resulting form of $u$ after the bounce can then be used to determine the form of the modes $v_k$ at the onset of inflation and then compute the primordial spectrum of scalar perturbations.\\

In the following sections, we study, analytically, the evolution of the perturbations using $u$ after setting the initial state of perturbations using $v$. In Appendix \ref{sec:zetaevol}, the time evolution of perturbations and the issue of the growth of perturbations in contracting spacetimes is also addressed~\cite{Creminelli:2004jg,Allen:2004vz,Brandenberger:2009rs}.
\subsection{Time evolution of $u$}
\label{sub:uevol}
Hereon, we restrict the analysis to the case of a symmetric bounce and in this section, we consider the evolution of the variable $u$ in the contracting phase, at the bounce and in the inflationary phase.  As shown in Ref.~\cite{Falciano:2008gt}, the potential (\ref{eq:upot}) is very well approximated by neglecting the curvature term $\mcl K/\mcl H^2$, neglecting the time dependence of the horizon flow functions, and joining a slow-roll exponentially contracting phase directly to a slow-roll inflationary phase.  This is true because in the range of wavenumbers of observational interest, the spatial curvature term and the small amplitude features in $V_u(\eta)$ near $\eta=0$ are always negligible compared to $k^2$.  Note that, as discussed in the previous section, this approximation does not hold true in the case of the variable $\tilde{v}$.  Let us also point out that while it can be neglected when studying the perturbation $u$, the contribution of $\mcl K$, at the level of the background cosmology, is crucial to obtain the bounce.

We now proceed as in Ref.~\cite{Falciano:2008gt}.  We neglect the spatial curvature term, in which case, at first order in slow roll, Eq.~(\ref{eq:upot}) simplifies to
\be
V_u(\eta)=\mcl{H}^2\lb\epsilon_1+\frac{\epsilon_2}{2}\rb=\frac{1}{x_{\pm}^2}\lb \epsilon_1+\frac{\epsilon_2}{2}\rb\,,
\label{eq:potVsr}
\ee
where we have used the expression
\be
\mcl{H}_{\pm}(\eta)= -\frac{1+\epsilon_1}{\eta-\eta_{\pm}}=\pm\frac{1+\epsilon_1}{x_{\pm}}.
\label{eq:Hbounce-sr}
\ee
In the last step, we defined $x_{\pm}=|\eta-\eta_{\pm}|$, where the ``$-$'' or the ``$+$'' sign is chosen according to whether one considers times $\eta$ before or after the bounce at $\eta=0$. Up to first order in slow roll, the equation of motion for $u$ therefore simply reads
\be\label{eq:uevolapprox}
u''+\lsb k^2-\frac{1}{x_{\pm}^2}\lb \epsilon_1+\frac{\epsilon_2}{2}\rb\rsb u=0\,,
\ee
where the subscript $k$ on the mode function $u$ is implicit.  Note that Eq.~(\ref{eq:uevolapprox}) applies for \emph{symmetric} bounces only because for an asymmetric bounce, the values of $\epsilon_1$ and $\epsilon_2$ cannot be assumed to be equal before and after the bounce. In Appendix \ref{sub:numex}, we give numerical examples showing that in cosmologies with strong departures from symmetry, additional features appear in the potential for $u$.

We now briefly discuss the meaning of the conformal time parameters $\eta_{-}$ and $\eta_{+}$.  The expressions for $\mcl H_{\pm}$ used in (\ref{eq:Hbounce-sr}) are the analogs of the familiar expression used in standard inflation where one has, to first order in slow roll,
\be
\mcl H(\eta)=-\frac{1+\epsilon_1}{\eta}\,,
\ee
with $\eta \in ]-\infty,0[$.  Here, instead, the conformal time intervals of the contracting and expanding phases in Eq.~(\ref{eq:Hbounce-sr}) are $]\eta_-,0\,]$ and $[\,0,\eta_+[$ respectively, so that $\eta_-$ and $\eta_+$ are the asymptotic values of the conformal time for early and late cosmic times, which we shall label $t_-$ and $t_+$, respectively.  For instance, in the de Sitter bounce discussed in Sec.~\ref{sub:desitterbounce}, one has $\eta_{\pm}=\pm\pi/2$.

We now consider Eq.~(\ref{eq:uevolapprox}) separately in the two eras before and after the bounce, where the solutions for $u$ take the form of Hankel functions of the first and second kind $H_{\nu}^{(1)}$ and $H_{\nu}^{(2)}$.   We can write the solutions explicitly as
\bea
u^-(\eta)=\sqrt{kx_-}\lsb U_1^-(k)H_{\nu}^{(1)}(kx_-)+U_2^-(k)H_{\nu}^{(2)}(kx_-)\rsb,\label{eq:uminus}\\
u^+(\eta)=\sqrt{kx_+}\lsb U_1^+(k)H_{\nu}^{(1)}(kx_+)+U_2^+(k)H_{\nu}^{(2)}(kx_+)\rsb,\label{eq:uplus}
\eea
where the order $\nu$ of the Hankel functions (equal before and after the bounce) is given by 
\beq\label{eq:defnu}
\nu=\frac{1}{2}+\epsilon_1+\frac{\epsilon_2}{2}\,,
\eeq
and where as before
\beq
x_-=|\eta-\eta_-|=\eta-\eta_-\,,\qquad x_+=|\eta-\eta_+|=\eta_+-\eta\,.\label{eq:xminus}
\eeq
It is important to realize that the arguments of the Hankel functions are different so that the expressions given in Eqs.~(\ref{eq:uminus}) and (\ref{eq:uplus}) are distinct. These two solutions (and their derivatives) can be matched at the time of the bounce $\eta=0$ using the standard procedure.  Doing so and expanding the Hankel functions for large values of $|kx_{\pm}|$ (that is, for $\eta\simeq 0$), one obtains
\bea
U_1^+&=&U_2^-\lb\sigma_k+i\rb e^{-i(k\Delta\eta-\pi\nu)}
\label{eq:U1plusapprox},\\
U_2^+&=&U_1^-\lb\sigma_k-i\rb e^{i(k\Delta\eta-\pi\nu)},
\label{eq:U2plusapprox}
\eea
where we retain only the leading order terms and where we have defined the parameter
\be
\sigma_k=\frac{2\epsilon_1+\epsilon_2}{k\Delta\eta}
\label{eq:defsigmak}
\ee
and rewritten the conformal time difference as
\beq\label{eq:defdeltaeta}
\Delta\eta=\eta_+-\eta_-\,.
\eeq
Note that the expression for $\sigma_{k}$ is first order in slow-roll. This quantity will be seen in Sec.~\ref{sub:primordialspectrum} to correspond to a slight tilt and amplitude modification of the standard form of primordial spectrum of scalar fluctuations. The quantity $\Delta\eta$ is the {\it conformal} cosmological bouncing scale which we shall discuss in some more detail in Sec.~\ref{sec:bouncingtimescale}. We shall see that it is interpretable as the angular frequency of oscillations induced by the bouncing phase in the primordial power spectrum of scalar and tensor perturbations.  In the case of a pure de Sitter bounce, $\Delta\eta=\pi$.
\subsection{Relating $u$ and $v$}
\label{sub:utov}
In this section, we compute the form of the Mukhanov-Sasaki variable at the onset of inflation in terms of its form prior to the bounce.  Once determined, the expression for the Mukhanov-Sasaki variable in the expanding phase can be used in the standard way to compute the primordial spectrum of scalar perturbations.

Let us recall that when expressed in terms of $u$, the time evolution of scalar perturbations is continuous across the bounce.  The coefficients of the two linearly independent solutions for $u^+$ can then be expressed in terms of those for $u^-$.  This calculation was the subject of Sec.~\ref{sub:uevol}.  On the other hand, the time evolution in terms of the Mukhanov-Sasaki variable $\tilde v$ is singular at the bounce and in fact $\tilde v$ itself is ill-defined at the bounce (see Sec.~\ref{sub:preliminaries}).  As a result, a calculation similar to the one performed in Sec.~\ref{sub:uevol} is not possible in the case of $\tilde v$. One may nevertheless easily determine the functional form of the Mukhanov-Sasaki variable in the deflationary and inflationary phases on either side of the bouncing phase because these are regions where $\mcl K/a^2\ll 1$ and where slow roll applies so that $\tilde v\simeq v$ and $V_{\tilde v}(\eta)\simeq V_v(\eta)$.  In these regions, the equation of motion for the mode functions $v$ is then given, at first order in slow roll, by the standard expression
\be
v''+\lsb k^2-\frac{1}{x_{\pm}^2}\lb 2+3\epsilon_1+\frac{3\epsilon_2}{2}\rb\rsb v=0\,,
\label{eq:evolv}
\ee
where the subscript $k$ on $v$ is implicit.  The solutions of Eq.~(\ref{eq:evolv}) are given by
\bea
v^-(\eta)&=&\sqrt{kx_-}\lsb V_1^-H_{\varrho}^{(1)}(kx_-)+V_2^-H_{\varrho}^{(2)}(kx_-)\rsb,
\label{eq:vminus}\\
v^+(\eta)&=&\sqrt{kx_+}\lsb V_1^+H_{\varrho}^{(1)}(kx_+)+V_2^+H_{\varrho}^{(2)}(kx_+)\rsb,
\label{eq:vplus}
\eea
where $\varrho=\nu+1$ with $\nu$ defined in Eq.~(\ref{eq:defnu}). From the relations in Eqs.~(\ref{eq:urelatev}) between $u$ and $v$, and using Eqs.~(\ref{eq:uminus}), (\ref{eq:uplus}), (\ref{eq:vminus}) and (\ref{eq:vplus}), one finds
\be
U_i^{\pm}=\pm\lb\frac{\kappa}{3}\rb^{1/2}k^{-1}V_i^{\pm},
\ee
so that the coefficients of $v^-$ and $v^+$ are simply related by
\bea
V_1^+(k)&=&V_2^-(k)\lb\sigma_k+i\rb e^{-i(k\Delta\eta-\pi\varrho)}\,,
\label{eq:V1+}\\
V_2^+(k)&=&V_1^-(k)\lb\sigma_k-i\rb e^{i(k\Delta\eta-\pi\varrho)}\,.
\label{eq:V2+}
\eea
We stress once more that although analogous to Eqs.~(\ref{eq:U1plusapprox}) and (\ref{eq:U2plusapprox}), the expressions (\ref{eq:V1+}) and (\ref{eq:V2+}) cannot be obtained using the procedure followed to relate $u^+$ and $u^-$ because, as discussed above, the potential $V_{\tilde{v}}(\eta)$ in Eq.~(\ref{eq:defvpot}) is singular at the bounce, and because $\tilde v$ is ill-defined at the bounce.

We shall now require that the canonical variable $v$ be quantized.  In regions where modes are sub-Hubble and spatial curvature is negligible, the standard quantization procedure of the canonical variable $v$ can be applied.  The consistency of the temporal mode functions $v$ with the commutation relations required for quantization is ensured provided $v$ satisfies the Wronskian condition~\cite{Mukhanov:1990me}
\be
(v)'\,v^*-(v^*)'\,v=i\,.
\ee
This imposes
\be
|V_1^{\pm}|^2-|V_2^{\pm}|^2=\mp\frac{\pi}{4}k^{-1}
\label{eq:wronskian}
\ee
on the coefficients of the Hankel functions in (\ref{eq:vminus}) and (\ref{eq:vplus}).
\subsection{Initial conditions for $v$}
\label{sub:initialstate}
Unlike for inflation, there exists no unique prescription to fix the initial conditions for the cosmological perturbations at the onset of the deflationary phase in a bouncing cosmology.  This is because, rather than starting sub-Hubble in their vacuum state, modes that enter the contracting phase have a prior history.  For instance, and although this is by no means the only possibility, they may have started sub-Hubble during a radiation- or matter-dominated contracting era connected to the deflationary phase that preceeds the bounce.  At the onset of inflation, therefore, modes cannot be expected to be in their vacuum state.  In this paper, we shall nevertheless assume, for definiteness, that $V_1^-$ and $V_2^-$ can be parameterized as
\be
V_1^-=\frac{\sqrt{\pi}}{2}\,\varsigma_1\,k^{-\alpha/2}e^{i\theta_1}
\qquad\text{and}\qquad
V_2^-=\frac{\sqrt{\pi}}{2}\,\varsigma_2 \,k^{-\beta/2}e^{i\theta_2}\,,
\label{eq:ansatzVminusinitial}
\ee
where $\varsigma_1,\,\varsigma_2\in\real$ and positive, $\alpha$ and $\beta$ are numbers while $\theta_1$ and $\theta_2$ are phase angles.  Given Eq.~(\ref{eq:ansatzVminusinitial}), one finds from Eq.~(\ref{eq:wronskian}) that
\be
\varsigma_2^2k^{-\beta}=\varsigma_1^2k^{-\alpha}-k^{-1}\,.
\label{eq:iniV-}
\ee
In order to satisfy Eq.~(\ref{eq:iniV-}), one has to choose $\alpha=\beta=1$ and hence $\varsigma_2^2=\varsigma_1^2-1$.  Dropping the subscript on $\varsigma_1$ so that $\varsigma_1\rightarrow \varsigma$, one finally has
\be
V_1^-=\frac{\sqrt{\pi}}{2}\,|\varsigma|\, k^{-1/2}e^{i\theta_1}
\qquad\text{and}\qquad
V_2^-=\frac{\sqrt{\pi}}{2}\,\left|1-\varsigma^2\right|^{1/2}\,k^{-1/2}e^{i\theta_2}\,,
\label{eq:ansatzVminusfinal}
\ee
where we have absorbed the choice of sign in going from $|V_1^-|^2$ and $|V_2^-|^2$ to $V_1^-$ and $V_2^-$ into the phases $\theta_1$ and $\theta_2$.  Additionally, it can be checked using Eqs.~(\ref{eq:V1+}) and (\ref{eq:V2+}) that for $\eta>0$ the normalization condition is verified up to first order in slow roll,
\be
|V_1^+|^2-|V_2^+|^2=-\frac{\pi}{4}\,k^{-1}\lb 1+\sigma_k^2\rb\,.
\ee
where $\sigma_k^2$ is second order in slow roll, as can be checked from (\ref{eq:defsigmak}).

The initial conditions for $v^-$ at times $\eta<0$ have a simple interpretation.  They represent, through the phases $\theta_1$ and $\theta_2$ and through the amplitude parameter $\varsigma$, a deviation from the Bunch-Davies vacuum, which can be recovered provided $\varsigma=1$ and $\theta_1=(\pi/2)\varrho+\pi/4$. Furthermore, provided one \emph{does} choose the adiabatic vacuum prescription, it can be seen from  Eqs~(\ref{eq:V1+}) and (\ref{eq:V2+}) that $v^+$ is given by a single mode function. Consequently, in this case and at first order in slow roll, we can expect to recover the standard form of the primordial spectrum of scalar fluctuations (see Sec.~\ref{sub:primordialspectrum}).

A comment on the order of magnitude of $\varsigma$ is also in order.  It is well known that cosmological perturbations on super-Hubble scales generically grow in contracting spacetimes~\cite{Creminelli:2004jg,Allen:2004vz,Brandenberger:2009rs}.  This is because while in an expanding phase, a perturbation can be split into a constant and a decaying mode, in a contracting phase, a perturbation is split into a constant and a growing mode. The amplitude of a perturbation with wavenumber $k$, which we parameterize by $\varsigma$, is therefore dependent on the rate and the duration of the contracting phase.  As discussed in Appendix \ref{sub:zetagrowth}, the amplitude of perturbations can be made to deviate only slightly from the amplitude of vacuum fluctuations provided the contracting phase is either slow, short, or both slow and short.  In the case at hand, the contracting phase is a fast and short deflationary phase ($\sim 60$ $e$-folds in a short conformal time interval) during which super-Hubble modes grow only very little.  Furthermore, fluctuations can be assumed to remain small in any contracting phase preceding deflation, provided the rate of contraction is small.  Finally, since $\varsigma$ turns up in the expression for the amplitude of the primordial, CMB and matter power spectra, that it remains close to 1 is evidently an observational necessity for the viability of the model.  We shall therefore assume that $\varsigma\simeq 1$ when we consider cosmological observables in Sec.~\ref{sec:obs}.  
\subsection{Primordial scalar perturbation spectrum}
\label{sub:primordialspectrum}
We shall now use the late time (super-Hubble) behaviour of $v^+$ to compute the primordial spectrum of scalar perturbations in the expanding phase, $\mcl P_{\zeta}$, where, in the subscript, the variable $\zeta=v/z$ is the curvature perturbation in the comoving gauge discussed in the Appendix.  The spectrum $\mcl P_{\zeta}$ is given by the usual expression
\be
\mcl P_{\zeta}=\frac{k^3}{2\pi^2}|\zeta|^2=\frac{k^3}{2\pi^2}\left|\frac{v^+}{z}\right|^2,
\label{eq:Pzeta}
\ee
and on super-Hubble scales, $v^+$ is given in terms of the asymptotic forms of the Hankel functions for $kx_+\ll 1$,
\bea
v^+&\simeq&\sqrt{kx_+}\lcb V_1^+(k)\lsb \frac{1}{\Gamma(1+\varrho)}\lb\frac{kx_+}{2}\rb^{\varrho}+\frac{i}{\Gamma(1-\varrho)\sin\pi\varrho}\lb\frac{kx_+}{2}\rb^{-\varrho}\rsb\right.\nonumber\\
&&\left.+V_2^+(k)\lsb \frac{1}{\Gamma(1+\varrho)}\lb\frac{kx_+}{2}\rb^{\varrho}-\frac{i}{\Gamma(1-\varrho)\sin\pi\varrho}\lb\frac{kx_+}{2}\rb^{-\varrho}\rsb\rcb\,.
\label{eq:vonset}
\eea
Given that, as long as slow roll is not violated, $\varrho>0$, we may keep only the second term in each line of (\ref{eq:vonset}).  From the growing modes of (\ref{eq:vonset}) and from
\be
z^{-1}=\lb\frac{\kappa}{2}\rb^{1/2}\frac{1}{a\sqrt{\epsilon_1}},
\ee
where, since $\mcl K/a^2\ll 1$ we have taken $\chi\simeq 1$, and where we have used the definition of the first horizon flow function, we find
\be
\mcl{P}_{\zeta}=\frac{2}{\mpl^2}\frac{k^3}{\pi}\frac{1}{a^2\epsilon_1}\frac{2^{2\varrho}}{\left[\Gamma(1-\varrho)\sin(\pi\varrho)\right]^{2}}\lb kx_+\rb^{1-2\varrho}|V_1^+-V_2^+|^2.
\label{eq:Pbounce}
\ee
It is convenient to split the explicit expression for $\mcl{P}_{\zeta}$ into a first standard (tilt) part $\mcl{P}_{\zeta}^{\mrm{std}}$ and a second part $\mcl{P}_{\zeta}^{\mrm{osc}}$, which will be seen to be oscillatory in $k$.  Let us thus write
\be
\mcl{P}_{\zeta}=\mcl{P}_{\zeta}^{\mrm{std}}\times\mcl{P}_{\zeta}^{\mrm{osc}}\,.
\label{eq:Pbouncesplit}
\ee
The standard part of the spectrum is given by
\be
\mcl{P}_{\zeta}^{\mrm{std}}=\frac{k^{2}}{2\mpl^2}\frac{1}{a^2\epsilon_1}\frac{2^{2\varrho}}{\left[\Gamma(1-\varrho)\sin(\pi\varrho)\right]^{2}}\lb kx_+\rb^{1-2\varrho}\nonumber
\ee
which, up to first order in slow roll, reads
\be
\mcl{P}_{\zeta}^{\mrm{std}}\simeq\frac{H^2}{\mpl^2\pi\epsilon_1}\lcb1-\lsb2(1+C)\epsilon_1+C\epsilon_2+(2\epsilon_1+\epsilon_2)\ln(kx_+)\rsb\rcb,
\label{eq:Pstd}
\ee
where $C=\gamma_{\mrm{E}}+\ln 2-2$ and $\gamma_{\mrm{E}}$ is Euler's constant.  Eq.~(\ref{eq:Pstd}) is the standard inflationary result for $\mcl P_{\zeta}$. The oscillatory part of the spectrum reads
\be
\mcl P_{\zeta}^{\mrm{osc}}\equiv \frac{4}{\pi}\,k\,|V_1^+-V_2^+|^2\,.
\ee
From (\ref{eq:V1+}) and (\ref{eq:V2+}), one has
\be
V_1^+-V_2^+=\mcl{A}\lb |V_1^-|e^{i\lb\theta_1-\phi\rb}-|V_2^-|e^{i\lb\theta_2+\phi\rb}\rb,
\label{eq:V1mV2}
\ee
where
\be\label{eq:defamplitude}
\mcl{A}=\lb 1+\sigma_k^2\rb^{1/2},\qquad\phi=\lsb(2\epsilon_1+\epsilon_2)\sigma_k^{-1} -\pi\varrho\rsb-\varphi,\qquad\varphi=\arctan \sigma_k^{-1},
\ee
with $\sigma_k$ defined by Eq.(\ref{eq:defsigmak}) and $\Delta \eta$ by Eq.~(\ref{eq:defdeltaeta}).  The expression in Eq.~(\ref{eq:V1mV2}) can also be rewritten as
\be
V_1^+-V_2^+=\mcl{A}\lsb |V_1^-|^2+|V_2^-|^2-2|V_1^-||V_2^-|\cos\lb \theta_1-\theta_2-2\phi\rb\rsb^{1/2} e^{i\Psi},
\ee
where
\be
\Psi=\arctan\lsb\frac{|V_1^-|\sin\lb \theta_1-\phi\rb-|V_2^-|\sin\lb \theta_2+\phi\rb}{|V_1^-|\cos\lb \theta_1-\phi\rb-|V_2^-|\cos\lb \theta_2+\phi\rb}\rsb.
\ee
Using these expressions, the oscillatory contribution to the spectrum $\Posc$ reads
\be\label{eq:Poscfinal}
\Posc=\frac{4}{\pi}\,\mcl{A}^2\,k\lsb |V_1^-|^2+|V_2^{-}|^2-2|V_1^-||V_2^-|\cos \lb\theta_1-\theta_2-2\phi\rb\rsb.
\ee
Up to this point, our calculations were exact to first order in slow-roll. We now make an approximation to simplify the cosine term in Eq.~(\ref{eq:Poscfinal}).  We first rewrite it as
\bea
\cos\lb \theta_1-\theta_2-2\phi\rb&=&\cos\lb \theta_1-\theta_2-2k\Delta\eta+2\pi\nu\rb\cos\lb 2 \arctan\sigma_k^{-1}\rb+\nonumber\\
&&\sin\lb \theta_1-\theta_2-2k\Delta\eta+2\pi\nu\rb\sin\lb 2 \arctan \sigma_k^{-1}\rb\,.\label{eq:sinterm}
\eea
Because $\sigma_k^{-1}\gg 1$, and as $k$ grows, the cosine and sine terms that have the arctangent as their argument rapidly go to $-1$ and $0$ respectively, so that one is simply left with just the term $\cos\lb \theta_1-\theta_2-2k\Delta\eta+2\pi\nu\rb$ in the first line of Eq.~(\ref{eq:sinterm}).  As for $\mcl{A}^2$, one can easily see from Eq.~(\ref{eq:defamplitude}) that it is equal to one at first order in slow-roll. Furthermore, expanding $\Posc$ at that order and using Eq.~(\ref{eq:ansatzVminusfinal}) one obtains
\bea
\Posc&\simeq&\varsigma^2+|1-\varsigma^2|-2\varsigma|1-\varsigma^2|^{1/2}\lsb\cos (2k\Delta\eta+\theta_1-\theta_2)\right.\nonumber\\
&&\left.-\pi (2\epsilon_1+\epsilon_2)\sin(2k\Delta\eta+\theta_1-\theta_2)\rsb\,.
\label{eq:Posc1st}
\eea
The $\Posc$ contribution to $\mcl P_{\zeta}$ thus contains both a constant part and an oscillatory part with frequency $2\Delta\eta$.  The spectrum is then given by the product of $\mcl P_{\zeta}^{\mrm{std}}$ in Eq.~(\ref{eq:Pstd}) and $\Posc$ in Eq.~(\ref{eq:Posc1st}).

Let us finally expand $H$, $\epsilon_1$ and $\epsilon_2$ around the pivot scale $k_\p^{-1}=a(x^+_\p) H(x^+_\p)$, taken to be the logarithmic mean of the range of observable scales.  At first order in the expansion, $\mcl P_{\zeta}^{\mrm{std}}$ then reads
\be
\mcl P_{\zeta}^{\mrm{std}}\simeq\frac{H_p^2}{\mpl^2\pi\epsilon_{1\,\mrm p}}\lcb 1-\lsb 2(1+C)\epsilon_{1\,\mrm p}+C\epsilon_{2\,\mrm p}+(2\epsilon_{1\,\mrm p}+\epsilon_{2\,\mrm p})\ln\left(\frac{k}{k_{\mrm{p}}}\right)\rsb\rcb,
\label{eq:Ptiltkstar}
\ee
where the $\epsilon_{i\,\mrm p}$ and $H_{\mrm p}$ are evaluated at $k_{\mrm p}$.  Assuming for simplicity that $\theta_1$ and $\theta_2$ are $k$-independent, the expansion of $\Posc$ around $k_{\mrm p}$ reads
\bea
\Posc&\simeq&\varsigma^{2}+\left|1-\varsigma^{2}\right|-2\varsigma\,\left|1-\varsigma^2\right|^{1/2}\times\label{eq:Posckstar}\\
&&\left[\cos\left(2k\Delta\eta+\theta_{1}-\theta_{2}\right)+\pi\left(2\epsilon_{1\,\mrm{p}}+\epsilon_{2\,\mrm{p}}\right)\,\sin\left(2k\Delta\eta+\theta_{1}-\theta_{2}\right)\right].\nonumber
\eea
At first order in slow roll, and in the limit $\varsigma\rightarrow 1$, the modes $v^-$ and $v^+$ are in the Bunch-Davies vacuum, and any modifications to $\mcl P_{\zeta}$ induced by the bouncing phase disappear.  This conclusion is also true for $\varsigma\rightarrow 0$ but this limit corresponds to a set of negative frequency modes $v^-$ and $v^+$ (see Sec.~\ref{sub:initialstate}).  For $\varsigma$ very different from $0$ or $1$, oscillations can appear at leading order in the spectrum, rather than as small corrections.  Note, finally that the form of $\mcl P_{\zeta}$ obtained is reminiscent of the primordial spectrum obtained in trans-Planckian inflationary models but with some important differences.  In trans-Planckian models~\cite{Martin:2003kp,Martin:2003sg}, the oscillatory term has a $\ln(k)$ dependence rather than a dependence linear in $k$.  In addition, the amplitude of oscillations in trans-Planckian models is suppressed by the product of the slow roll parameters with the ratio $H_{\mrm{inf}}/M_{\mrm{c}}\ll 1$ where $H_{\mrm{inf}}$ is the Hubble parameter during inflation and $M_{\mrm{c}}$ is a new mass scale of order $\mpl$. Here instead, the amplitude depends only on the deviation in amplitude from the vacuum state and is not suppressed by factors of order $\epsilon_i$ or by some new mass scale.
\section{Tensor perturbations}
\label{sec:tensors}
\subsection{Tensor perturbation variable}
\label{sub:tensorperts}
In this section and the next, we compute the transfer of tensor perturbations through the bouncing phase and the primordial spectrum of gravitational waves.  The evolution equation for the tensor modes $h$ (where here $h$ is not to be confused with the reduced Hubble parameter $h\simeq 0.7$ appearing in quantities corresponding to distances) reads
\be
h''+ \left[ k^{2}-V_{h}(\eta)\right]h=0 \qquad \text{with} \qquad V_{h}(\eta)= \mcl H'+\mcl H^2-2\mcl{K},
\label{eq:tensorevol}
\ee
where the subscript $k$ on the mode functions $h$ is implicit.  Because tensor perturbations are not subject to gauge issues, $h$ is everywhere well-defined and its evolution equation is regular.  As for $V_u(\eta)$ in Sec.~\ref{sub:uevol}, we neglect the spatial curvature term and use the approximate form
\be
V_{h}(\eta)\simeq\mcl{H}^{2}\lb 2-\epsilon_{1}\rb\,.
\ee
Using Eq.~(\ref{eq:Hbounce-sr}) the equation for $h$ becomes
\be
h''+\lb k^{2}-\frac{2+3\epsilon_{1}}{x_{\pm}^{2}}\rb h=0\,.
\ee
The solutions of this this equation before and after the bounce read
\bea
h^{-}&=&\sqrt{kx_{-}}\left[T_{1}^{-}(k)\,H_{\nuT}^{(1)}\left(kx_{-}\right)+T_{2}^{-}(k)\,H_{\nuT}^{(2)}\left(kx_{-}\right)\right],
\label{eq:muminus}\\
h^{+}&=&\sqrt{kx_{+}}\left[T_{1}^{+}(k)\,H_{\nuT}^{(1)}\left(kx_{+}\right)+T_{2}^{+}(k)\,H_{\nuT}^{(2)}\left(kx_{+}\right)\right],
\label{eq:muplus}
\eea
where, keeping terms up to first order in slow roll,
\be
\nuT=\frac{3}{2}+\epsilon_{1}\,.
\label{eq:nut}
\ee
and where the coefficients $T_1^{\pm}$ and $T_2^{\pm}$ are unknown functions of the wavenumber $k$.  Matching the coefficients of $h^+$ and $h^-$ at $\eta=0$, expanding the Hankel functions for $|kx_{\pm}| \ll 1$, and retaining terms in the expansion suppressed by negative powers of $k\Delta\eta$ up to linear order, one obtains
\bea
T_1^+&=&T_2^-\lb\tau_k+i\rb e^{-i(k\Delta\eta-\pi\nuT)},
\label{eq:TU1plusapprox}\\
T_2^+&=&T_1^-\lb\tau_k-i\rb e^{i(k\Delta\eta-\pi\nuT)},
\label{eq:TU2plusapprox}
\eea
where
\be
\tau_{k}=\frac{2+3\epsilon_{1}}{2k\Delta\eta}.
\ee
Contrary to the expressions for $U_1^+$ and $U_2^+$ in Eqs.~(\ref{eq:U1plusapprox}) and (\ref{eq:U2plusapprox}) which were found to be exact at first order in slow roll, the truncated expressions for $T_1^+$ and $T_2^+$ in Eqs.~(\ref{eq:TU1plusapprox}) and (\ref{eq:TU2plusapprox}) are not exact at first order in slow roll since, in principle, higher powers $\tau_k^p$, with $p\in\entier$, contribute constant terms and terms of order $\epsilon_1$ to the expansion.  At first order in slow roll, these terms read $\tau_k^p=(1+3p\,\epsilon_1/2)/(k\Delta\eta)^p$.  These higher order terms $\tau_k^p$ ($p>1$) can be safely neglected only if the constraint $\tau_k^p\ll \mcl O(\epsilon_1)$ is satisfied.  It can be verified that for $\Delta\eta\geq \pi$ ($\Delta\eta=\pi$ corresponding to a de Sitter bounce) and for any $\epsilon_1<1$ this constraint holds true for any $n\gsim 20$ [recall that $k^2=n(n+2)$ for $\mcl K>0$].

Consistency with quantization again requires that the Wronskian condition be imposed on the tensor modes.  As for the scalar modes, the Wronskian condition imposed prior and after the bounce reads
\beq
\left|T_{1}^{\pm}\right|^{2}-\left|T_{2}^{\pm}\right|^{2}=\mp\frac{\pi}{4}\,k^{-1}.
\label{eq:tensorwronskian}
\eeq
As for $v$, we now assume that $T_1^-$ and $T_2^-$ are parameterized as
\be
T_1^-=\displaystyle\frac{\sqrt{\pi}}{2}\,\vartheta_1\,k^{-\alpha_{_{\mrm T}}/2}e^{i\psi_1}\qquad\text{and}\qquad T_2^-=\frac{\sqrt{\pi}}{2}\,\vartheta_2\,k^{-\beta_{_{\mrm T}}/2}e^{i\psi_2}\,.
\label{eq:initensor}
\ee
Eq.~(\ref{eq:tensorwronskian}) then imposes $\alpha_{_{\mrm T}}=\beta_{_{\mrm T}}=1$ and $\vartheta_2^2=\vartheta_1^2-1$.  Again we drop the subscript on $\vartheta_1$ so that $\vartheta_1\rightarrow \vartheta$, which gives
\be\label{eq:tensorinitialparams}
T_1^-=\frac{\sqrt{\pi}}{2}\,|\vartheta| \,k^{-1/2}e^{i\psi_1}\qquad\text{and}\qquad T_2^-=\frac{\sqrt{\pi}}{2}\,\left|1-\vartheta^2\right|^{1/2}\,k^{-1/2}e^{i\psi_2}\,.
\ee
From Eqs.~(\ref{eq:TU1plusapprox}) and (\ref{eq:TU2plusapprox}), the Wronskian condition on $h^+$ reads
\be
\left|T_{1}^{+}\right|^{2}-\left|T_{2}^{+}\right|^{2}=-\frac{\pi}{4}\,k^{-1}\left(1+\tau_{k}^{2}\right)\,.
\label{eq:conservedwronskian}
\ee
Unlike for scalar modes, the Wronskian condition for $h^+$ only holds if $\tau_k^p\ll \mcl O(\epsilon_1)$.
\subsection{Primordial tensor perturbation spectrum}
For tensor modes, the primordial power spectrum $\PT$ reads
\beq\label{eq:tensorspectrum}
\PT=\frac{k^{3}}{2\pi^{2}}\frac{64\pi}{\mpl^{2}}\left|\frac{h^{+}}{a}\right|^{2}\,.
\eeq
Using the late-time expansion of the $h^{+}$ solution in Eq.~(\ref{eq:muplus}), this becomes
\beq\label{eq:Ptinserted}
\PT=\frac{k^3}{2\pi^2}\frac{64\pi}{\mpl^2}\frac{1}{a^2}\frac{2^{\nuT}}{\lsb\Gamma\lb 1-\nuT\rb\sin\lb\pi\nuT\rb\rsb^2}\lb kx_+\rb^{1-2\nuT}|T_1^+-T_2^+|^2\,.
\eeq
Once again splitting Eq.~(\ref{eq:Ptinserted}) into a standard part and a bounce-induced oscillatory part, $\PT=\PT^{\mrm{std}}\times\PT^{\mrm{osc}}$, we have
\bea
\PT^{\mrm{std}}&=&\frac{k^{2}}{2}\frac{16}{\mpl^{2}}\frac{1}{a^{2}}\frac{2^{2\nuT}}{\left[\Gamma(1-\nuT)\sin(\pi\nuT)\right]^{2}}\lb kx_+\rb^{1-2\nuT}\,,\\
\PT^{\mrm{osc}}&=&\frac{4}{\pi}\,k\,|T_1^+-T_2^+|^2\,.
\eea
Expanding at first order in $\epsilon_{1}$, one recovers, from $\PT^{\mrm{std}}$, the standard result
\beq
\PT^{\mrm{std}}=\frac{16}{\pi}\frac{H^{2}}{\mpl^{2}}\left[1-2\left(C+1\right)\epsilon_{1}-2\epsilon_{1}\ln\left(kx_{+}\right)\right],
\label{eq:spctensorstd}
\eeq
while $\PT^{\mrm{osc}}$ is analogous to the form of $\mcl{P}_{\zeta}^{\mrm{osc}}$ with the substitution $V_{i}^{+}\rightarrow T_{i}^{+}$ ($i=1,\,2$). Using Eqs.~(\ref{eq:tensorinitialparams}) and (\ref{eq:nut}) and proceeding as in the scalar case, we find, up to first order in slow-roll,
\bea
\PT^{\mrm{osc}}&\simeq&|\vartheta|^{2}+|1-\vartheta^{2}|-2\,|\vartheta|\,|1-\vartheta^{2}|^{1/2}\nonumber\\
&&\qquad\times\left[\cos(2k\Delta\eta+\psi_{1}-\psi_{2})-2\pi\epsilon_{1}\,\sin(2k\Delta\eta+\psi_{1}-\psi_{2})\right]\,.\label{eq:Ptoscfinal}
\eea
As for the scalar perturbations, Eqs.~(\ref{eq:spctensorstd}) and (\ref{eq:Ptoscfinal}) can then easily be expanded around the pivot scale $k_\p$.
\subsection{Tensor-to-scalar ratio}
The tensor-to-scalar-ratio $r$ also splits into a standard and a non-standard part, $r=r^{\mrm{std}}\times r^{\mrm{osc}}$.  The ratio of the standard parts of the tensor and scalar spectra yields the usual result at leading order, $r^{\mrm{std}}=\PT^{\mrm{std}}/\mcl{P}_{\zeta}^{\mrm{std}}=16\epsilon_{1}$, while the non-standard parts of the spectra yield the additional ratio
\be
r^{\mrm{osc}}=\frac{\PT^{\mrm{osc}}}{\Posc}=\frac{|T_1^+-T_2^+|^2}{|V_1^+-V_2^+|^2}.
\label{eq:STratio}
\ee
Given the choice in Eqs.~(\ref{eq:ansatzVminusinitial}) and (\ref{eq:initensor}), this ratio depends on the values of $\varsigma$ and $\vartheta$, on the phases $\theta_1$, $\theta_2$, $\psi_1$ and $\psi_2$, and it is scale-dependent through the $2k\Delta\eta$ term in the sine and cosine arguments of $\mcl P_{\zeta}$ and $\PT$.  The tensor-to-scalar ratio~(\ref{eq:STratio}) therefore does not satisfy the standard consistency relation of single field inflationary models, for which one has $r=-8n_{\mrm T}$ with $n_{\mrm T}=-2\epsilon_1$ the tensor spectral index.
\section{The characteristic time scale $\Delta\eta$}
\label{sec:bouncingtimescale}
In previous sections we saw that the term $2k\Delta\eta$ plays a crucial role in the modified spectra after the bounce, appearing in the arguments of the sine and cosine functions in both $\Posc$ and $\PT^{\mathrm{osc}}$.  We shall therefore now discuss the significance of the time scale $\Delta\eta$ defined in Eq.~(\ref{eq:defdeltaeta}). We first review the relevant field space dynamics as a function of the number of $e$-folds in slow roll inflation and then turn to the discussion on $\Delta\eta$.
\subsection{Inflationary field space dynamics}
In inflation, the evolution of $\varphi$ as a function of the number of $e$-folds is a known function, depending only on the parameters of the potential and on the initial condition for $\varphi$ at the onset of inflation.  Indeed, as long as the slow-roll conditions are satisfied, \ie when $\epsilon_{i}\ll1$, the horizon flow functions introduced in Sec.~\ref{subsec:quasidS} can be written in terms of field derivatives of the potential~(\ref{eq:potential}),
\beq
\epsilon_1\simeq  \frac{1}{2\kappa}\lb\frac{V_{,\varphi}}{V}\rb^2\,,\qquad\epsilon_2\simeq  \frac{2}{\kappa}\lsb \lb\frac{V_{,\varphi}}{V}\rb^2-\lb\frac{V_{,\varphi \varphi}}{V}\rb\rsb\,,
\label{eq:eps1}
\eeq
which using the form of $V(\varphi)$ in Eq.~(\ref{eq:potential}) gives
\be
\epsilon_1\simeq\frac{8\varphi^2}{\kappa\lb \mu^2-\varphi^2\rb^2}\,,\qquad \epsilon_2\simeq\frac{8\lb \mu^2+\varphi^2\rb}{\kappa\lb\mu^2-\varphi^2\rb^2}\,.
\label{eq:eps2}
\ee
Thus, if the primordial spectrum is to be evaluated at some pivot scale $k_{\mrm{p}}$, the values  of $\epsilon_1$ and $\epsilon_2$ (and therefore that of the scalar spectral index $n_{\mrm{s}}$) depend only on $\mu$ and on the value of the field, $\varphi_{\mrm{p}}$, when modes of wavenumber $k_{\mrm{p}}$ exit the Hubble radius.

The end of the inflationary expansion is defined by $\epsilon_1=1$ at $\varphi=\varphi_{\mathrm{e}}$. This yields
\be\label{eq:phie}
\varphi_{\mrm{e}}=\sqrt{\frac{2}{\kappa}}\lsb \lb 1+\frac{\kappa}{2}\,\mu^2\rb^{1/2}-1\rsb.
\ee
The slow-roll trajectory is then found from
\be\label{eq:srtrajectory}
N(\varphi)=-\int_{\varphi_{\mathrm{i}}}^{\varphi}\dd\varphi\,\frac{\kappa V}{V'}
=\frac{\kappa}{8}\,\mu^2\lsb\lb\frac{\varphi_{\mrm i}}{\mu}\rb^2-\lb\frac{\varphi}{\mu}\rb^2+2\ln\lb\frac{\varphi}{\varphi_{\mrm i}}\rb\rsb,
\ee
where $\varphi_{\mrm i}$ is the value of $\varphi$ at some initial time. Inverting this expression, one can obtain the field value at Hubble exit of modes corresponding to the largest observable scales,
\be
\frac{\varphi_{\star}}{\mu}=\sqrt{-W_0\lcb -\lb\frac{\varphi_{\mrm{e}}}{\mu}\rb^2\exp\lsb-\lb\frac{\varphi_{\mrm{e}}}{\mu}\rb^2-\frac{8N_{\star}}{\kappa\mu^2}\rsb\rcb},
\label{eq:phistar}
\ee
where $W_0(x)$ is the principal branch of the Lambert function and $\varphi_{\star}$ is the field value evaluated at $N_{\star}$, the number of $e$-folds between the Hubble exit of the largest observable modes and the end of inflation.  Given a value of $\mu$ and a value of $N_{\star}$ (typically $40<N_\star<60$) $\varphi_\star$ is obtained from Eq.~(\ref{eq:phistar}).  This can then be used to determine the value of $M$ by adjusting the amplitude of the spectrum of scalar perturbations to the COBE normalization.  For simplicity, we neglect the minor modifications to the expression for the COBE normalization that are introduced by the new form of $\mcl P_{\zeta}$ and by $\Omega_{\mcl K} \ne 0$ and simply write the normalization in the usual way, see \eg ref.~\cite{Martin:2006rs},
\be
M^4=487\epsilon_1\lsb 1-\lb\frac{\varphi_{\star}}{\mu}\rb^2\rsb^{-2}\times 10^{-12}\,\mpl^4.
\ee
Let us take $\varphi_{\mrm i}$ to be the field value at the onset of inflation (in standard inflation, $\varphi_{\mrm i}$ is given as an initial condition). The total number of $e$-folds of inflation $N_{\mathrm{inf}}$ is obtained when $\varphi_{\mrm{e}}$ of Eq.~(\ref{eq:phie}) is inserted into Eq.~(\ref{eq:srtrajectory}).  It is useful, as a first approximation, to consider the case $\varphi\ll\mu$ for which $N_{\mathrm{inf}}$ can be approximated by
\be\label{eq:Ninf}
N_{\mathrm{inf}}=N(\varphi_{\mathrm{e}})\simeq
\frac{\kappa}{4}\,\mu^2\ln\lb\frac{\varphi_{\mathrm{e}}}{\varphi_{\mrm i}}\rb\,.
\ee
Inflation should last \emph{at least} 50 to 60 $e$-folds in order to solve the flatness, isotropy and horizon problems of Big Bang cosmology.  Together with the value of $\varphi_{\mathrm{e}}$ in Eq.~(\ref{eq:phie}), one may then use Eq.~(\ref{eq:Ninf}) to determine an upper bound for $\varphi_{\mrm i}$ (in a small field model, the value of $\varphi$ during inflation increases), which yields
\beq
\varphi_{\mrm i}\leq\varphi_{\mrm{e}}\,\exp\left(-\frac{4N_{\mrm{inf}}}{\kappa\mu^{2}}\right)\,.
\label{eq:phiiconstraint}
\eeq
This restricts $\varphi_{\mrm i}$ to very small values. Recall once again that in standard inflation, $\varphi_{\mrm i}$ is given as an initial condition.  We see from Eqs.~(\ref{eq:phistar}) and (\ref{eq:Ninf}) that the field evolution between $\varphi_{\star}$ and $\varphi_{\mrm{e}}$ is independent of $\varphi_{\mrm i}$ but the total number of $e$-foldings of inflation is a function of $\varphi_{\mrm i}$.
\subsection{Computing $\Delta\eta$}
Let us now turn to the case of the bouncing cosmology and the calculation of $\Delta\eta$.  The conformal time interval $\Delta\eta$ is obtained by integrating
\be
\Delta\eta=\eta_+-\eta_-=\int_{\eta_-}^{\eta_+}\dd\eta.
\label{eq:D-eta-int}
\ee
Because the slow-roll contracting and expanding phases (sr) and the non-singular phase close to the bounce (b) cannot be approximated by a unique expression, it follows that for a symmetric bounce
\be
\Delta\eta=\Delta\eta_{\mrm{sr}}^-+\Delta\eta_{\mrm{b}}+\Delta\eta_{\mrm{sr}}^+=\Delta\eta_{\mrm{b}}+2\Delta\eta_{\mrm{sr}}=\int_{\eta_{\mrm{e}}^-}^{\eta_{\mrm{i}}^+}\mrm{d}\eta_{\mrm{b}}+2\int_{\eta_{\mrm{i}}^+}^{\eta_+}\mrm{d}\eta_{\mrm{sr}}\,.
\ee
Here, $\eta_{\mrm{e}}^-$ and $\eta_{\mrm{i}}^+$ are the values of $\eta$ at which the cosmology respectively exits and enters the slow-roll regime in the contracting and expanding phases.  The slow-roll parts of the integral (\ref{eq:D-eta-int}), evaluated in the interval from $\eta_{\mrm{i}}^+$ to $\eta_{+}$ is given by
\be
\Delta\eta_{\mrm{sr}}=\int_{N_{\mrm{i}}}^{N_{\mrm{e}}}\frac{\dd N}{\dd \mcl H(N)}\,,
\ee
where $N_{\mrm{i}}=\ln(a_{\mrm{i}}/\ab)$ and $N_{\mrm{e}}=\ln(a_{\mrm{e}}/\ab)$. In these expressions, $a_{\mrm i}$ is the scale factor at the onset of inflation and $a_\mrm{e}$ is the scale factor at the end of inflation.  Expanding $\ln\mcl H(N)$ around $N_{\mrm{p}}$ where $N_{\mrm{p}}=\ln(a_{\mrm{p}}/\ab)$ is the number of $e$-folds from the bounce up to the pivot scale, with $N_{\mrm{i}}<N_{\mrm{p}}<N_{\mrm{e}}$, gives
\bea
\ln\mcl H(N)&=&\ln\mcl H(N_{\mrm{p}})+\left.\frac{\dd\ln\mcl H}{\dd N}\right|_{N_{\mrm{p}}}\lb N-N_{\mrm{p}}\rb+\frac{1}{2}\left.\frac{\dd^2\ln\mcl H}{\dd N^2}\right|_{N_{\mrm{p}}}\lb N-N_{\mrm{p}}\rb^2+\dots\nonumber\\
&\simeq&\ln \mcl H_\p+\lb 1-\epsilon_{1\,\p}\rb \lb N-N_{\mrm{p}}\rb-\frac{1}{2}\epsilon_{1\,\p}\,\epsilon_{2\,\p}\lb N-N_{\mrm{p}}\rb^2
\eea
At linear order in $\epsilon_{i}$, one thus has
\be
\mcl H(N)=\mcl H_\p e^{\lb 1-\epsilon_{1\,\p}\rb\lb N-N_{\mrm{p}}\rb},
\ee
so that
\be
\Delta\eta_{\mrm{sr}}^{+}=\frac{e^{\lb 1-\epsilon_{1\,\p}\rb N_{\mrm{p}}}}{\mcl H_\p(\epsilon_{1\,\p}-1)}\lsb e^{\lb\epsilon_{1\,\p}-1\rb N}\rsb_{N_{\mrm{i}}}^{N_{\mrm{e}}}.
\ee
Since $N_{\mrm{e}}-N_{\mrm{p}}=N_{\star}$, with $40<N_\star<60$, we finally have
\be
\Delta\eta_{\mrm{sr}}^{+}\simeq \frac{1}{\mcl H_{\mrm{p}}\lsb 1-\epsilon_{1\,\p}\rsb}e^{\lb 1-\epsilon_{1\,\p}\rb\lb N_{\mrm{p}}-N_{\mrm{i}}\rb},
\ee
while from Eq.~(\ref{eq:srtrajectory}) one also has
\be
N_{\mrm{p}}-N_{\mrm{i}}=\frac{\kappa}{8}\mu^2\lsb \lb\frac{\varphi_{\mrm{i}}}{\mu}\rb^2-\lb\frac{\varphi_{\mrm{p}}}{\mu}\rb^2+2\ln\lb\frac{\varphi_{\mrm{p}}}{\varphi_{\mrm{i}}}\rb\rsb.
\ee
Here, the crucial point is that the field value at the onset of inflation $\varphi_{\mrm{i}}$, although it is constrained through Eq.~(\ref{eq:phiiconstraint}), is this time not given as an initial condition but depends on the detail of the bounce dynamics that preceed inflation and is an unknown function of $\ab$ and $\mu$. This is directly related to the fact that the exact dynamics from the bouncing phase intermediate between the deflationary and inflationary phases is not a known (or easily approximated) function of $\ab$ and $\mu$.  As a result, $\Delta\eta_{\mrm{sr}}$ but also $\Delta\eta_{\mrm{b}}$ and the initial field value $\varphi_{\mrm i}$, and finally $\Delta\eta$ are quantities that cannot be obtained analytically.

The value of $\Delta\eta$ can however easily be obtained by integrating the background cosmology numerically for any value of $\ab$ and $\mu$.  We do so in Fig.~\ref{fig:DeltaEta1D} where we plot, in the full lines, the quantity $a_0\Delta\eta$, which is the {\it theoretical} frequency of oscillations for physical wave numbers, for the renormalized potential parameter $\mu$ in the range $2.0\leq\mu\leq 4.5$, over the allowed range for $\ab$ [see Eq.~(\ref{eq:ab_and_M})] and in the range $10^{-8}\leq|\Omega_{\mcl K}|\leq 10^8$.

Furthermore, remembering that in the case of a $\mcl K=+1$ universe, the wavenumber $k$ is discrete with $k=\sqrt{n\lb n+2\rb}$ for integer $n$, the arguments of the sine and cosine functions in Eq.~(\ref{eq:Posckstar}), which are relevant for the oscillatory features in the primordial spectra, read
\be
2k\Delta\eta\simeq 2\lb n+1-\frac{1}{2n}\rb\lb m\pi+\delta\eta\rb\qquad (n\gg 1).
\ee
Here, $m$ is an integer and hence $0<\delta\eta<\pi$, such that, in Eq.~(\ref{eq:Posckstar}), one can make the substitution
\be
2k\Delta\eta\rightarrow 2\lb\frac{n+1}{a_0}\rb a_0\delta\eta=k_{\mrm{phys}}(2a_0\delta\eta)
\ee
for large $n$, where $k_{\mrm{phys}}$, as previously, is the physical wavenumber and $a_{0}$ is the scale factor today.  The {\it effective} frequency is therefore $2a_0\delta\eta$ rather than $2a_0\Delta\eta$.  The quantity $a_0\delta\eta$ is shown in the circles of Fig.~\ref{fig:DeltaEta1D}. The quantity $\delta\eta$ itself is shown in a two-dimensional plot in Fig.~\ref{fig:DeltaEta2D} in the ranges $3.0\leq\mu\leq 4.0$ and $2.04\leq\Upsilon\leq 2.08$ and for $10^{-9}\leq|\Omega_{\mcl K}|\leq 10^3$.  This reduced range for $\mu$ corresponds to $n_s\simeq 0.95$, in agreement with the observationally constrained spectral tilt of the primordial power spectrum of scalar perturbations, while the range for $\Upsilon$ was chosen in order to satisfy observational bounds on $\Omega_{\mcl K}$.

Fig.~\ref{fig:DeltaEta1D} demonstrates that for $\mu=3.5$ and $\mu=3.0$ the {\it effective} frequency drops significantly in some narrow bands of $\Upsilon$ [or equivalently bands of $a_{\mrm{b}}$, see Eq.~(\ref{eq:ab_and_M})] while Fig.~\ref{fig:DeltaEta2D} shows the existence of this band for the case $3.0\leq\mu\leq 4.0$ in more detail.  These bands correspond to values of $\Delta\eta\simeq 2\pi$ with $\delta\eta\ll 1$.  The existence of these narrow bands is crucial.  This is because, given that $a_0$ is at the very least equal to $20h^{-1}$ Gpc, the {\it effective} frequency of oscillations, $a_0\delta\eta$, can only be small enough for such features to be observable in the CMB or in the matter power spectrum if $\delta\eta\ll 1$.

In fact, the results of Figs.~\ref{fig:DeltaEta1D} and \ref{fig:DeltaEta2D} suggest a new possibility for an indirect determination of the spatial curvature $\Omega_{\mcl K}$ should oscillations linear in $k$ and attributable to a bouncing cosmology of the type described in this paper be measured in the temperature anisotropy map and/or in the matter density power spectrum.  Given accurate measurements of $n_{\mrm{s}}$ and $a_0\delta\eta$, a good estimate of $\mu$ can be obtained through Eqs.~(\ref{eq:eps1}) and (\ref{eq:eps2}).  Given an accurate measurement of $a_0\delta\eta$, the value of $\Upsilon$ is known so that the value of the scale factor at the bounce is recovered.  Furthermore, given $V(\varphi)$, the value of $\mu$, say of order $3.0$, and the value of $\ab$ are enough to obtain $\Delta\eta$ numerically.  Finally, given the knowledge of $\Delta\eta$ and that of $a_0\delta\eta$, the value of $a_0$ is easily obtained.  Combined with $H_0$, it yields $\Omega_{\mcl K}$.

At last, one can estimate the number of oscillations per decade $\mfk D$ in the wavenumber $k$ from
\be
\mcl N_{\mrm{osc}}\lb\mfk D\rb=\frac{\delta\eta}{\pi}\,10^{i(\mfk D)},
\ee
where $i(\mfk D)$ is an integer.  For example, for $\mu=3\,\mpl$ [see Eq.~(\ref{eq:potential})], $\ab\simeq 2.5\times 10^5\lP$ and setting $\delta \eta=0.001$, the observable range $10^{-3}\,h\,\mrm{Mpc}^{-1}<k_{\mrm{phys}}<10\,h\,\mrm{Mpc}^{-1}$ is $10^2\lsim k \lsim 10^6$ and $|\Omega_{\mcl K}|\lsim 10^{-3}$.  This gives a total of four decades, $\mfk D=1,2,3,4$, with $i(\mfk D)=3,4,5,6$. In this case, $\mcl N_{\mrm{osc}}\lb\mfk D\rb$ ranges from $0.3$ (for $\mfk D=1$) to $3\times 10^2$ (for $\mfk D=4$).  If instead, $\delta\eta=0.01$, then $\mcl N_{\mrm{osc}}\lb\mfk D\rb$ ranges from $3$ (for $\mfk D=1$) to $3\times 10^3$ (for $\mfk D=4$).  To see that these estimates are correct, see Sec.~\ref{sub:mps}.
\FIGURE{
\includegraphics[height=7.85cm]{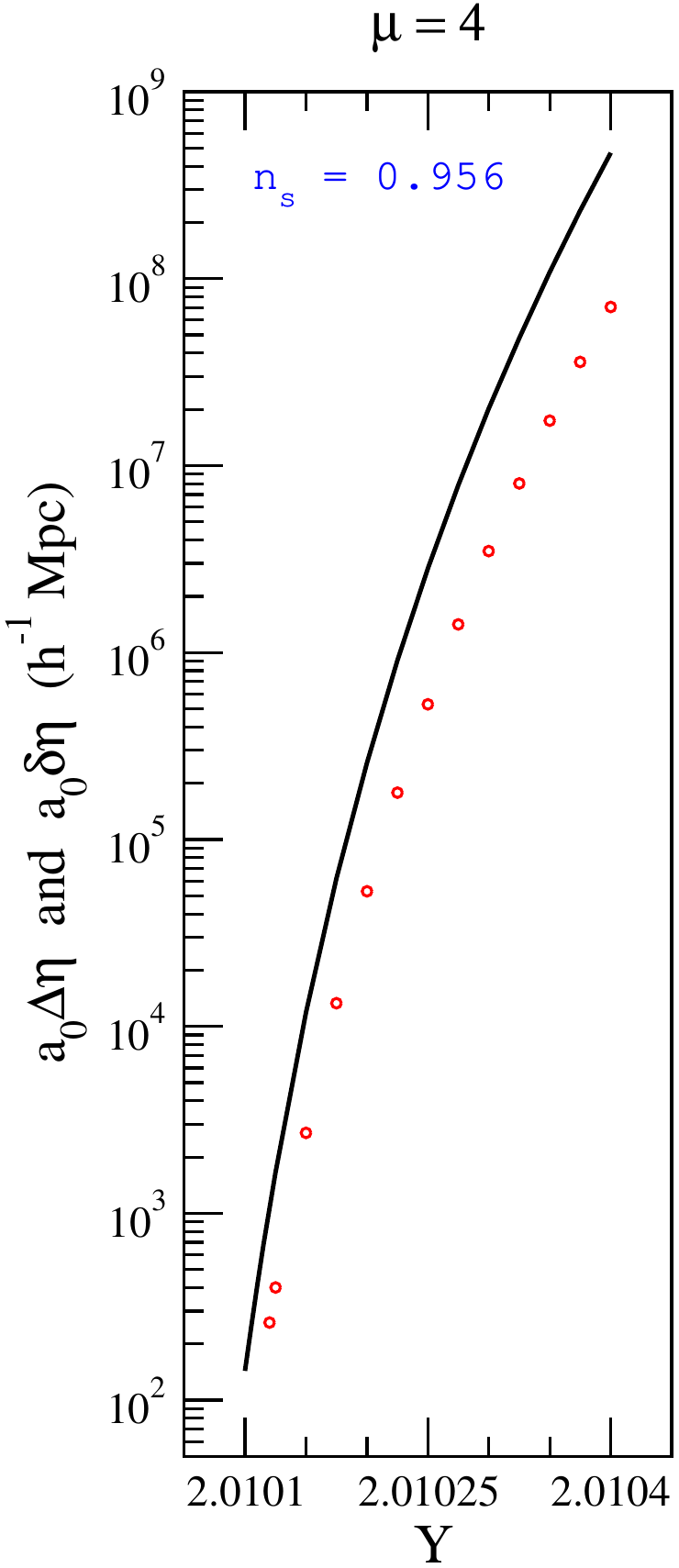}
\includegraphics[height=7.85cm]{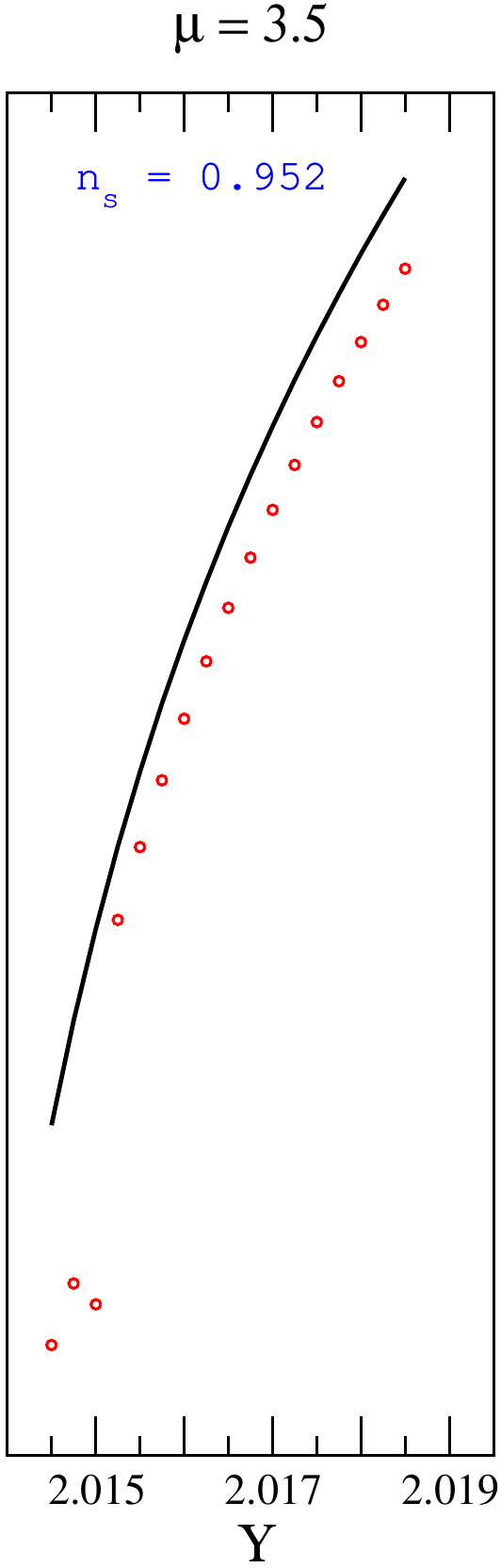}
\includegraphics[height=7.85cm]{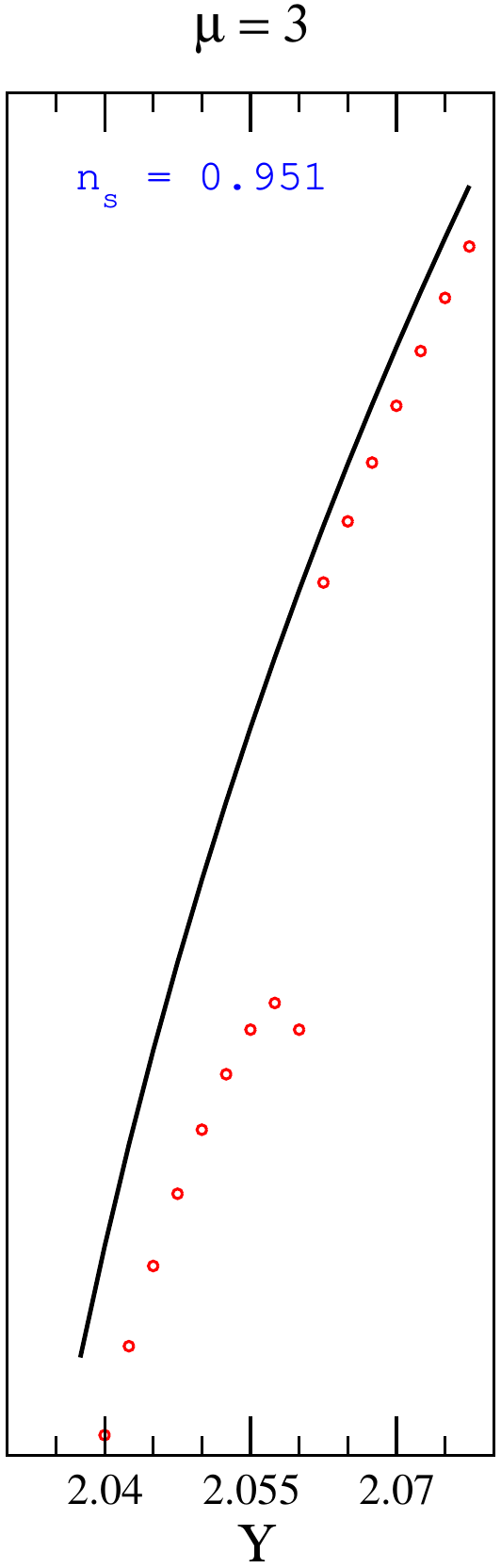}
\includegraphics[height=7.85cm]{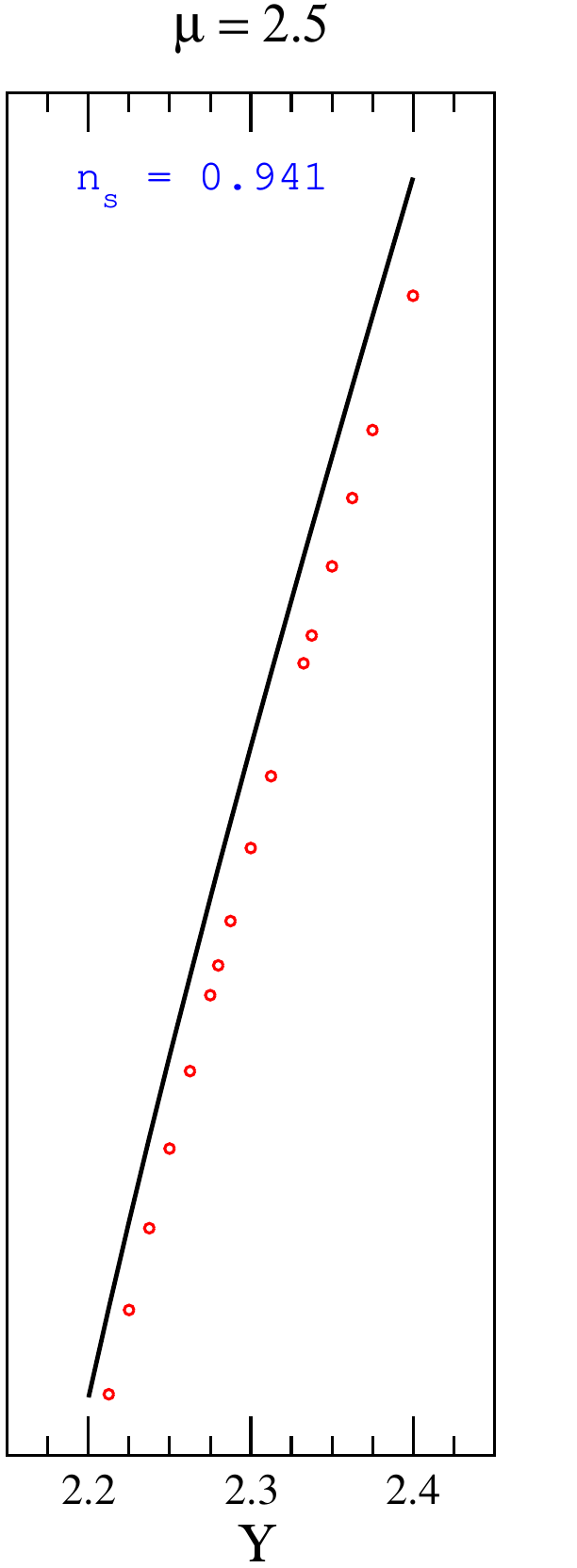}
\includegraphics[height=7.85cm]{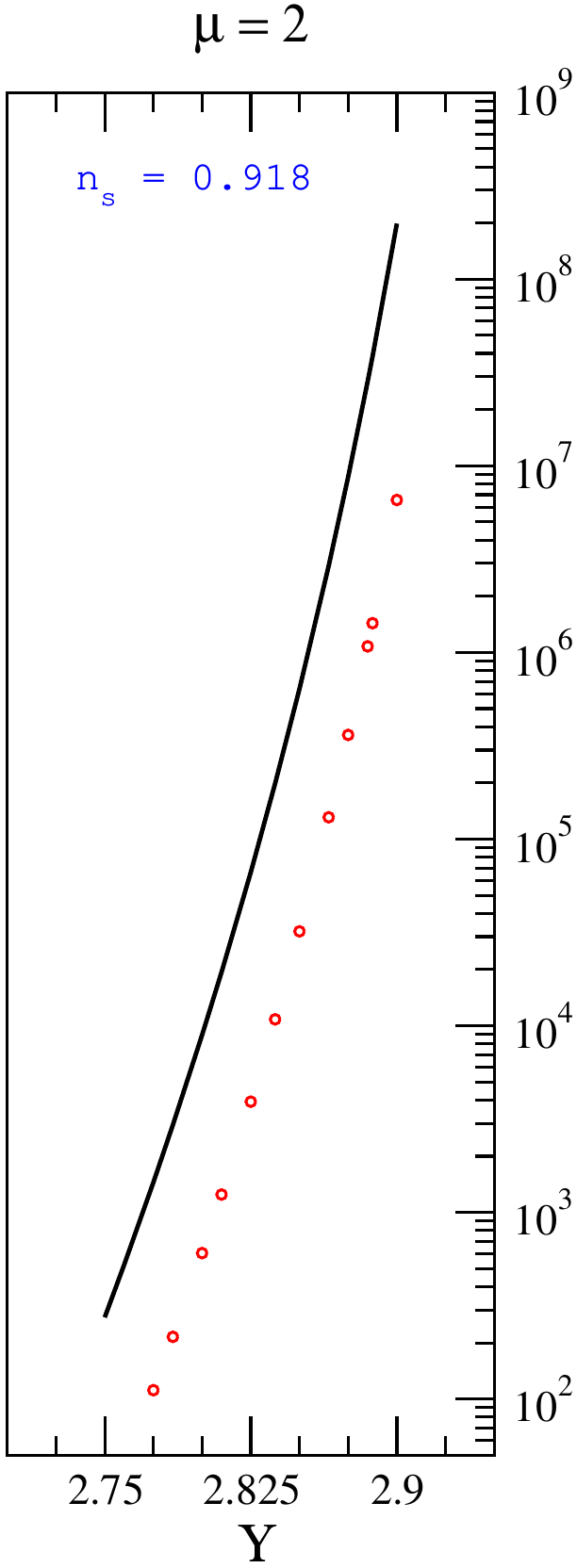}
\caption{Numerical integration of $a_0\Delta\eta$ and $a_0\delta\eta$.  The full lines are $a_0\Delta\eta$, while the red circles are $a_0\delta\eta$.  For $\mu\geq 3$, $n_{\mrm s} > 0.95$ and $\Delta\eta\simeq 2\pi$ so that there exist regions in which $a_0\delta \eta$ drops to significantly lower values, see Fig.~\ref{fig:DeltaEta2D}.}
\label{fig:DeltaEta1D}
}
\FIGURE{
\includegraphics[width=10cm]{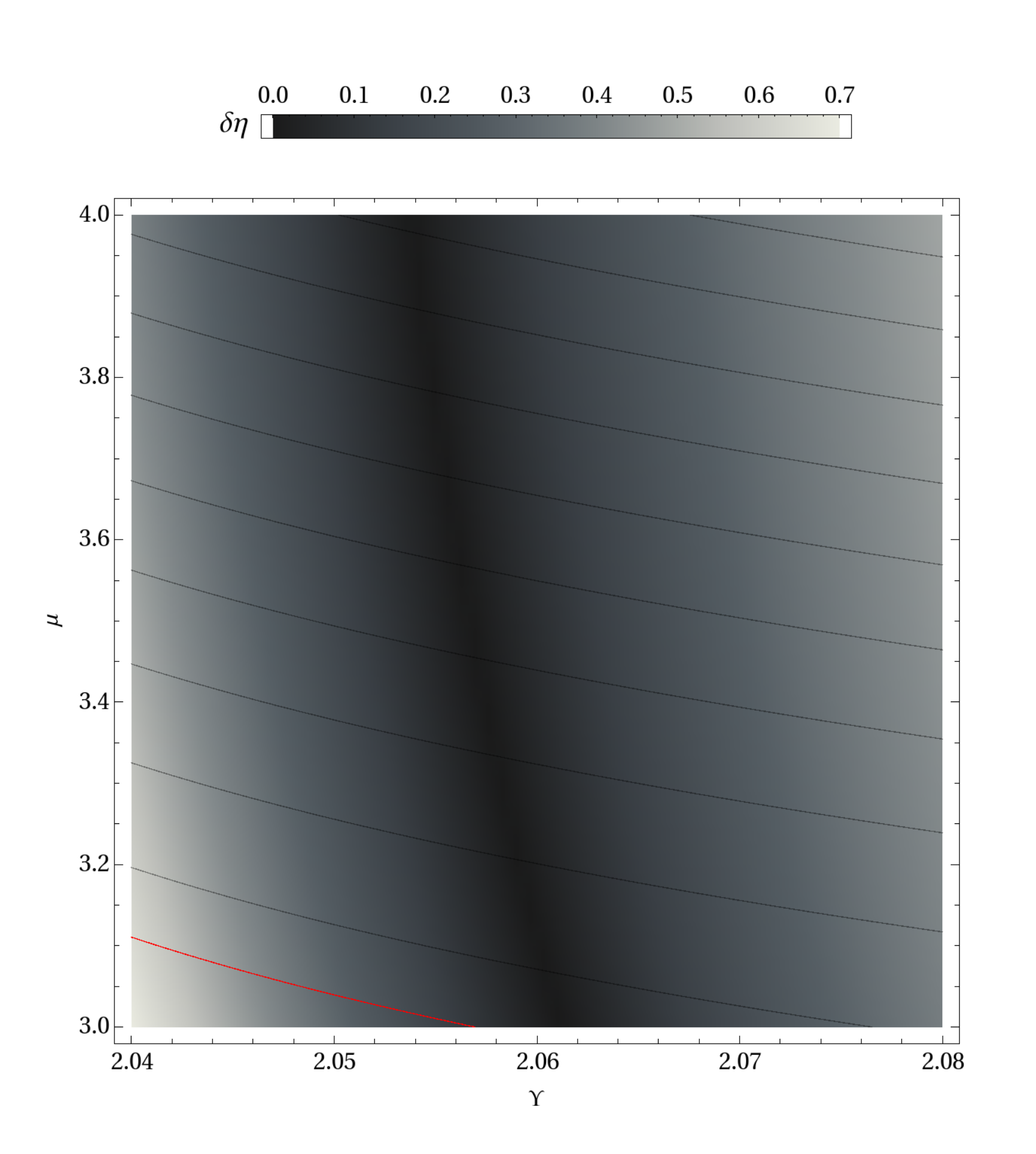}
\caption{Numerical integration of $\delta\eta$.  The thin lines represent $|\Omega_{\mcl K}|$ isolines on a logarithmic scale, the red one denoting $|\Omega_{\mcl K}|=2\times 10^{-3}$.  In the range $3\leq\mu\leq 4$, $n_{\mrm{s}}\simeq 0.95$.}
\label{fig:DeltaEta2D}
}
\section{CMB and matter power spectra}
\label{sec:obs}
In previous sections, we derived the scalar and tensor primordial power spectra and showed that oscillations induced by the combined effect of the deviation of incoming perturbations from the vacuum state and by the characteristic time scale $\Delta\eta$ appear in $\mcl P_{\zeta}$ and $\mcl P_h$.  In addition, we found that measuring the scalar spectral index $n_{\mrm s}$ and the frequency of oscillations makes possible, within the scope of this model, an indirect determination of the values of $\ab$ and of $\Omega_{\mcl K}$ today.

The oscillations can however only be measured if they are transferred to the angular power spectrum of the CMB (the $\mcl C_{\ell}$'s) and/or to the matter power spectrum $\mcl P_\delta$.  The transfer function relating $\mcl P_\delta$ to $\mcl P_\zeta$ is simply a $k$-dependent multiplicative function and oscillations can be expected to be transfered from $\mcl P_\zeta$ to $\mcl P_\delta$ unchanged, see Sec.~\ref{sub:mps}.  The $\mcl C_{\ell}$'s, on the other hand, are related to $\mcl P_{\zeta}$ through a complicated expression so that the transfer of oscillations to the angular power spectrum of the CMB is by no means obvious.  This question is investigated in this section.  We first write down the analytic expressions for the angular power spectra in the range $1\leq \ell \leq 10$. We then give numerical examples of $\mcl C_{\ell}^{\mrm{s}}$ over the entire $\ell$ range using the \texttt{CAMB} code~\cite{Lewis:1999bs}.
\subsection{Scalar anisotropy spectrum}
\label{sub:scalarCl}
For $|\Omega_{\mcl K}|\ll 1$ and for $\ell\lsim 10$ but for $\mcl K=+1$, the scalar multipole moments $\mcl C_{\ell}^{\mrm{s}}$ can be related to the primordial scalar curvature perturbation spectrum $\mcl{P}_{\zeta}$ by~\cite{Lehoucq:2002wy,Riazuelo:2002ct,PPJPU}
\be
\mcl C_{\ell}=\frac{2\pi^2}{25}\sum_{n=2}^{\infty}\frac{\mcl P_{\zeta}M_{n}\left|P_{-1/2+n}^{-1/2-\ell}\lsb\cos\lb\chi\rb\rsb\right|^2}{n\lb n+1\rb\lb n+2\rb\sin\lb\chi\rb},
\ee
with
$$
M_{n}=\lcb
\begin{array}{lcr}
\displaystyle\prod_{i=0}^{\ell}\lb i+1\rb^2-n^2 & \qquad & \text{for } \ell\leq n\\
0 & \qquad & \text{for } \ell>n
\end{array}
\right.
\label{eq:ClSum}
$$
and where $P_{-1/2+n}^{-1/2-\ell}$ are the associated Legendre polynomials while $\chi=\eta_0-\eta_{\mrm{lss}}\simeq \ell_{H}/a_0=\sqrt{|\Omega_{K}|}$ is the (conformal) radial distance to the last scattering surface, with $\eta_0$ and $\eta_{\mrm{lss}}$, the conformal times today and at the time of last scattering, respectively.  With the exception that the integral over $k$ should be replaced by a sum over $n$, the textbook expression used in the case $\mcl K=0$, see \cite{PPJPU,Durrer:2001gq}, is a very good approximation to Eq.~(\ref{eq:ClSum}) for $|\Omega_{\mcl K}|\ll 1$.  The expression for the scalar $\mcl C_{\ell}$'s can be carried over to the case of the $\mcl C_{\ell}$'s for tensor perturbations.

 The presence of the slowly oscillating sinusoidal functions in $\mcl P_{\zeta}$ introduces an oscillation in the kernel of the sum in Eq.~(\ref{eq:ClSum}).  In particular, one can in general expect constructive or destructive interference between the oscillations of the $P_{-1/2+n}^{-1/2-\ell}$'s having as a characteristic scale $\chi$ with the oscillations of the cosine and sine terms in $\mcl P_{\zeta}$ of characteristic frequency $2\Delta\eta$. Summed over $n$, this induces slow oscillations in the $\mcl C_{\ell}$'s at small $\ell$ values.  The precise shape and locations of the deviations from the standard $\mcl C_{\ell}$'s induced by the oscillations depend on $\Omega_{\mcl K}$ via $\chi$ and on $\delta\eta$ while their amplitude depends on $\varsigma$.  

The COBE normalization can be expected to be modified in two ways.  First, note that for $\varsigma > 1$ the constant term in $\mcl P_{\zeta}^{\mrm{osc}}$ is equal to $2\varsigma^2-1$.  The amplitude of the spectrum is therefore modified by this first factor.  Secondly, the amplitude of the spectrum will also be modified by a contribution coming from the oscillatory part.  This contribution depends, once again, on $\varsigma$, on $\chi$ and on $\Delta\eta$.

In Figure \ref{fig:Cl}, we present the results of a numerical simulation using the \texttt{CAMB} code~\cite{Lewis:1999bs}.  The quantities shown in the dashed red and full blue lines are defined by
\be
\Delta \mcl C_{\ell}=\mcl C_{\ell}^{\mrm{THEORY}}-\mcl C_{\ell}^{\mrm{WMAP}}
\ee
where the theoretical $\mcl C_{\ell}$'s are the $\mcl C_{\ell}$'s obtained, respectively, either from an inflationary cosmology or from the bouncing cosmology described in this paper.  The error bars are the WMAP error bars for the measured $\mcl C_{\ell}$'s. The $\Delta \mcl C_{\ell}$'s were computed for $\varsigma=1.01,\,1.1$ and $1.2$ and for $\delta\eta=0.001$ and $0.01$.  These results were obtained for $\mu=3$, $N_{\star}=50$, $\Upsilon\simeq2.055$.  For this choice of parameters, $\epsilon_1=0.0025$, $\epsilon_2=0.045$, $N_{\mrm{inf}}=65$, and for a period of radiation and matter domination (neglecting reheating) lasting 65 $e$-folds~\cite{Liddle:2003as}, such that $\Omega_{\mcl K}=-0.002$.  The figure mainly illustrates that at the frequency $\delta\eta=0.01$, and for $\varsigma=1.01$, the oscillatory features in the primordial spectrum appear to improve the fit to the WMAP data in comparison with what is obtained in the absence of oscillations in $\mcl P_{\zeta}$.  For larger values of $\varsigma$, such as 1.1 and 1.2, the effect of the oscillations worsens the fit near the first accoustic peak at $\ell\simeq 200$ over a widening range of $\ell$ values as $\varsigma$ is increased.  At higher frequency, for $\delta\eta=0.01$, the resulting $\mcl C_{\ell}$'s have large fluctuations that appear to be phase-shifted with respect to the result obtained in the inflationary case.  At low $\ell$, oscillations are not visible but there remains a change in amplitude with respect to the $\Delta \mcl C_{\ell}$'s obtained from inflation.  This could be expected from Eq.~(\ref{eq:ClSum}) for the values of $\varsigma$ and $\delta\eta$ chosen.
\FIGURE{\includegraphics[width=7.25cm]{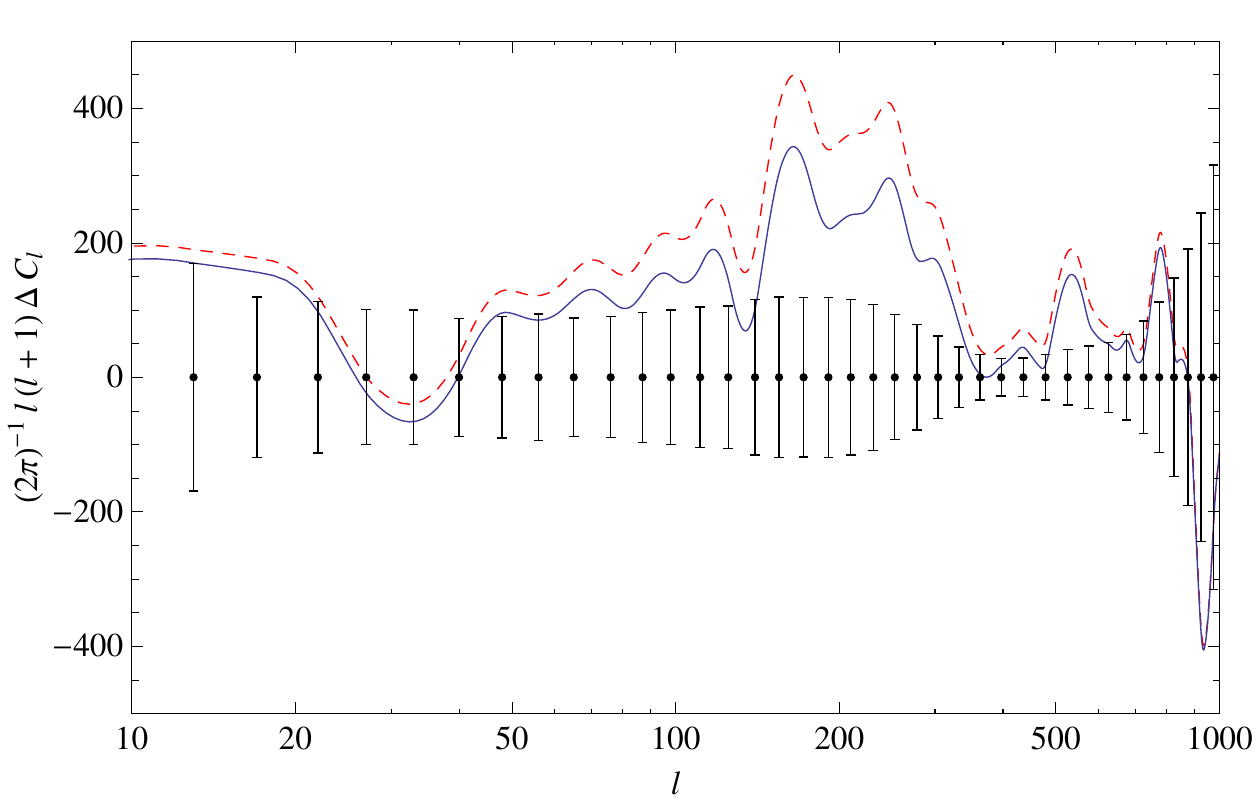}\hspace{0.25cm}\includegraphics[width=7.25cm]{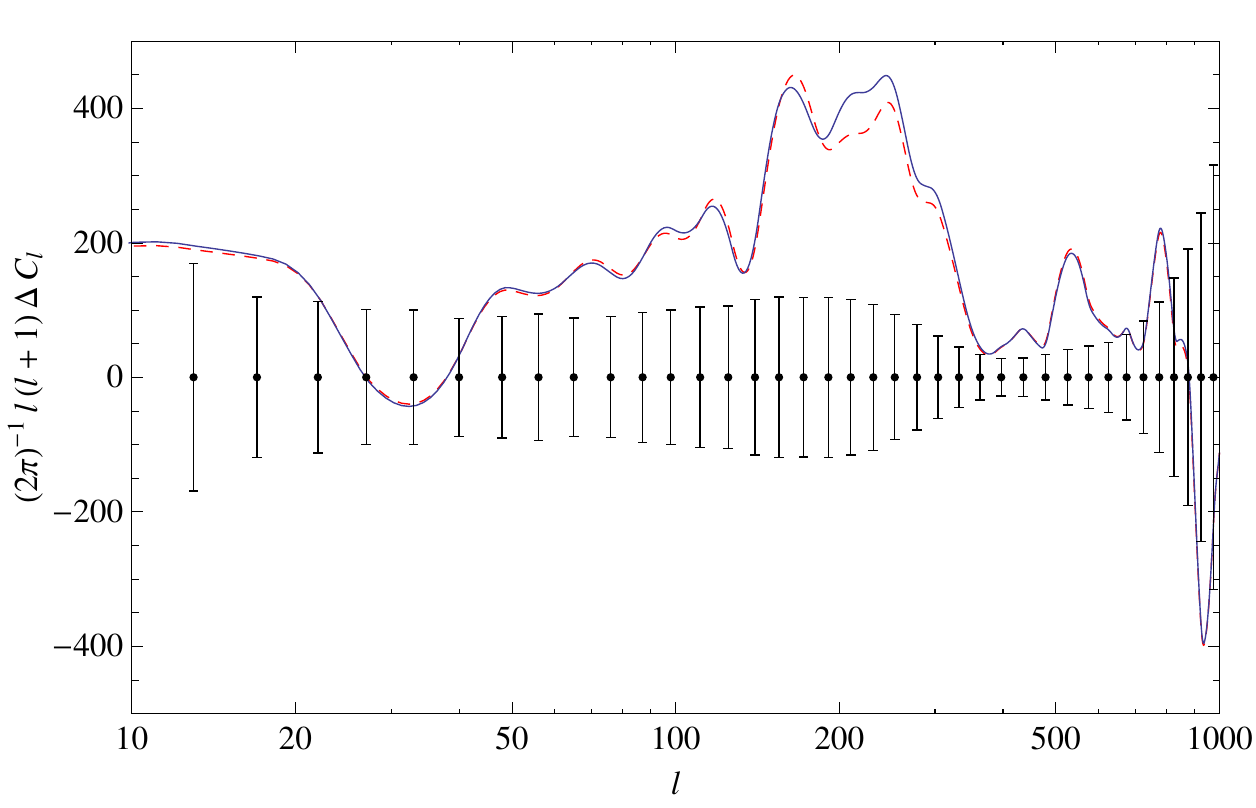}\\
\includegraphics[width=7.25cm]{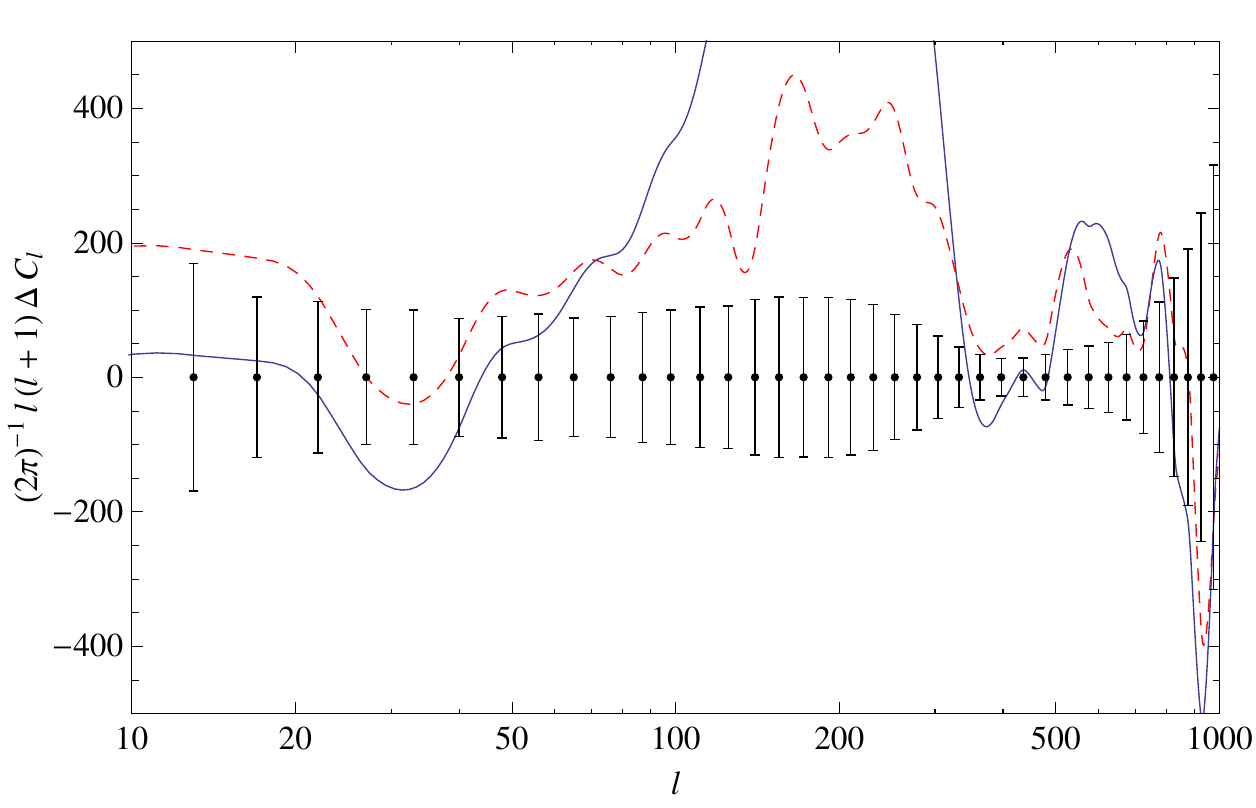}\hspace{0.25cm}\includegraphics[width=7.25cm]{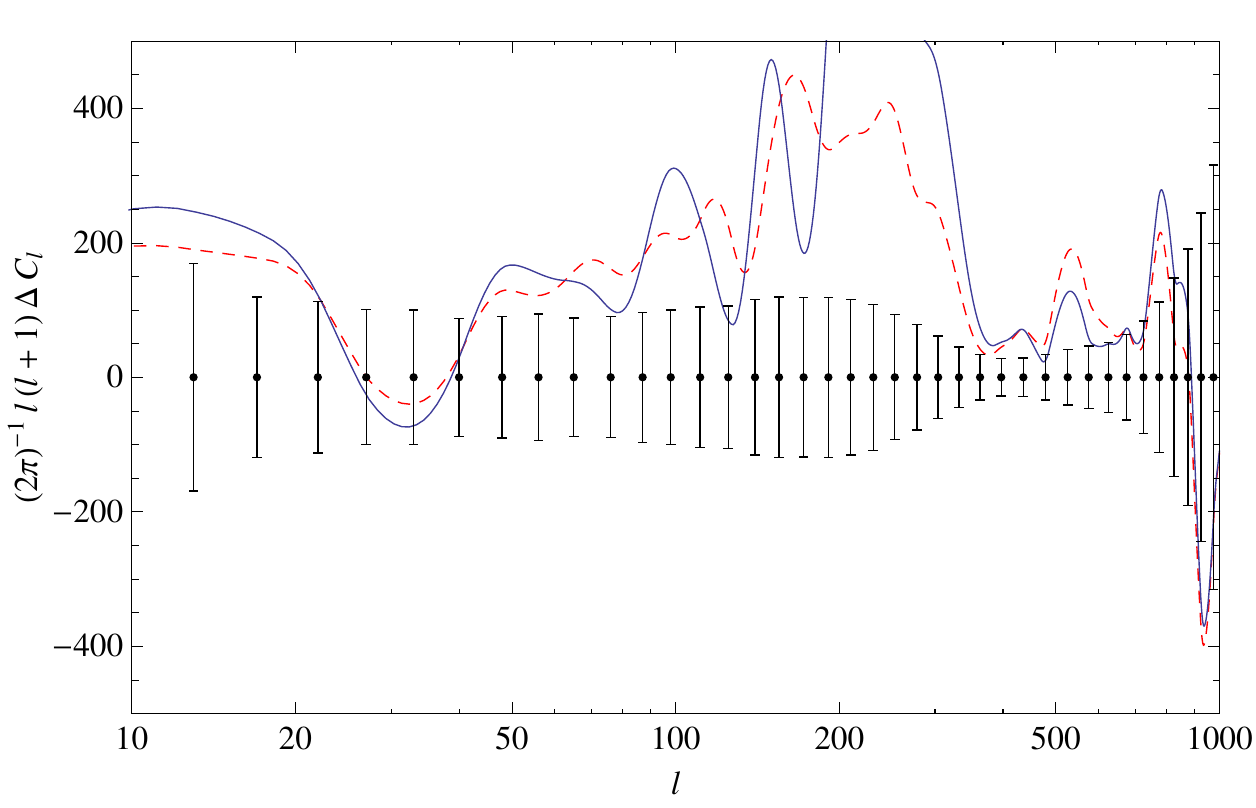}\\
\includegraphics[width=7.25cm]{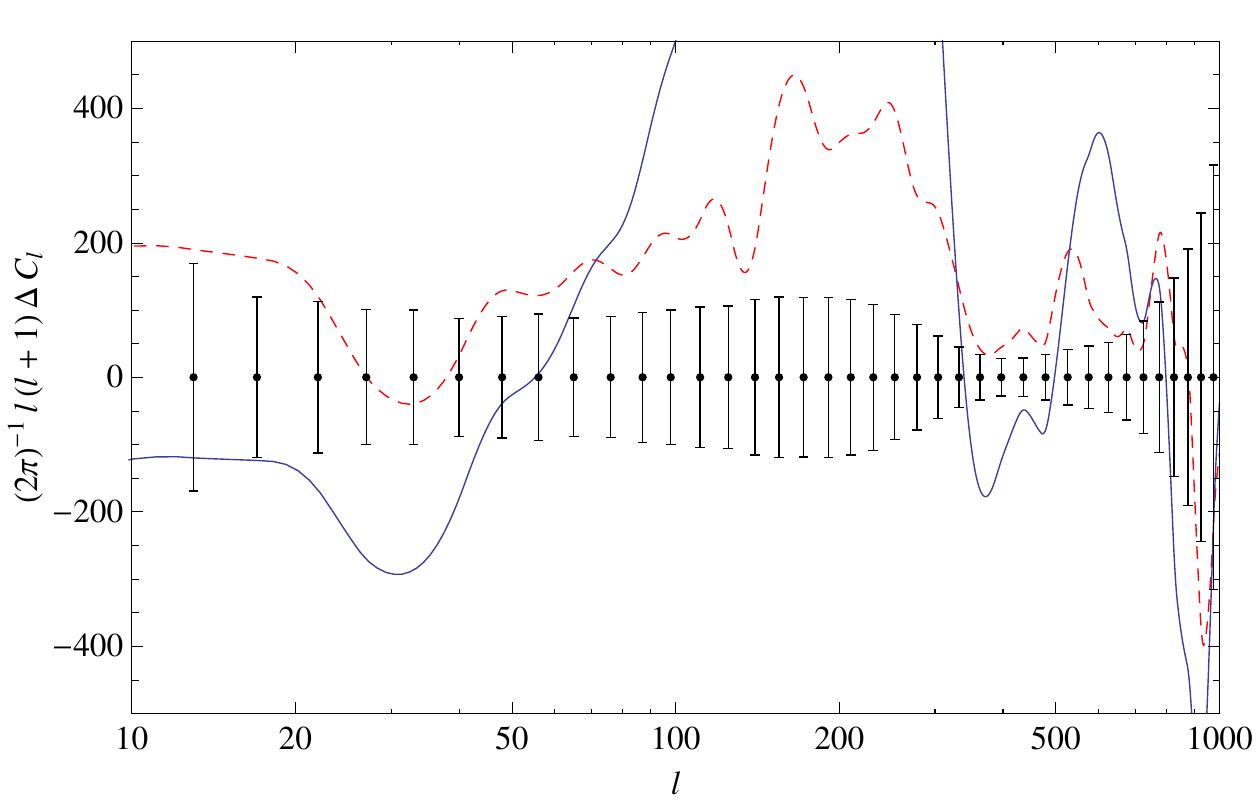}\hspace{0.25cm}\includegraphics[width=7.25cm]{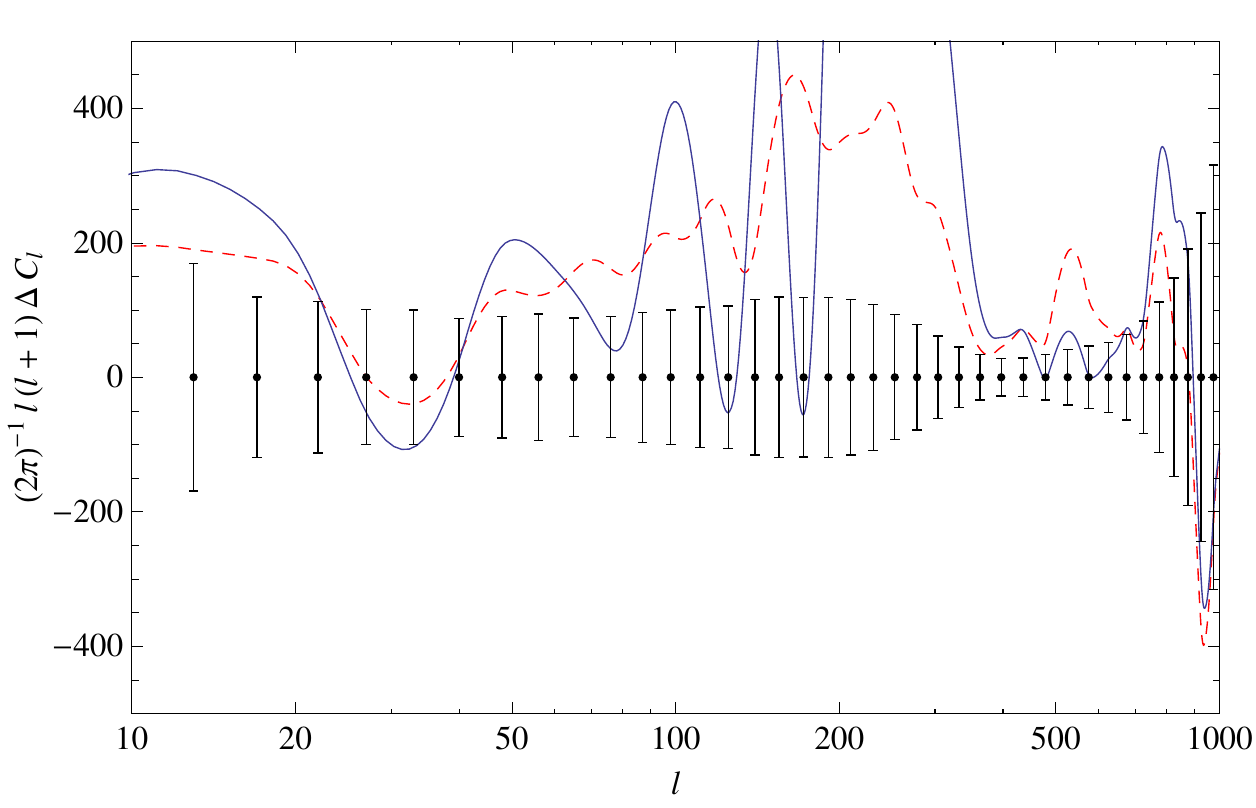}
\caption{$\Delta \mcl C_{\ell}$ spectra for $\varsigma=1.01$, $1.1$ and $1.2$, from top to bottom respectively and for $\delta\eta=0.001$ and $\delta\eta=0.01$ from left to right, generated using \texttt{CAMB}. The error bars are the WMAP error bars on the measured values of the $\mcl C_{\ell}$'s, the red dashed lines are the $\Delta \mcl C_{\ell}$'s for an inflationary cosmology and the blue line in full represents the $\Delta \mcl C_{\ell}$'s for the bouncing cosmology presented in this work.}
\label{fig:Cl}}
\subsection{Matter density power spectrum}
\label{sub:mps}
The matter power spectrum $\mcl P_{\delta}$ is related to the primordial power spectrum $P_\zeta(k)$ by a $k$-dependent transfer function $T(k)$, \ie one can write $P_{\delta}(k) = T(k)\,P_\zeta(k)$.  Roughly speaking, $T(k)\simeq 1$ for small $k$, and $T(k)\propto k^{-2}$ at large wavenumbers.  The oscillating behaviour observed in the primordial perturbation spectrum is therefore expected to be transmitted to the matter power spectrum.  Again, using the \texttt{CAMB} code~\cite{Lewis:1999bs} with standard cosmological parameter values, we have computed the theoretical matter power spectra for the values of the parameters specified in Section~\ref{sub:scalarCl}.  The results are shown in blue in Fig.~\ref{fig:matter}.  The theoretical spectra were then convolved with the observational window functions of the Sloan Digital Sky Survey (SDSS)~\cite{Tegmark:2003ud} and are shown by the red dots in Fig.~\ref{fig:matter}.  These theoretical results are compared with the SDSS data, shown in black in the figure.  The convolution evidently smoothes out the oscillations, and it is clear from Fig.~\ref{fig:matter} that there exist values of $\varsigma$ and $\Delta\eta$ such that the resulting matter spectra are fully degenerate with those obtained from a standard slow roll power spectrum.  The range of scales displayed is centered on the second decade ($\mfk D=2$) of observable scales.  The number of oscillations in this decade is about 3 or 4 for $\delta\eta=0.001$ and about 20 to 30 for $\delta\eta=0.01$.  These are the estimates that were found in Sec.~\ref{sec:bouncingtimescale}.
\FIGURE{\includegraphics[width=7.25cm]{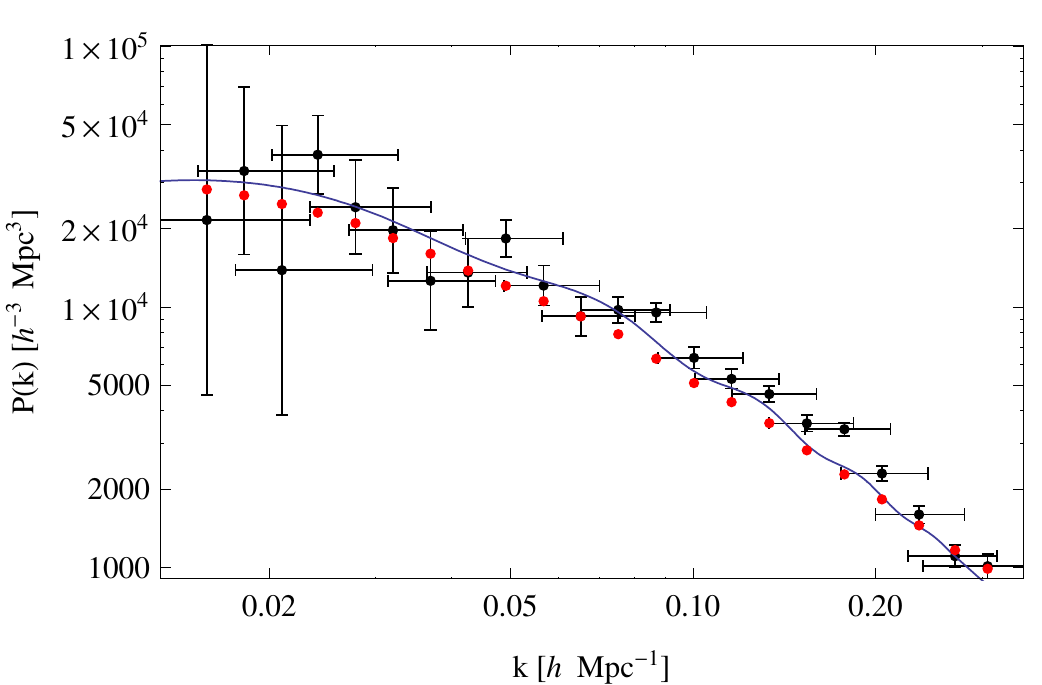}\hspace{0.25cm}\includegraphics[width=7.25cm]{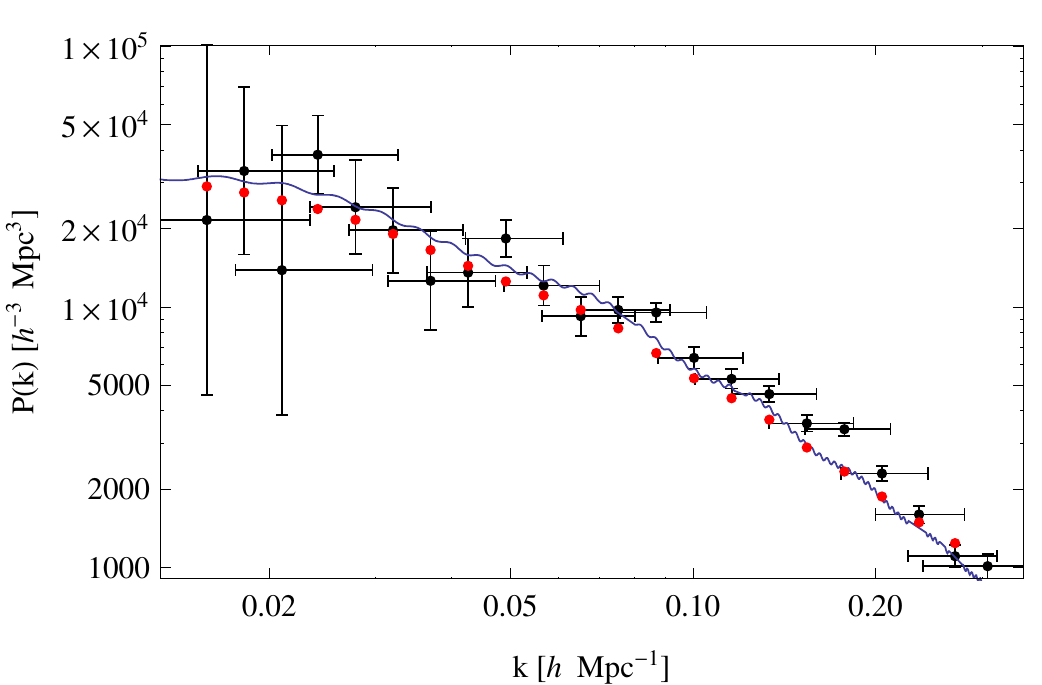}\\
\includegraphics[width=7.25cm]{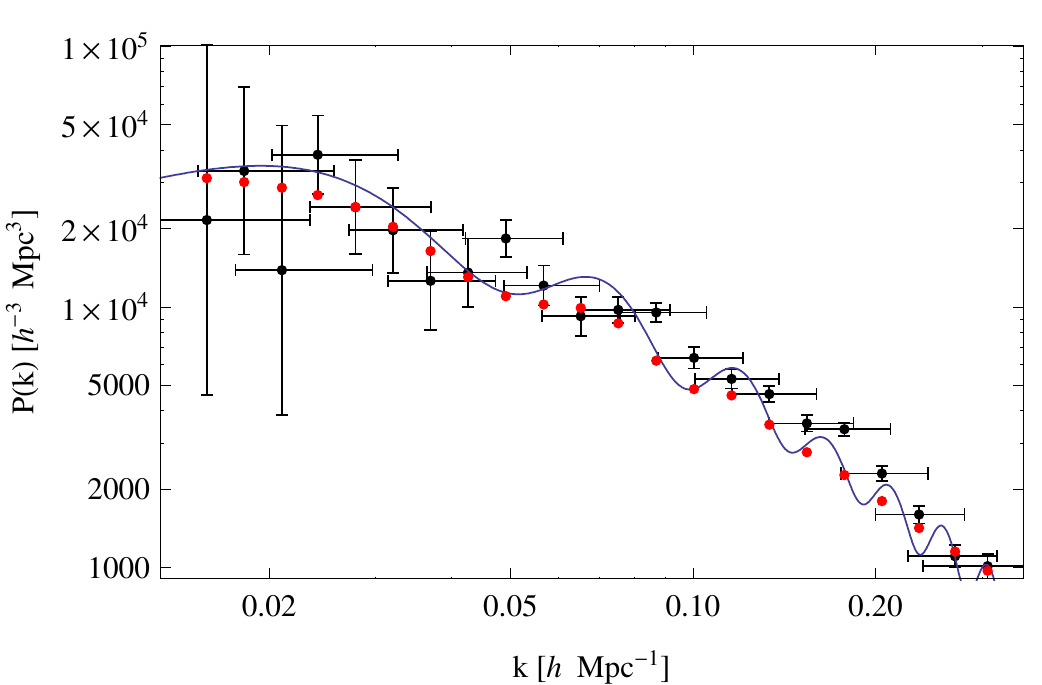}\hspace{0.25cm}\includegraphics[width=7.25cm]{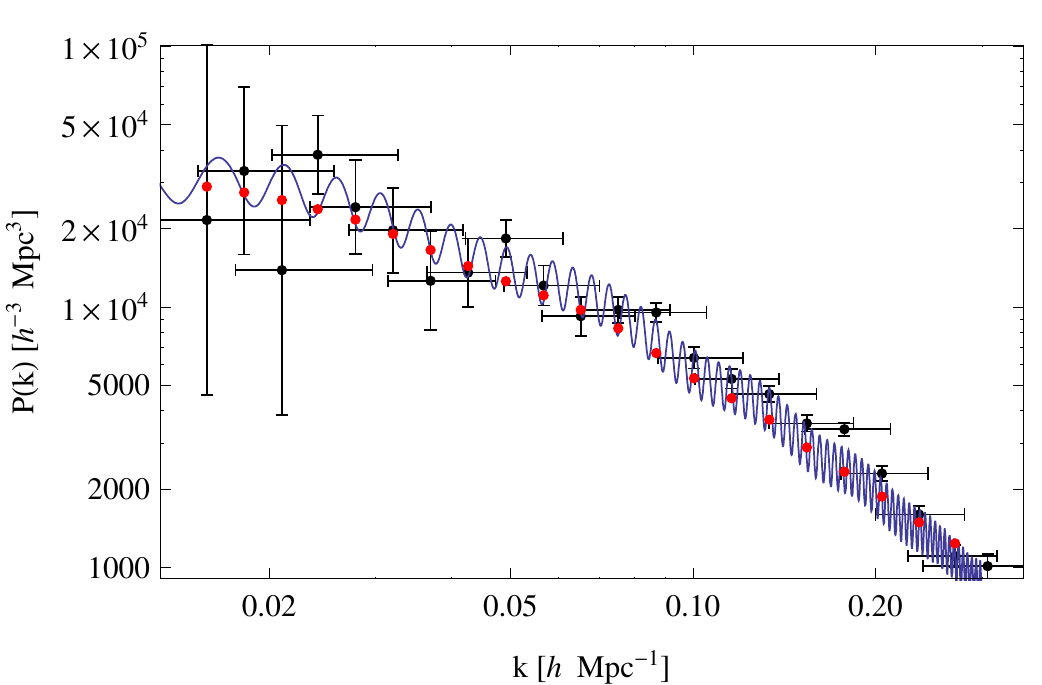}\\
\includegraphics[width=7.25cm]{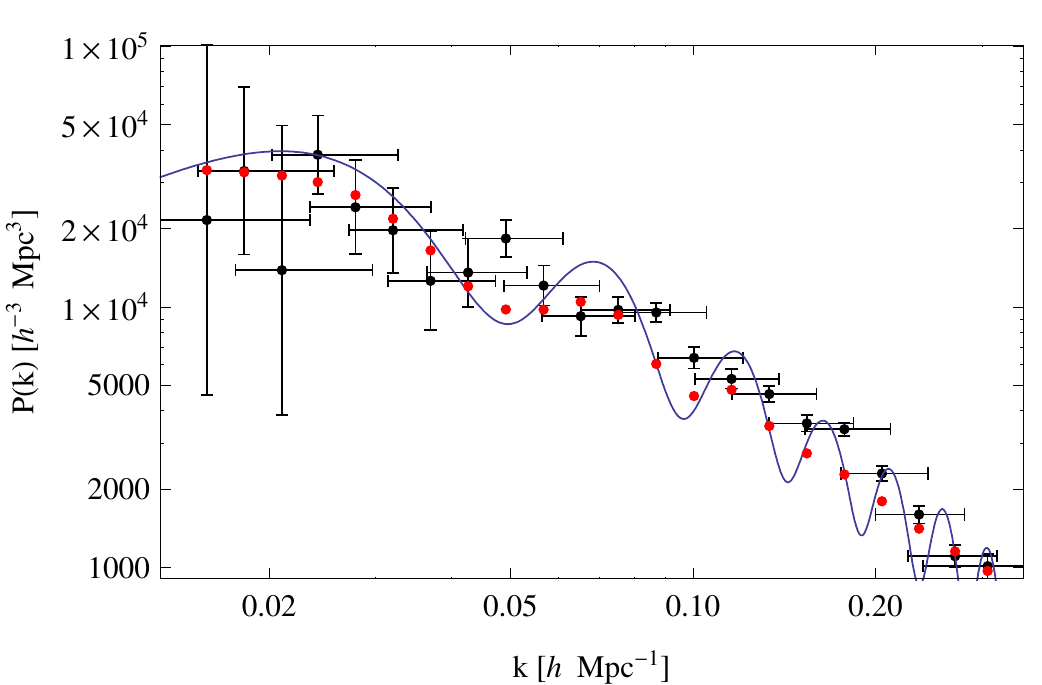}\hspace{0.25cm}\includegraphics[width=7.25cm]{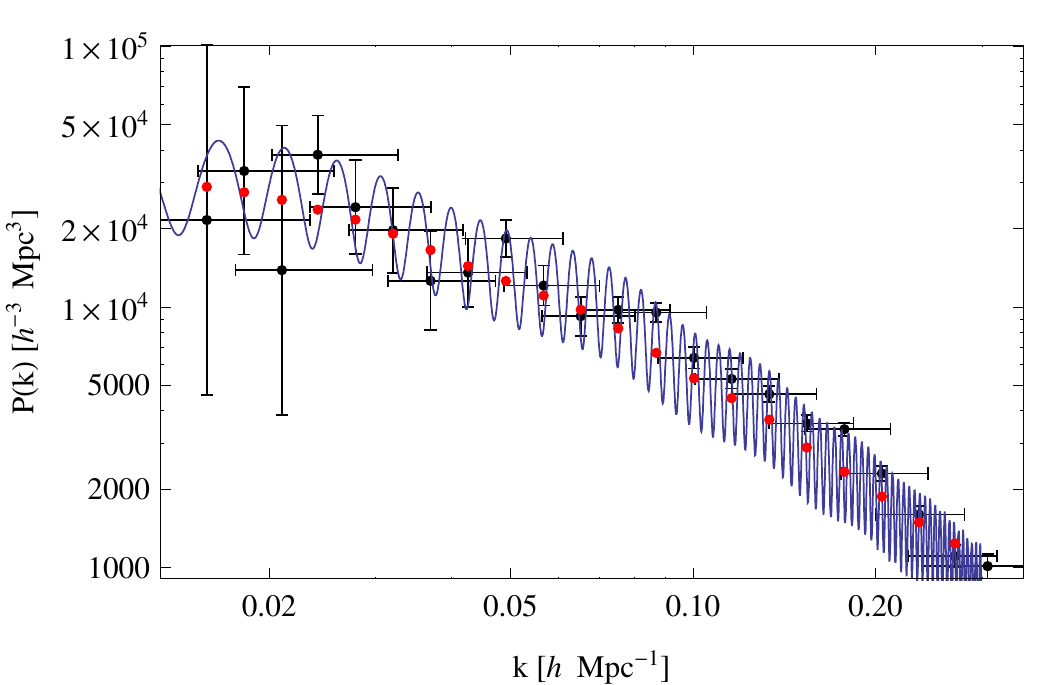}\\
\caption{Matter power spectrum for $\varsigma=1.01$, $1.1$ and $1.2$, from top to bottom respectively and for $\delta\eta=0.001$ and $\delta\eta=0.01$ from left to right. The black dots with error bars are the data points from the Sloan Digital Sky Survey (SDSS)~\cite{Tegmark:2003ud}, the blue line is the theoretical prediction of the bouncing cosmological model obtained using \texttt{CAMB}, and the red dots are the simulated data points obtained by convolving the blue line with the SDSS window functions.}
\label{fig:matter}}
\section{Conclusions}
\label{sec:conclusions}
This paper refines and completes the analysis begun in Ref.~\cite{Martin:2003sf} and continued in Ref.~\cite{Falciano:2008gt}.  In Ref.~\cite{Martin:2003sf}, the authors focused on the immediate vicinity of the bouncing phase and conducted a detailed analysis of the bounce-inducing background cosmology and on the transfer of fluctuations through the bounce by modifying the exact solution obtained when considering a de Sitter universe with closed spatial sections.  In Ref.~\cite{Falciano:2008gt}, the framework used in Ref.~\cite{Martin:2003sf} was used to show that for symmetric or quasi-symmetric general relativistic bouncing cosmologies with $\mcl K=+1$ the peak in the potential for the variable $u$ could never be large so that the metric perturbation is always sub-Hubble in the vicinity of the bounce.  A cosmology smoothly connecting a contracting phase and a bouncing phase to an inflationary phase was then proposed and analyzed at both the background and linear perturbations level.  In the present work, we have provided a detailed summary of the background cosmology exploited in Ref.~\cite{Falciano:2008gt} and have performed a much more detailed calculation of the transfer of perturbations through the contracting, bouncing and inflationary phases of the cosmological background discussed in Ref.~\cite{Falciano:2008gt}.  We have parameterized the initial state of perturbations prior to the bounce in terms of the Bunch Davies vacuum state and used the solution for the Mukhanov-Sasaki variable (far from the high spatial curvature region) in terms of Hankel functions, to reduce the number of unknown parameters needed to define the initial state of first order perturbations, see Sec.~\ref{sub:initialstate}.  We then computed the effects of the bounce and the choice of initial conditions on the scalar and tensor mode primordial spectra $\mcl P_{\zeta}$ and $\mcl P_h$. We discussed the modified low multipoles of the CMB angular power spectrum, the modified COBE normalization and also provided numerical evidence, using \texttt{CAMB}, that the $\mcl C_{\ell}$'s and matter power spectrum $\mcl P_{\delta}$ are affected by the combined effect of the choice of initial conditions and the bouncing cosmology.

The crucial point is the appearance of oscillations in the power spectra.  These oscillations mainly depend on the initial state of the perturbations through the free parameter $\varsigma$ and on a new cosmological scale $\Delta\eta$.  The former sets the amplitude of the oscillations while the latter sets the oscillatory frequency.  As shown in Sec.~\ref{sec:obs}, there exists values of $\varsigma$ and $\Delta\eta$ for which the oscillations induced by the bouncing phase and by the choice of initial conditions do not conflict with the WMAP and SDSS data.  In fact, there appears to be parameter ranges in which the $\mcl C_{\ell}$'s and the $\mcl P_{\delta}$ derived from $\mcl P_{\zeta}$ are fully degenerate with those obtained from a standard slow-roll primordial spectrum.  The tensor-to-scalar ratio is also modified. It departs from the standard result in two ways, first by a modification of the overall amplitude and secondly by the existence of a scale-depedent oscillation, see Eq.~(\ref{eq:STratio}).

We also identified a new way of indirectly measuring the spatial curvature of the universe, in Sec.~\ref{sec:bouncingtimescale}, assuming the oscillations can be attributed to a bouncing cosmology of the kind described in this paper and can indeed be measured.  The scale factor at the bounce as well as the model parameter $\mu$ determine both $\Omega_{\mcl KK}$ and $\Delta\eta$, thereby establishing a one-to-one relationship between the two. If the frequency of oscillations could be measured together with the spectral index of $\mcl P_{\zeta}$, then $\Omega_{\mcl K}$ is determined, as well as $\ab$.

In Appendix \ref{sec:zetaevol}, we discussed the issue of the growth of perturbations in the contracting phase and pointed out the possible necessity of studying asymmetric bouncing cosmologies.  This will be the subject of future work.  Finally, a well-known problem of contracting spacetimes with spatial curvature is the instability of the background evolution~\cite{Falciano:2008gt,Starobinsky:1980te,Barrow:1980en}.  Again, this will be the subject of future work~\cite{Sebastien}.
\acknowledgments
The authors would like to thank J.~Martin, P.~Peter, C.~Ringeval and D.~Steer for fruitful discussions, useful suggestions, and careful proof-reading of the manuscript. M.~L.~also wishes to thank L.~Sriramkumar for valuable discussions as well as Louvain University and the Harish Chandra Research Institute for their hospitality in the final stages of this work.  M.~L.~is supported by ANR grant SCIENCE DE LISA, ANR--07-BLANC-0339. L.~L.~is partially supported by the Belgian Federal office for Science, Technical and Cultural Affairs, under the Inter-university Attraction Pole Grant No.~P6/11.  S.~C.~is supported by the Belgian Fund for research (F.R.I.A.) and the Belgian Science Policy (IAP VI-11).
\bibliography{references}
\begin{appendix}
\section{Time evolution of curvature perturbations}
\label{sec:zetaevol}
In this Appendix, we discuss the time evolution of curvature perturbations.  We first provide some numerical examples for symmetric as well as for non-symmetric bounces and then discuss the growth of perturbations in the contracting phase.
\subsection{Numerical examples}
\label{sub:numex}
Before gauge fixing, the scalar part of the perturbed metric is given by~\cite{Mukhanov:1990me}
\be
\dd s^2=a^2\lb\eta\rb\lsb -\lb 1+2A\rb\dd\eta^2+\lb1+2C\rb\gamma_{ij}\dd x^i\dd x^j\rsb,
\ee
where $A$ and $C$ are small but gauge-dependent scalar perturbation variables.  The gauge-invariant scalar degrees of freedom are~\cite{Mukhanov:1990me}
\bea
\Phi&=&A+\mcl H\lb B-E'\rb+\lb B-E'\rb',\\
\Psi&=&-C-\mcl H\lb B-E'\rb
\eea
where $B$ and $E$ are the scalar parts of the $g_{0i}$ and $g_{ij}$ components of the perturbed metric and where $\Phi$ and $\Psi$ are equal in the absence of anisotropic stress in the stress-energy tensor. The curvature perturbation on comoving and on constant energy density hypersurfaces, denoted by $\mcl R$ and $\zeta$, respectively, are given by~\cite{BST83,Wands:2000dp}
\be
\mcl R=C-\mcl H\frac{q}{\rho+P}\qquad\text{and}\qquad\zeta=C-\mcl H\frac{\delta\rho}{\rho'}\,,
\ee
where $q$ is the scalar potential of the momentum density $q_i$ of the perturbed energy-momentum tensor, defined as $q_i=\nabla_iq+\overline q^i$ with $\nabla_i\overline q^i=0$.  Both $\mcl R$ and $\zeta$ can be expressed in terms of gauge-invariant quantities as
\be
\mcl R=\Psi+\frac{2\mcl H}{3\lb\mcl H^2+\mcl K\rb}\frac{\Psi'+\mcl H\Phi}{1+w}
\ee
and
\be
\zeta=\Phi-\frac{2\mcl H}{3(1+w)(\mcl H^2+\mcl K)}\lcb \Phi'+\lsb 1-\frac{\mcl K}{\mcl H^2}+\frac{1}{3}\lb\frac{k}{\mcl H}\rb^2\rsb \mcl H\Phi\rcb\,.
\label{eq:zetabst}
\ee
On super-Hubble scales and in the absence of isocurvature perturbations, $\zeta'=0$ if $\Phi'+\mcl H\Phi$ does not diverge (this can be seen from the equation of motion for $\Phi$), and $\mcl R\simeq -\zeta$ provided $\mcl K\simeq 0$. Hence, both $\zeta$ and $\mcl R$ are conserved after Hubble horizon crossing.  In the absence of isocurvature perturbations, this holds in very general circumstances. These two quantities are therefore of primary cosmological interest.
For scalar field matter, $\mcl R$ can also be written in Newtonian gauge as
\be
\mcl R=-\Phi-\mcl H\frac{\delta\varphi}{\varphi'}
\ee
and it can be seen from (\ref{eq:MSV}), taking $\mcl K=0$, that  $\mcl R$ is related to $v$ by $v=-z\mcl R$.  If $\mcl K/a^2 \ll k^2$, we may work with $v$ and $z$ rather than with $\tilde v$ and $\tilde z$.  In this case, $\zeta\simeq -\mcl R=v/z$, and knowledge of $v$ is directly usable to obtain $\zeta$.

In Figure~\ref{fig:simulations}, we show the results of a numerical integration of the background cosmology, see (\ref{eq:F1}) to (\ref{eq:F2}), and curvature perturbation on equal density hypersurfaces $\zeta$, by integrating (\ref{eq:uevol}) and then using (\ref{eq:phitou}) and (\ref{eq:zetabst}), for three sets of initial conditions at the bounce with which the bouncing phase gets progressively more asymmetric.
\FIGURE{\includegraphics[width=4.8cm]{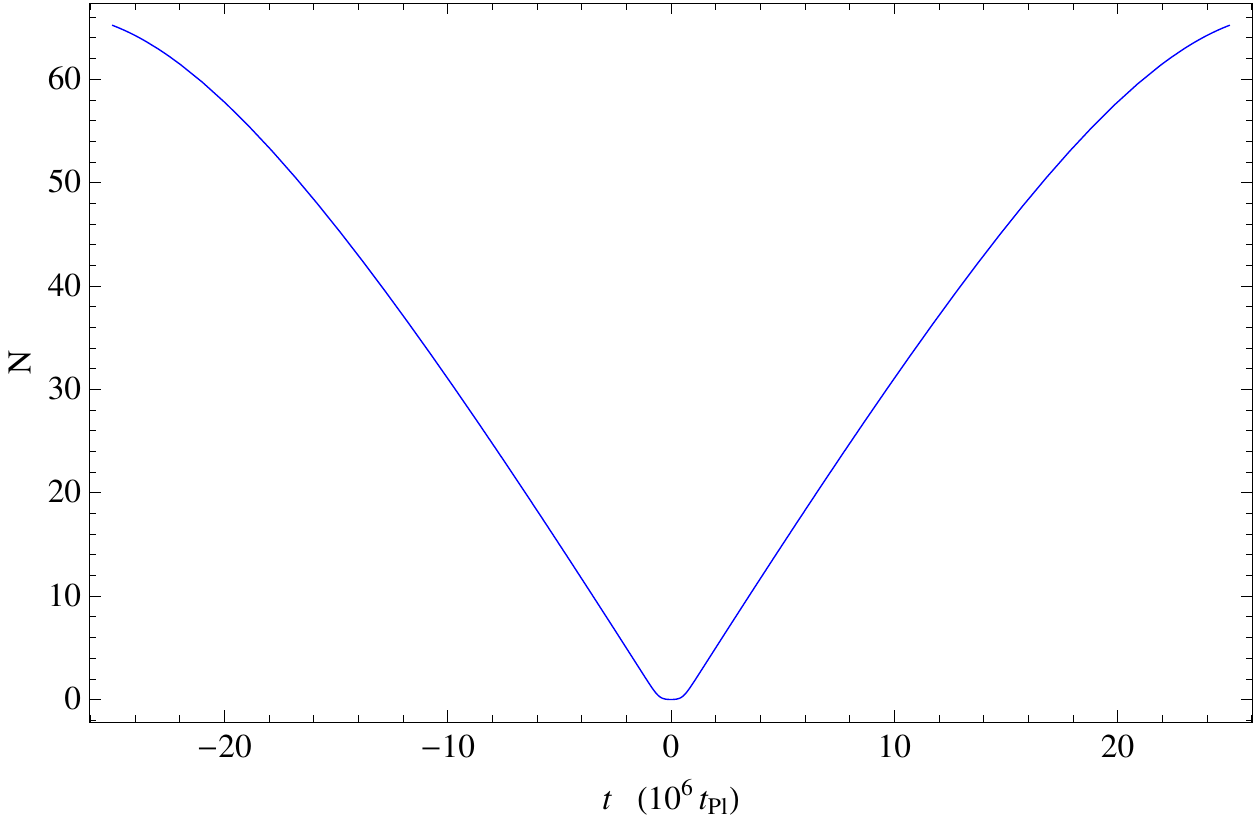}\hspace{0.25cm}\includegraphics[width=4.8cm]{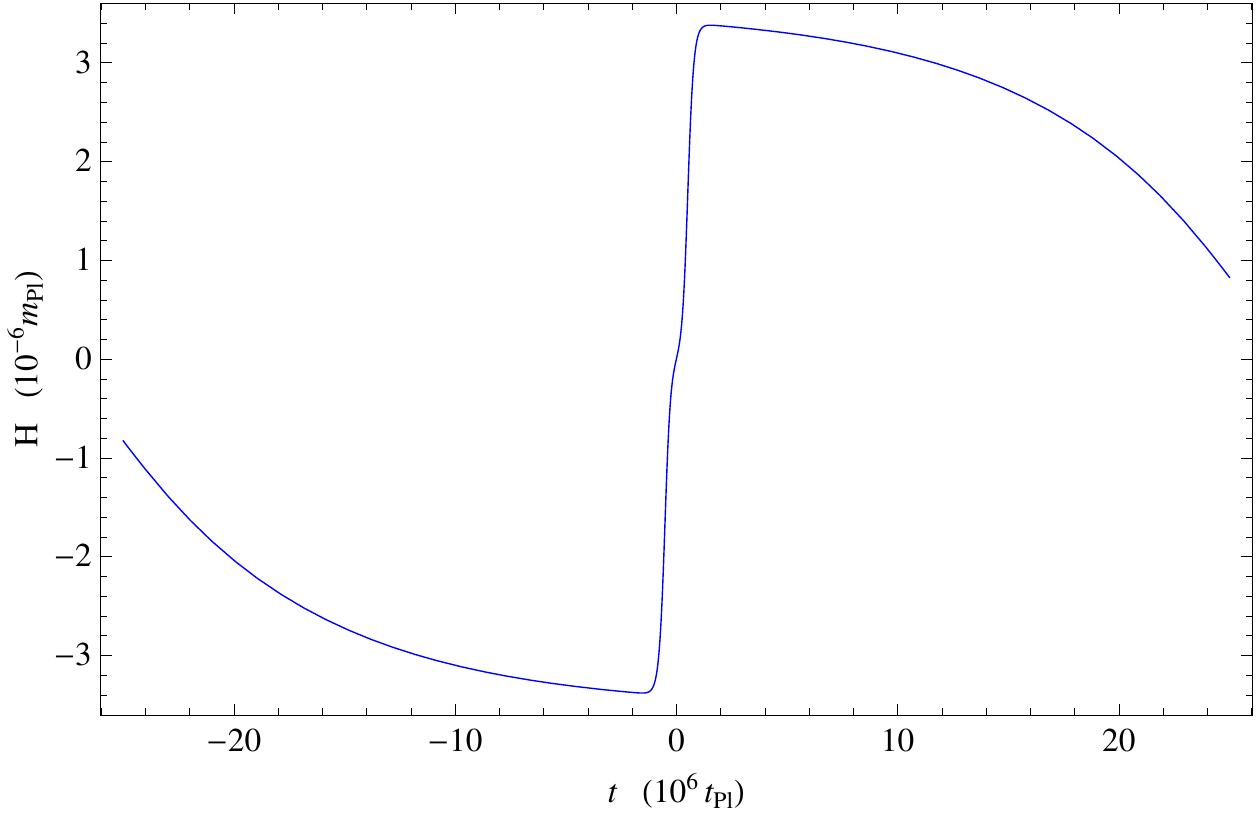}\hspace{0.25cm}\includegraphics[width=4.8cm]{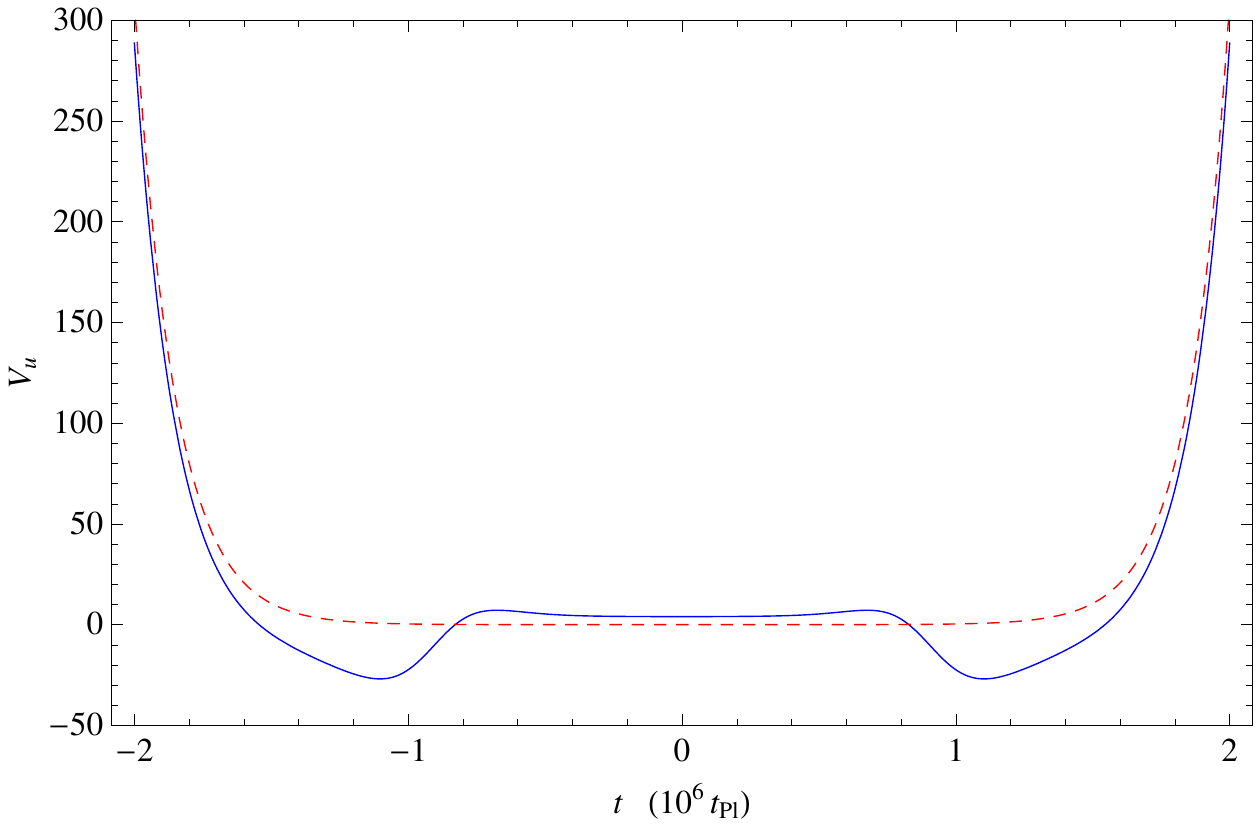}
\vspace{0.075cm}
\includegraphics[width=4.8cm]{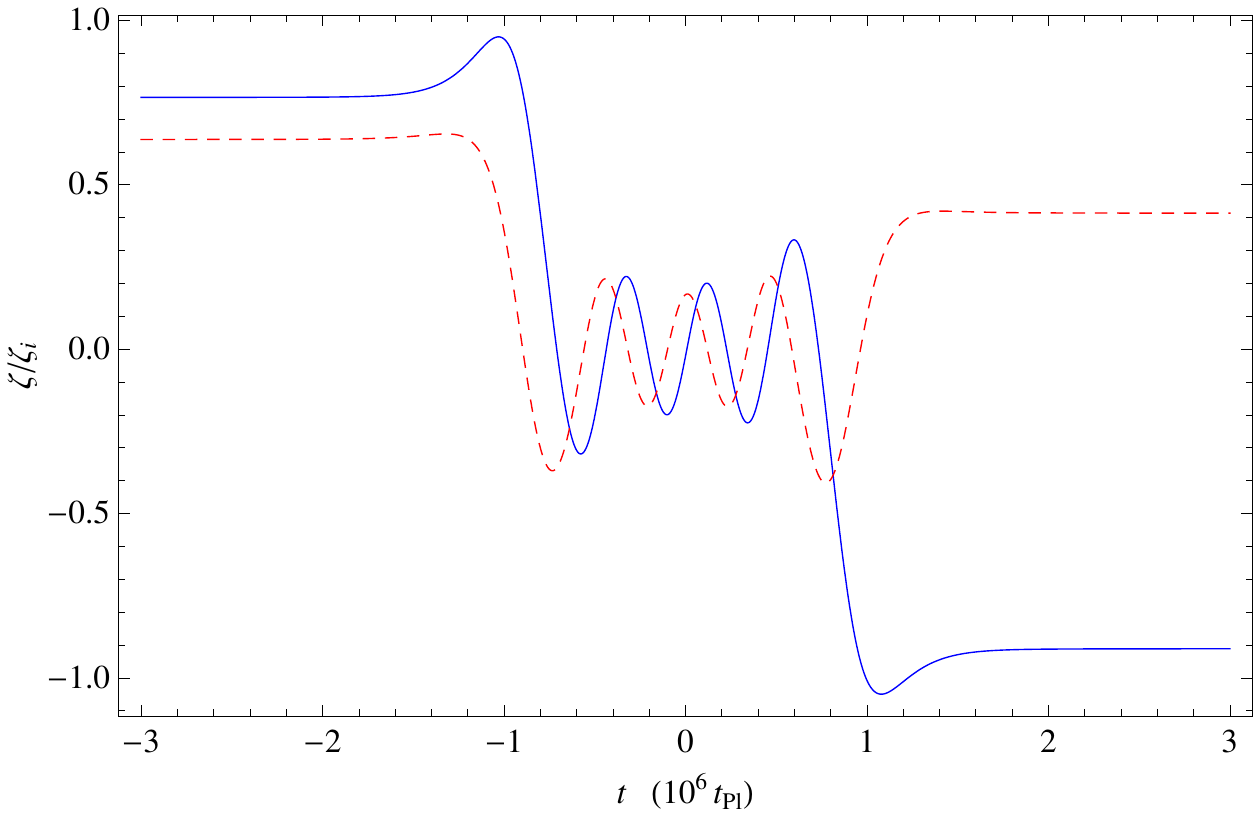}\hspace{0.25cm}\includegraphics[width=4.8cm]{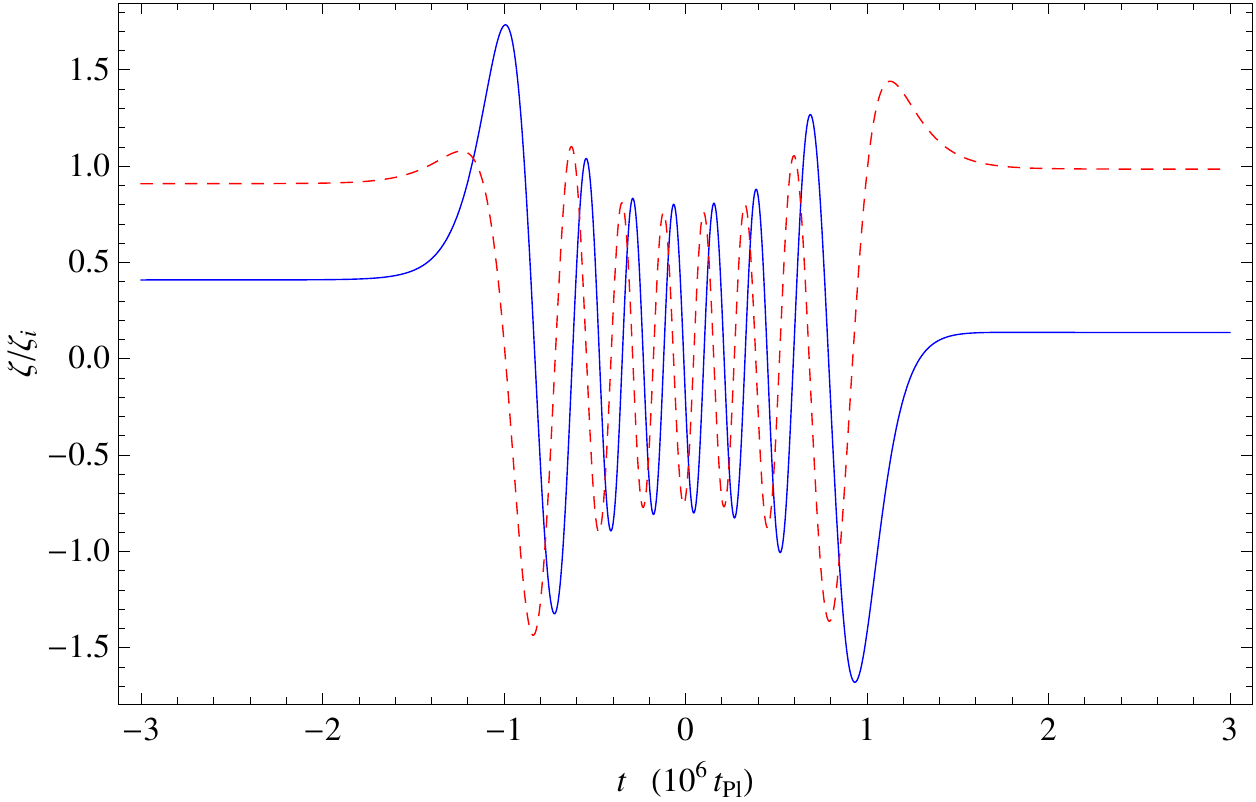}\hspace{0.25cm}\includegraphics[width=4.8cm]{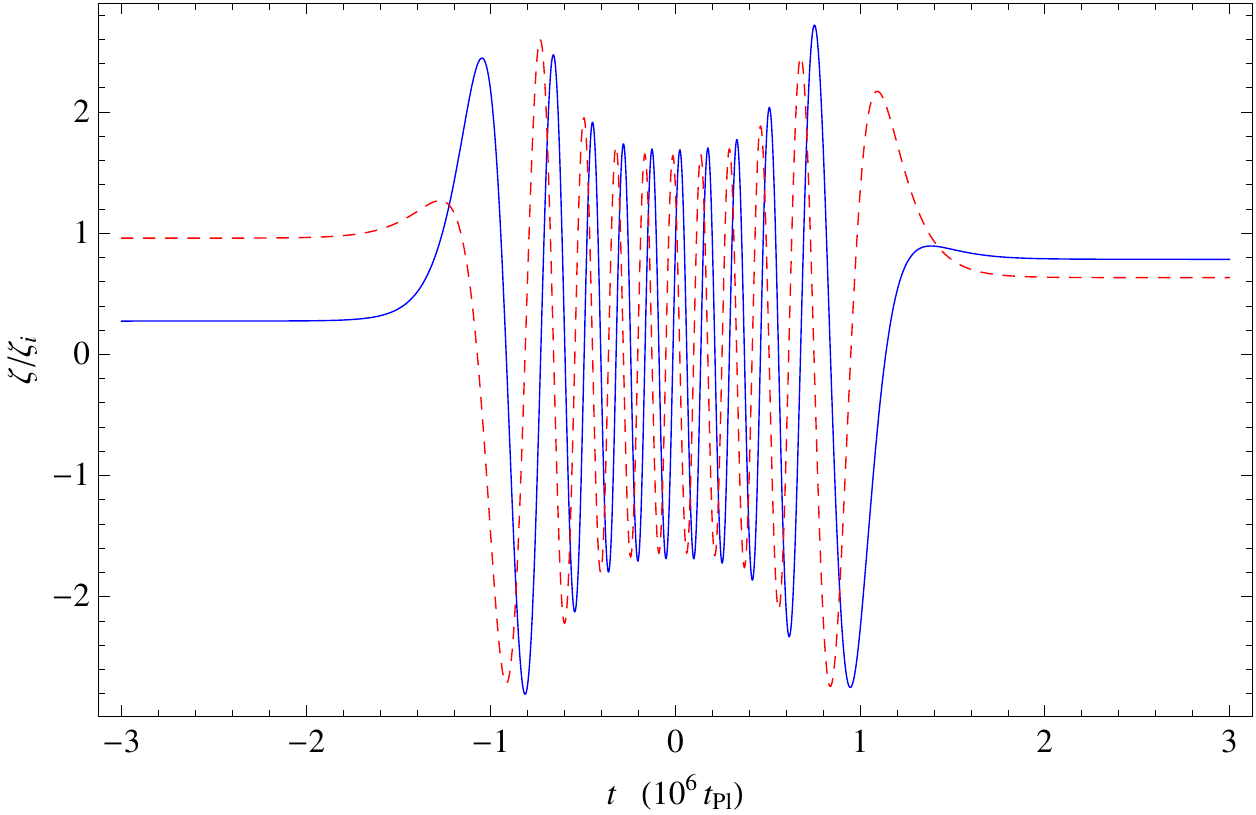}
\vspace{0.075cm}
\includegraphics[width=4.8cm]{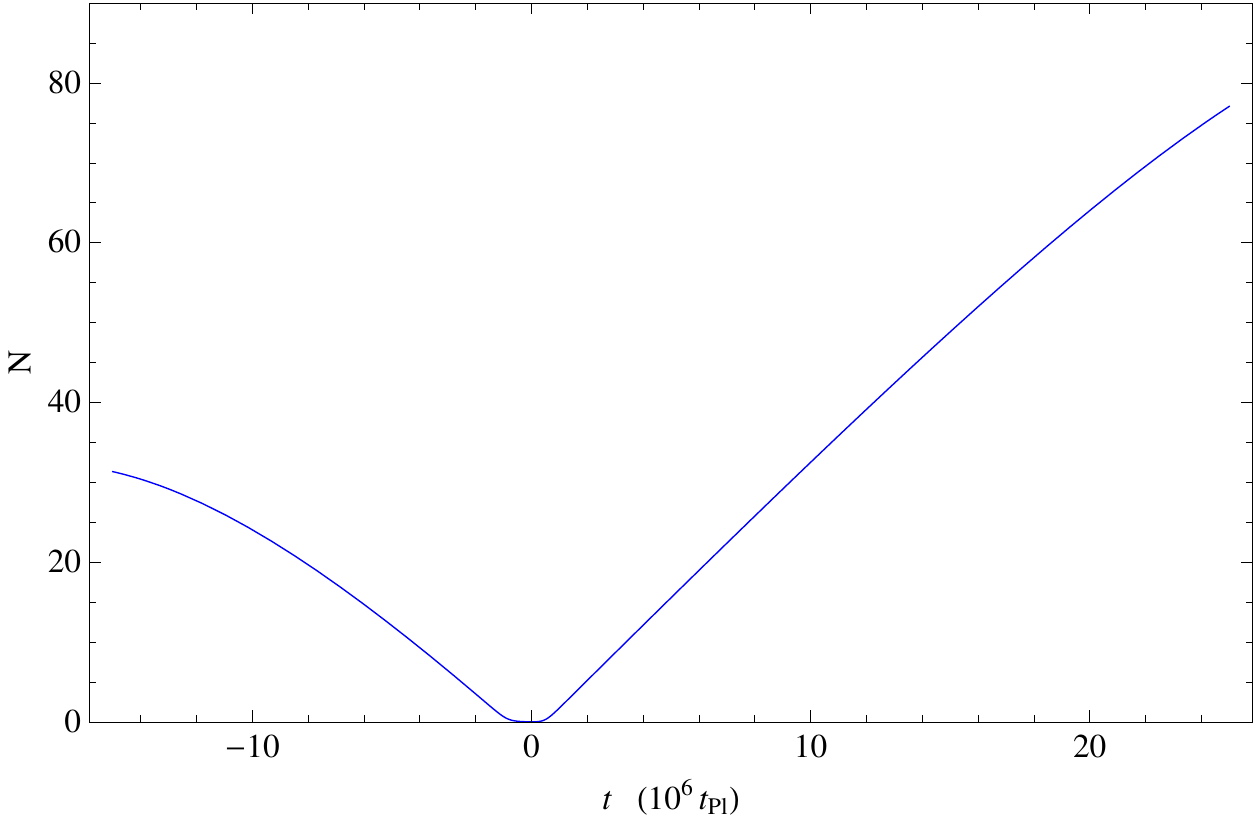}\hspace{0.25cm}\includegraphics[width=4.8cm]{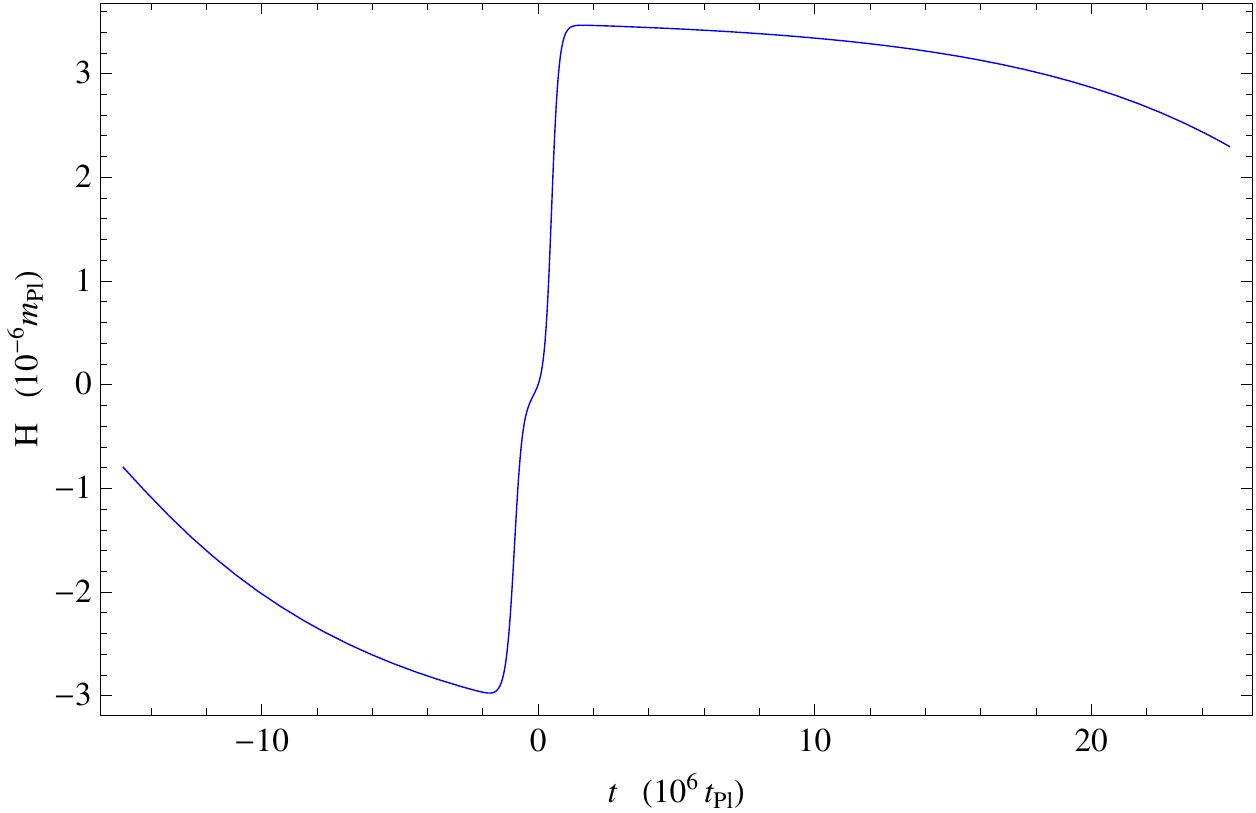}\hspace{0.25cm}\includegraphics[width=4.8cm]{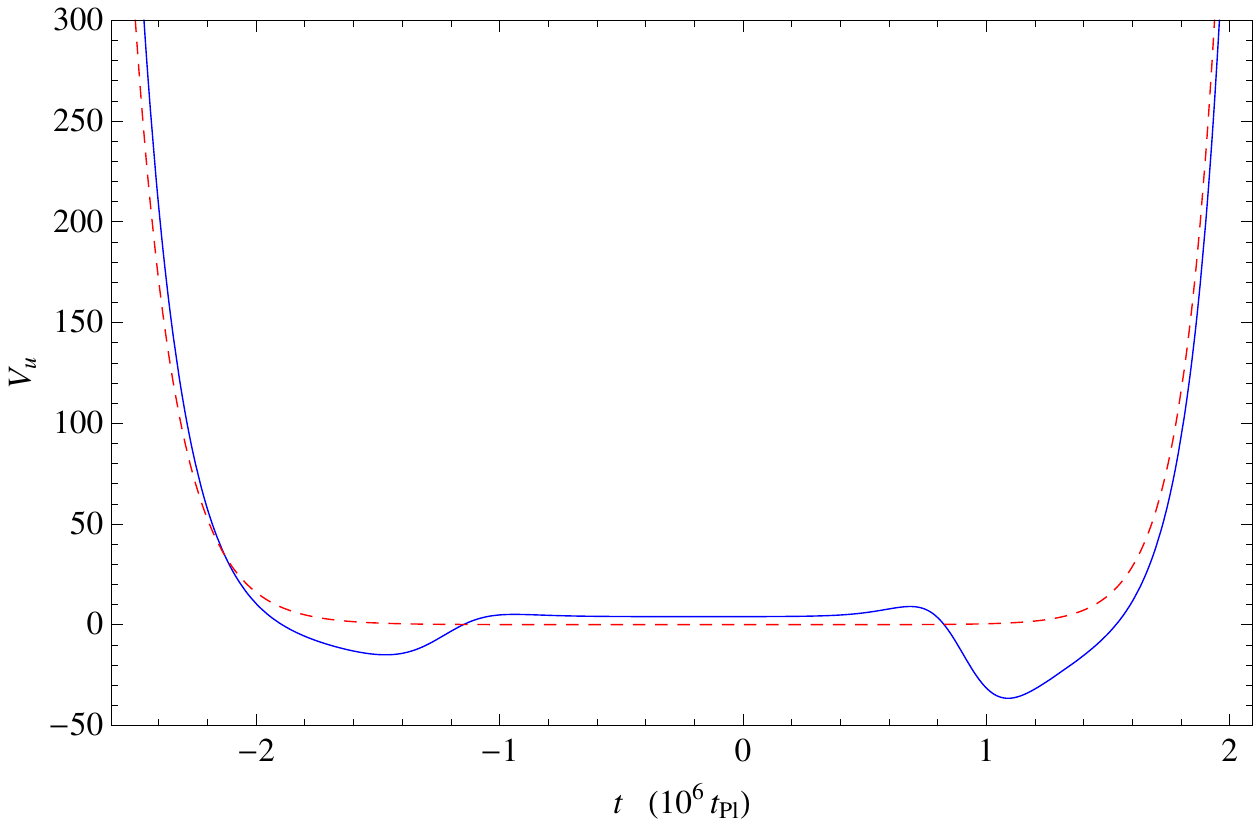}
\vspace{0.075cm}
\includegraphics[width=4.8cm]{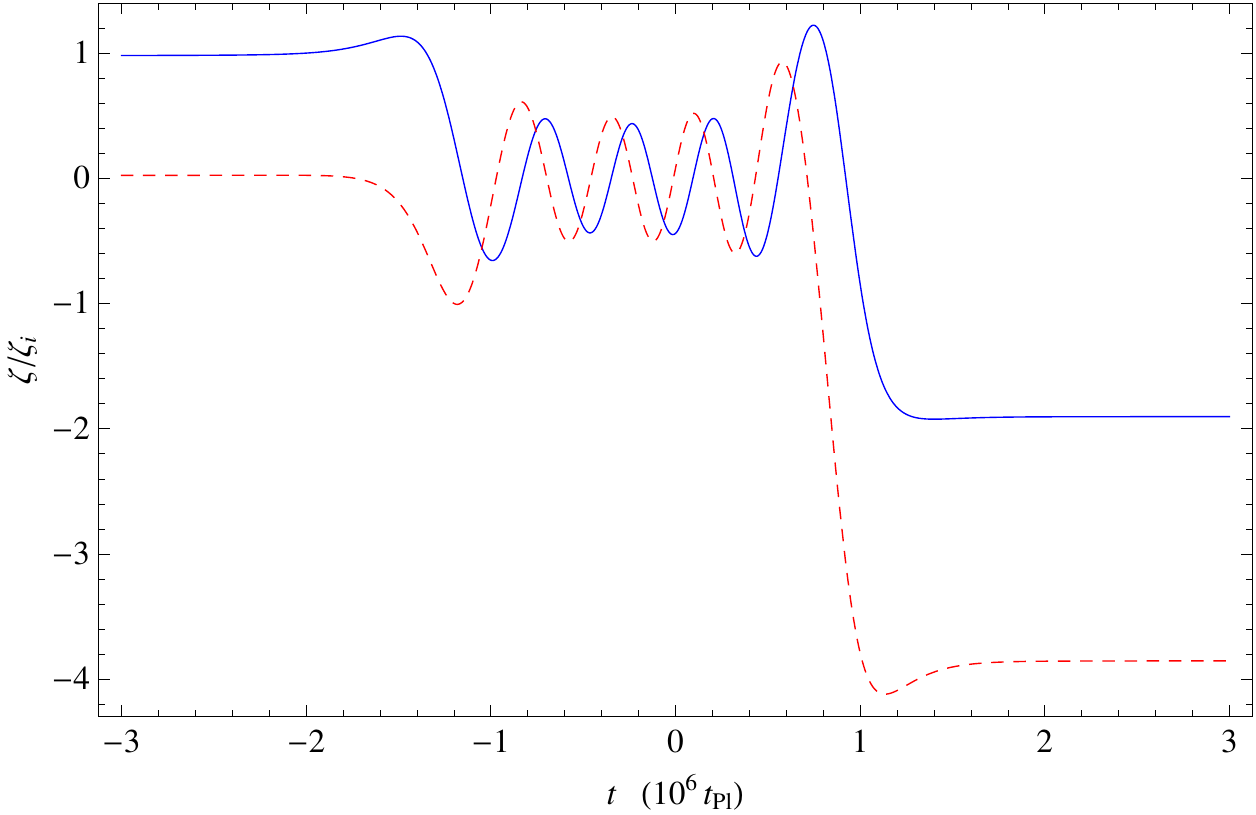}\hspace{0.25cm}\includegraphics[width=4.8cm]{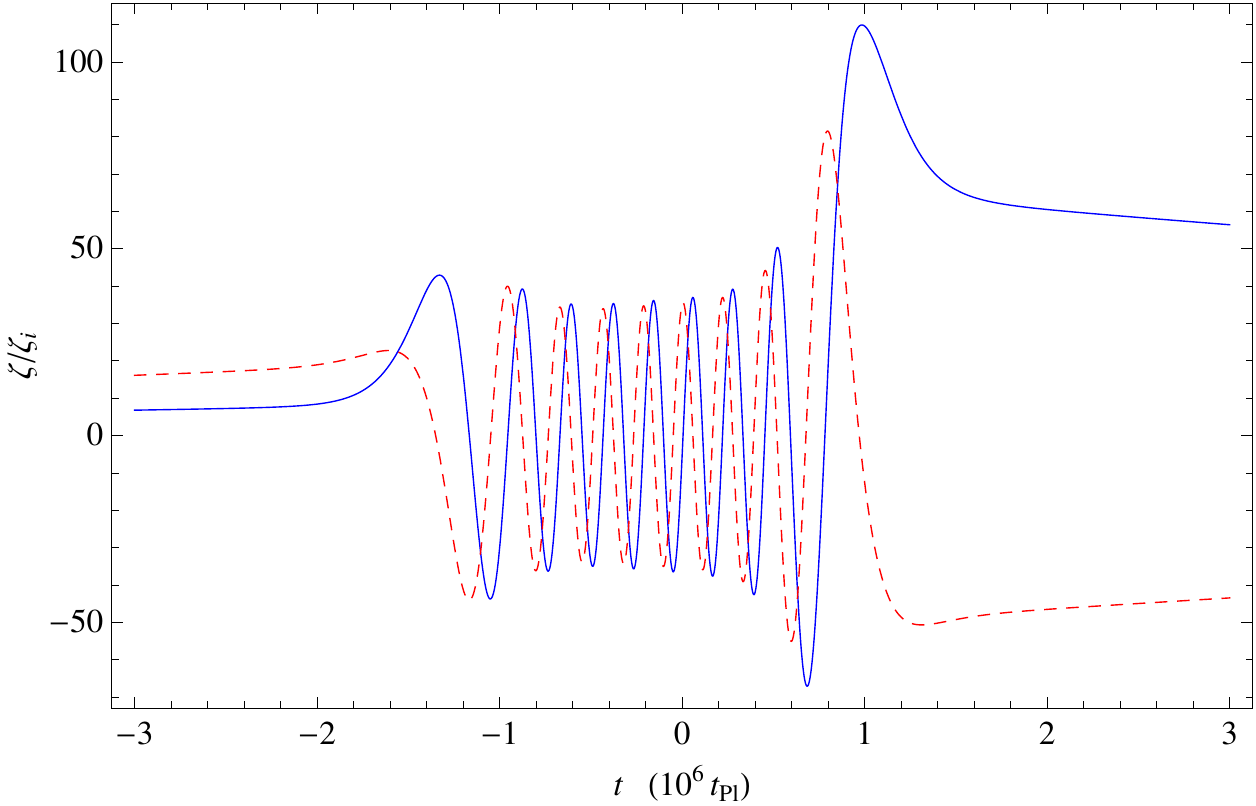}\hspace{0.25cm}\includegraphics[width=4.8cm]{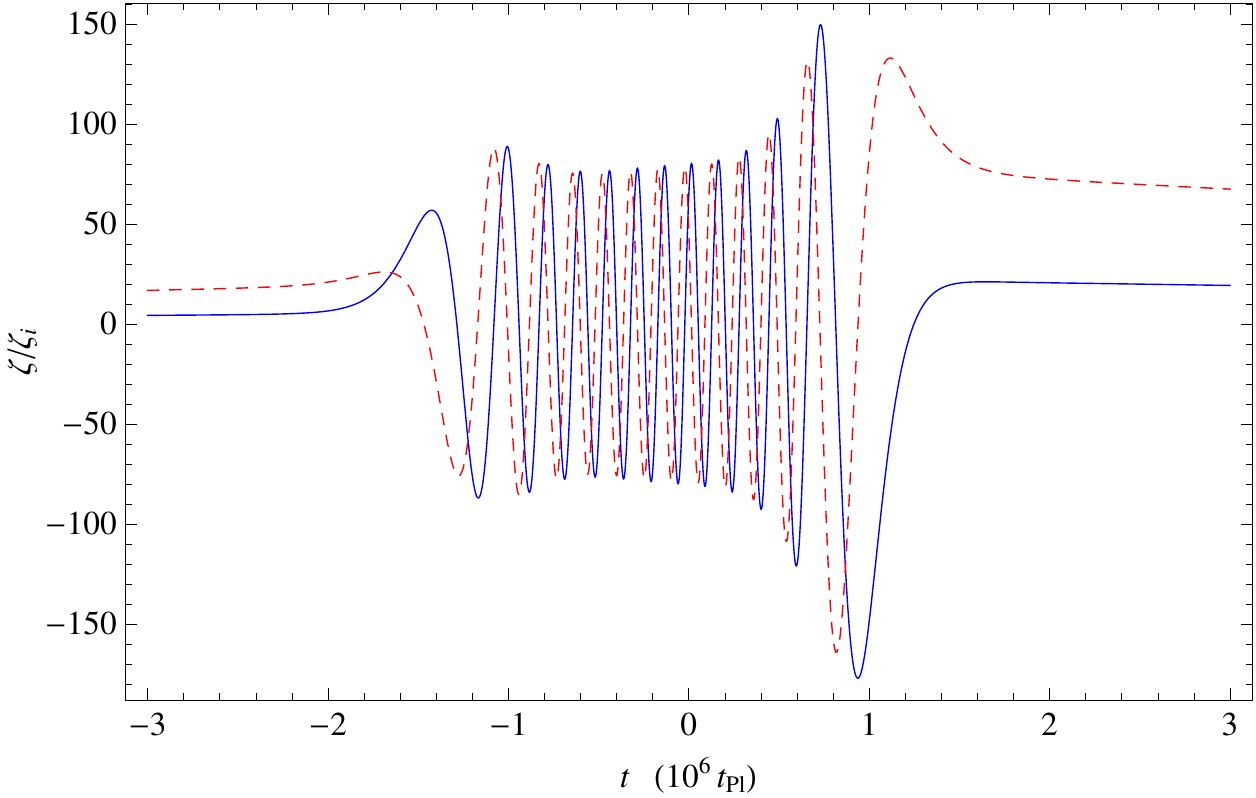}
\vspace{0.075cm}
\includegraphics[width=4.8cm]{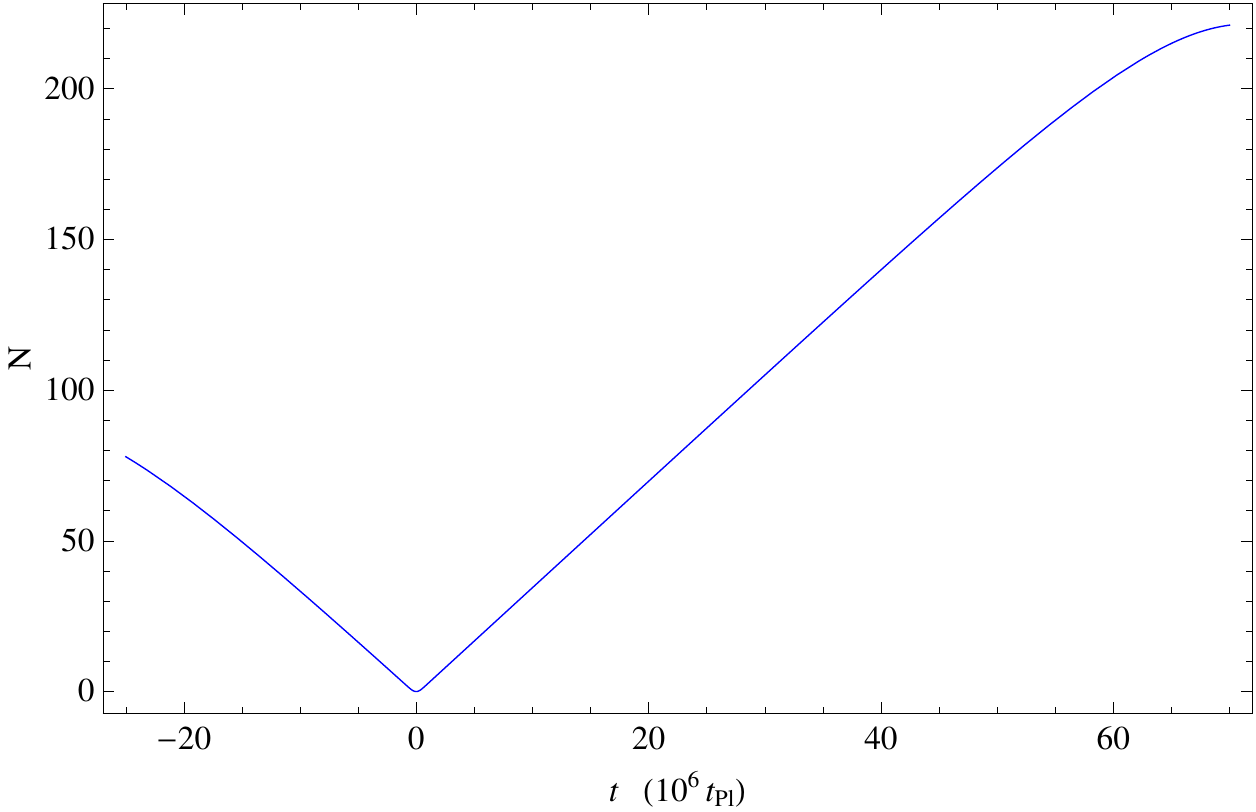}\hspace{0.25cm}\includegraphics[width=4.8cm]{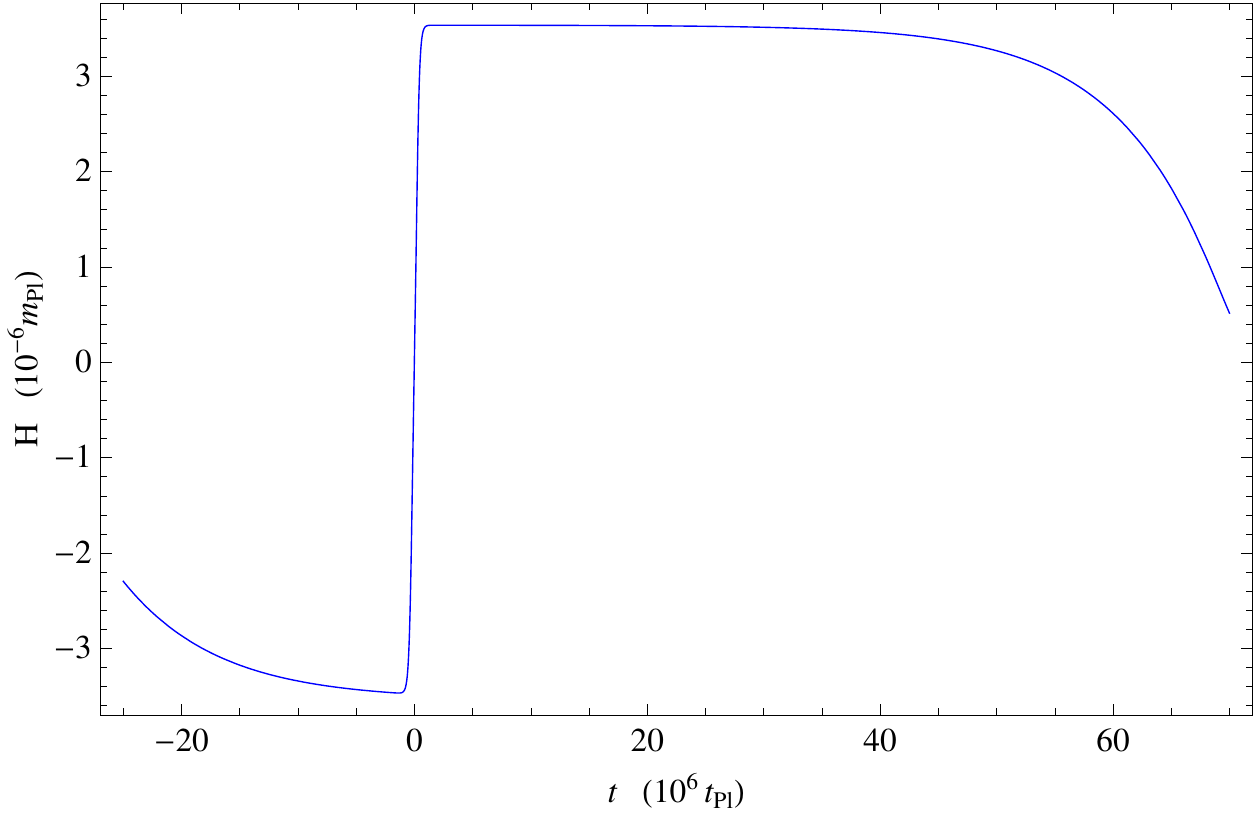}\hspace{0.25cm}\includegraphics[width=4.8cm]{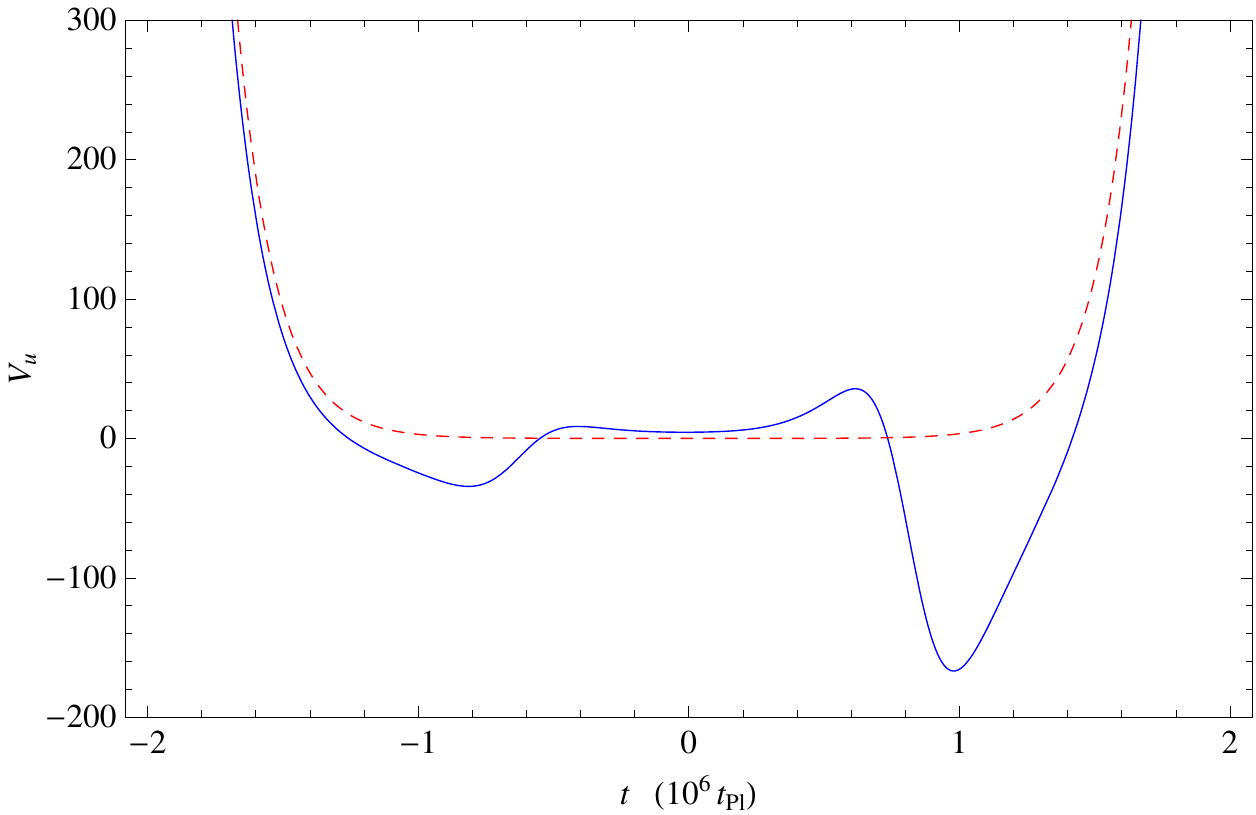}
\vspace{0.075cm}
\includegraphics[width=4.8cm]{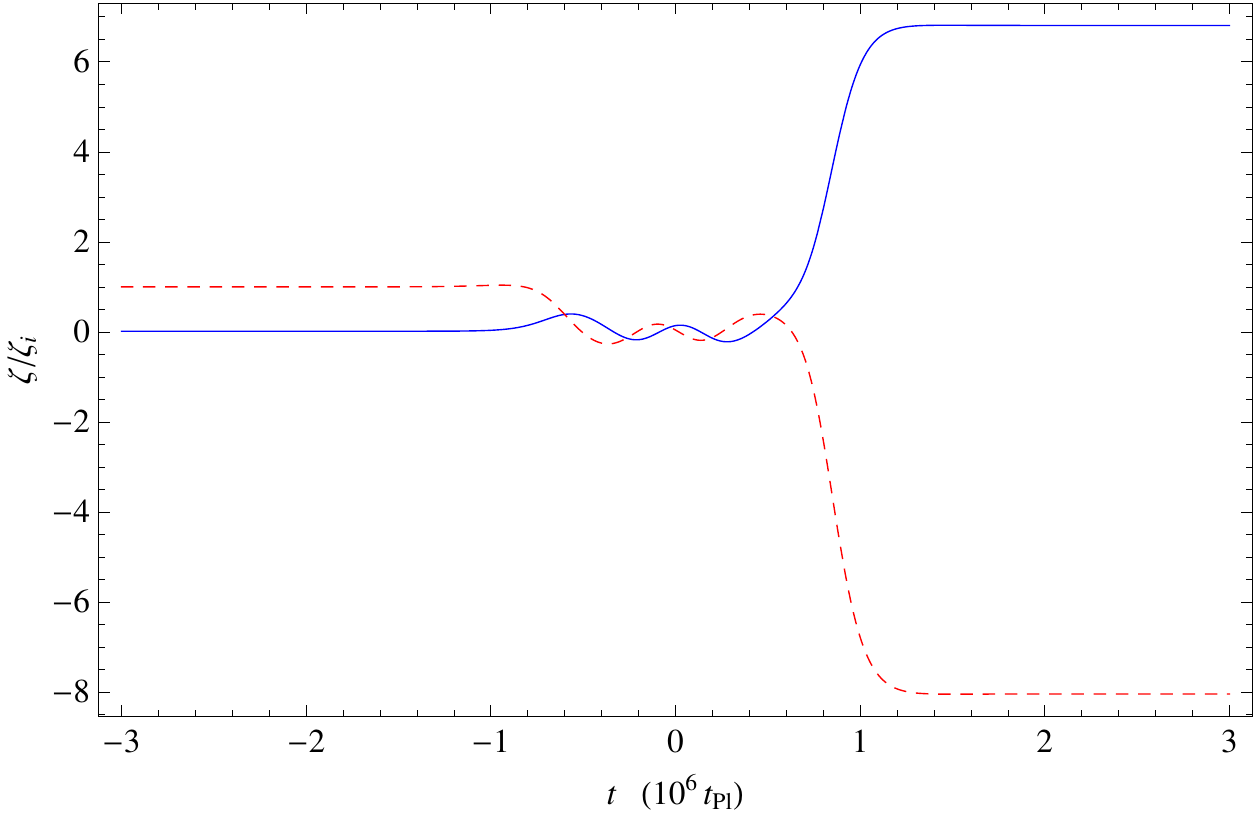}\hspace{0.25cm}\includegraphics[width=4.8cm]{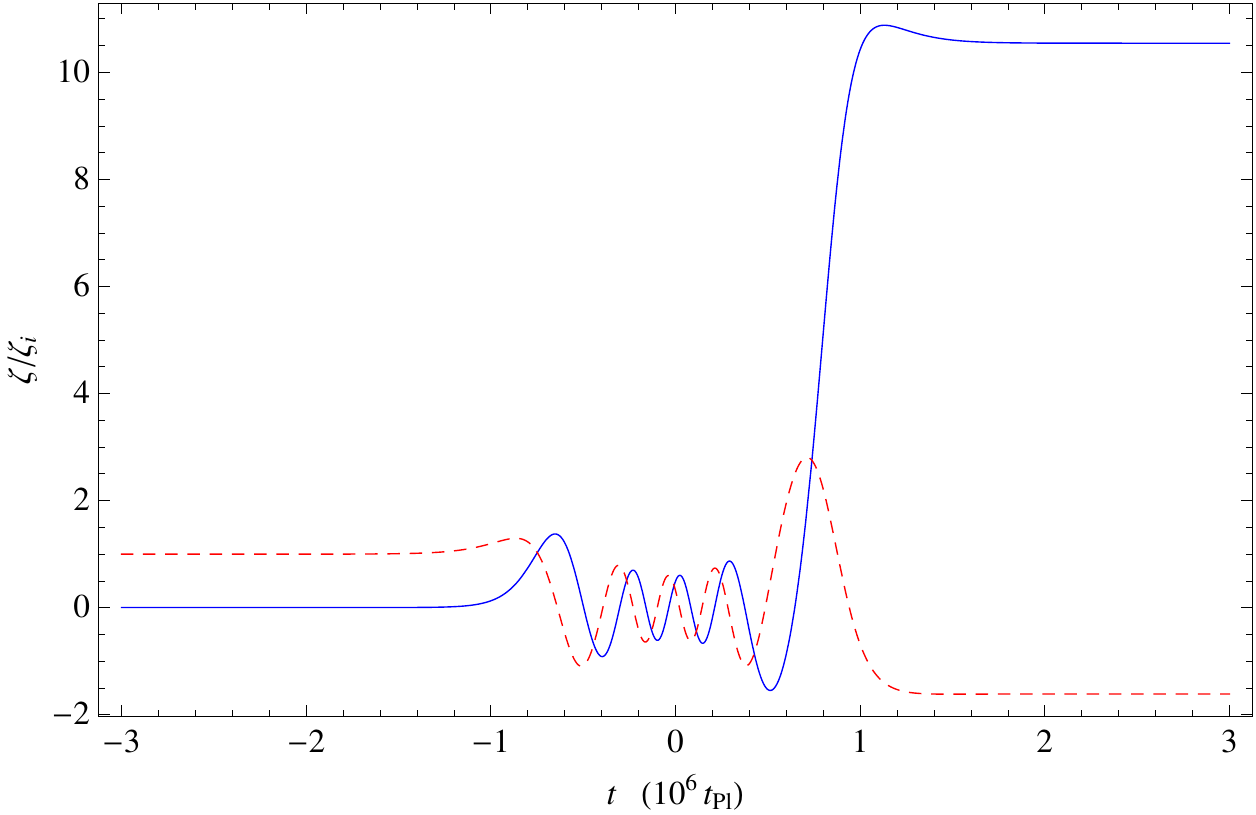}\hspace{0.25cm}\includegraphics[width=4.8cm]{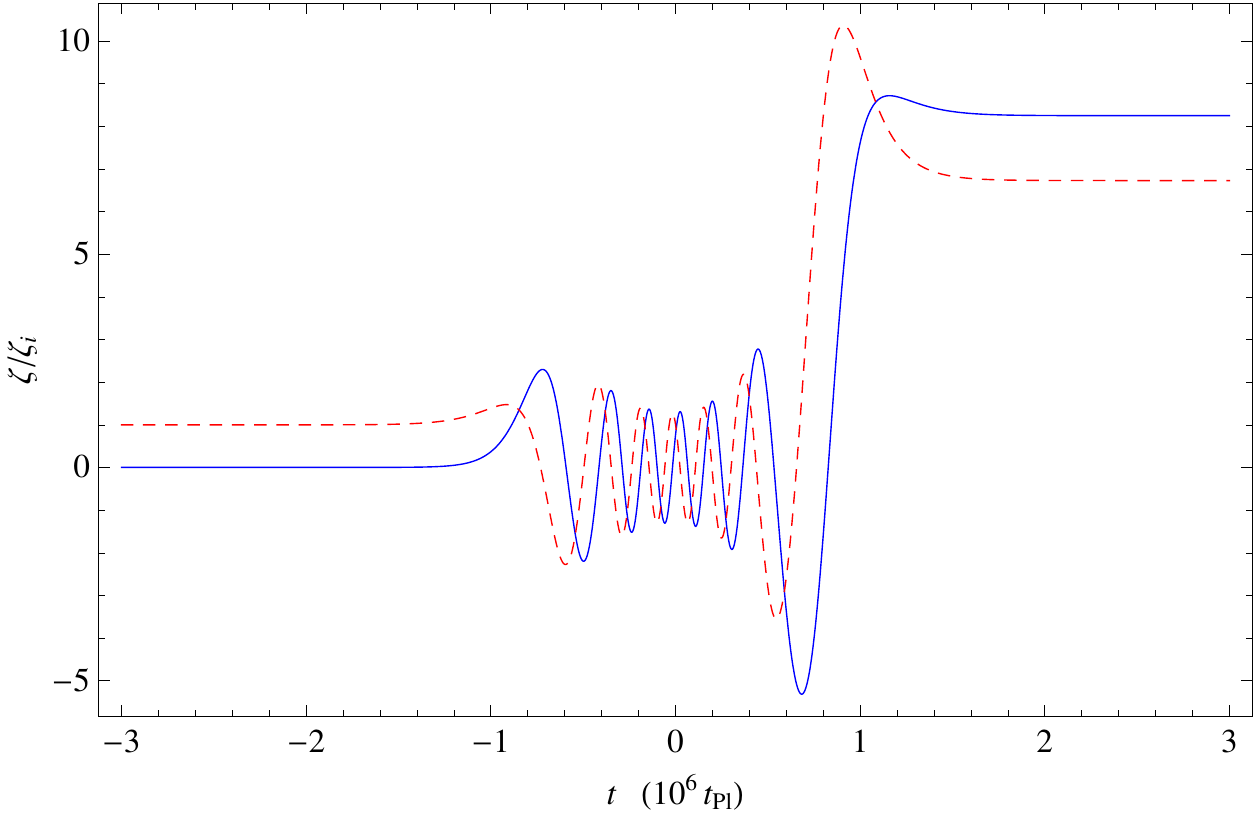}
\caption{From left to right, number of $e$-foldings $N$, Hubble parameter $H$, potential $V_u$ [true potential (blue), approximate potential (red dashed)] in rows 1, 3 and 5.  The normalized curvature perturbation $\zeta/\zeta_i$ [real part (blue), imaginary part (red dashed)] for $n=5,\,10,\,15$ in rows 2, 4 and 6, with $\zeta_i=\zeta(t_i)$ the curvature perturbation at time $t_i$.  From top to bottom : $\Upsilon=2.05$, $\varphi_{\mrm{b}}=0$ (top rows), $\Upsilon=2.05$, $\varphi_{\mrm{b}}=-\mpl/6$ (center rows), and $\Upsilon=2.5$, $\varphi_{\mrm{b}}=-\mpl/6$ (bottom rows).}
\label{fig:simulations}}
\subsection{Growth of perturbations in the contracting phase}
\label{sub:zetagrowth}
We now discuss the issue of the growth of perturbations in contracting spacetimes as discussed in Ref.~\cite{Creminelli:2004jg}, see also \cite{Brandenberger:2009rs}.  Let us consider the simple example of power-law contraction of the universe in which $a(\eta)\sim X^{p/(1-p)}$ where $p=2/3(1+w)$ is positive and $X=|\eta-\eta_{\mrm{c}}|$ for some characteristic conformal time $\eta_{\mrm{c}}<0$.  If $p<1$, $\eta<\eta_{\mrm{c}}$ and if $p>1$, $\eta>\eta_{\mrm{c}}$ so that for any $p>0$ the universe contracts as $\eta$ goes from $-\infty$ to $0$.  In the power law case, it is easy to solve the evolution equations for either $\Phi$ or $\mcl R$.  The solutions can be expressed, on super-Hubble scales, as the sum of two modes, one constant and one time-dependent,
\be
\Phi_k\sim\displaystyle C_{\Phi,1} X^{(p+1)/(p-1)}+ C_{\Phi,2}
\ee
and
\be
\mcl R_k\sim\displaystyle C_{\mcl R,1} X^{(3p-1)/(p-1)}+C_{\mcl R,2}\,.
\ee
When $\eta\rightarrow 0$ as the bounce is approched, in a matter- or radiation-dominated contracting phase, $p=2/3$ or $p=1/2$, the $C_{\Phi,1}$ mode diverges as $X^{-5}$ and $X^{-3}$ respectively, while the $C_{\mcl R,1}$ mode diverges as $X^{-3}$ and $X^{-1}$ respectively.  If $p\simeq 0$ as in the ekpyrotic scenario~\cite{KOST01,Khoury:2001zk}, the $C_{\Phi,1}$ mode grows as $X^{-1-2p}$ but the $C_{\mcl R,1}$ decays as $X^{1-2p}$.  If $w<-1/3$, corresponding to $\ddot a>0$, then $p> 1$ and $X$ grows.  For $w\simeq -1$, $p\gg 1$ and the $C_{\Phi,1}$ and $C_{\mcl R,1}$ modes grow as $X^{1+2/p}$ and $X^{3+2/p}$ respectively.  However, in this case, $X\sim 1/a$ and $a\sim e^{Ht}$, so that the conformal time interval that a mode remains super-Hubble in this regime is exponentially small so that both modes remain approximately constant.\\

It is clear that the perturbation variables $\Phi_k$ and $\mcl R_k$ do not evolve with the same power in $X$.  Furthermore, they are related to each other through their derivatives, so that, generally, the time-dependent mode of $\Phi$ does not correspond to the time-dependent mode $\mcl R$, and vice versa.  The example given above also implies that in a given cosmological setup, perturbation theory may be preserved or the possibility that it be violated be alleviated provided the right choice of perturbation variable is made~\cite{Creminelli:2004jg,Allen:2004vz}.  In an explicit model where modes evolve in and out of the Hubble radius, and where the equation of state parameter varies, as is the case for scalar field matter, the issue of whether modes grow large or not evidently becomes more subtle.  Let us ask which variables must remain small. 
In longitudinal gauge, the scalar part of the perturbed metric is diagonal, $B=E=0$, so that the line element takes the form given in (\ref{eq:metric}).  It is legitimate to ask whether $\Psi$ and $\Phi$ should remain small, or whether it is $A$ and $C$ that should remain small.  If the latter choice is made, then the gauge must be suitably chosen so that $A$ and $C$ do remain small in the cosmological evolution being studied.  As in the example above, the longitudinal gauge may not be appropriate if $\Phi$ is a growing function of conformal time.  In order to decide whether there exists a gauge in which perturbation theory remains valid, it is interesting to look at the intrinsic curvature of spatial sections and at the extrinsic curvature of comoving surfaces.  The intrisic curvature of spatial sections in the comoving gauge is given by
\be
\,^{(3)}R=\frac{6 \mcl K}{a^2}+\frac{4}{a^2}\lb\Delta+3\mcl K\rb\mcl R,
\ee
while the scalar part of the extrinsic curvature on comoving hypersurfaces is given by
\be
K_j^i=\frac{\mcl H}{a}\lb 1+2\mcl R\rb\delta_j^i+\lsb\frac{1}{a\mcl H}\partial_k\partial_j \mcl R-\frac{3\lb 1+w\rb}{2}\frac{\partial_k\partial_j}{\Delta}\mcl R'\rsb\delta^{ik},
\ee
where $\Delta^{-1}$ is the inverse Laplacian.  The evolution of each term in $\,^{(3)}R$ and $K_j^i$ as a function of $X$ is shown in Table \ref{tab:Xdep}.  In all cases but the ekpyrotic scenario, perturbative terms to $\,^{(3)}R$ and $K_j^i$ grow faster than background terms.  In order for perturbation theory to be preserved, the contracting phase must therefore be short -- this points to the study of strongly asymmetric bouncing cosmologies \cite{Brandenberger:2009rs} -- or the contracting phase must be induced by a potential for $\varphi$ giving rise to very large $w$; see however \cite{Barrow:2010rx}.  In this work, see Sec.~\ref{sub:initialstate}, we assume that the deflationary phase is short enough that perturbations remain small in the rapid contracting phase.   Figure \ref{fig:simulations} shows that this is in fact a realistic assumption. We further assume that the amplitude of fluctuations at the onset of the deflationary phase is small.  We thus choose initial conditions for the Mukhanov-Sasaki variable $v$ with an amplitude deviating only slightly from that of vacuum fluctuations, see Sec.~\ref{sub:initialstate}.
\TABLE{
\begin{tabular}{cccllllll}
Type & $p$ & $X$ & $\mcl K/a^2$ & $\mcl H/a$ & $\mcl R/a^2$ & $\mcl H\mcl R/a$ & $\mcl R/a\mcl H$ & $\mcl R'$\\
\hline\hline
Ekpyrotic & $\ll 1$ & $\searrow$ & $X^{-2p}$ & $X^{-1-p}$& $X^{1-4p}$ & $X^{-3p}$ & $X^{2-3p}$ & $X^{-2p}$ \\
Radiation & $1/2$ & $\searrow$ & $X^{-2}$ & $X^{-2}$ & $X^{-3}$ & $X^{-3}$ & $X^{-1}$ & $X^{-2}$ \\
Matter & $2/3$ & $\searrow$ & $X^{-4}$ & $X^{-3}$ & $X^{-7}$ & $X^{-6}$ & $X^{-4}$ & $X^{-4}$ \\
Exponential & $\gg 1$ & $\nearrow$ & $X^{2+2/p}$ & $X^{1/p}$ & $X^{5+4/p}$ & $X^{3+3/p}$ & $X^{5+3/p}$ & $X^{2+2/p}$\\
\hline
\end{tabular}
\caption{Super-Hubble conformal time evolution for each term in the intrinsic and extrinsic curvatures in the comoving gauge.}
\label{tab:Xdep}}
\end{appendix}
\end{document}